\documentclass[12pt]{article}

\usepackage[
  top    = 1in,
  bottom = 1in,
  left   = 1in,
  right  = 1in,
  marginparwidth = 0.5in
]{geometry}

\usepackage{setspace}
\usepackage[utf8]{inputenc}
\usepackage[english]{babel}

\usepackage{amsmath}
\usepackage{amssymb}
\usepackage{amsfonts}
\usepackage{amsthm}
\usepackage{mathrsfs}
\usepackage{mathtools}
\usepackage{bbm}

\theoremstyle{plain}

\theoremstyle{definition}

\theoremstyle{remark}

\usepackage{graphicx}          
\usepackage{epstopdf}
\usepackage{subcaption}
\usepackage{adjustbox}

\usepackage[dvipsnames]{xcolor}
\usepackage{color}
\usepackage{colortbl}

\definecolor{Gray}{gray}{0.95}
\definecolor{lavander}{cmyk}{0,0.48,0,0}
\definecolor{violet}{cmyk}{0.79,0.88,0,0}
\definecolor{burntorange}{cmyk}{0,0.52,1,0}
\def\lav{lavander!90}
\def\oran{orange!30}

\usepackage{booktabs}
\usepackage{tabularx}
\usepackage{multirow}
\usepackage{makecell}
\usepackage{longtable}
\usepackage{threeparttable}
\usepackage{pdflscape}

\usepackage{tikz}
\usepackage{tikz-3dplot}
\usetikzlibrary{
  arrows,
  arrows.meta,
  calc,
  chains,
  decorations.pathmorphing,
  decorations.pathreplacing,
  fit,
  matrix,
  positioning,
  shapes,
  shapes.geometric,
  shapes.symbols,
  backgrounds,
  trees
}

\tikzset{
  base/.style = {
    rectangle, rounded corners=4pt,
    text centered, font=\small,
    minimum height=1.1cm, text width=3.2cm, inner sep=6pt
  },
  annode/.style    = {base, fill=blue!10,   draw=blue!40,   thick},
  durnode/.style   = {base, fill=teal!10,   draw=teal!40,   thick},
  valnode/.style   = {base, fill=orange!10, draw=orange!40, thick},
  bynode/.style    = {base, fill=gray!8,    draw=gray!30,
                      text width=13.5cm, minimum height=0.7cm},
  bridgenode/.style= {base, fill=gray!8,    draw=gray!30,
                      text width=6cm,    minimum height=0.7cm},
  hdrnode/.style   = {base, minimum height=0.8cm, text width=13.5cm,
                      font=\small\bfseries, inner sep=5pt},
  anhdr/.style     = {hdrnode, fill=blue!15,   draw=blue!50,   thick},
  durhdr/.style    = {hdrnode, fill=teal!15,   draw=teal!50,   thick},
  valhdr/.style    = {hdrnode, fill=orange!15, draw=orange!50, thick},
  arr/.style       = {-{Stealth[length=5pt]}, thick,      gray!60},
  fatarr/.style    = {-{Stealth[length=6pt]}, very thick, gray!50},
  >=stealth',
  punktchain/.style = {
    rectangle, rounded corners, draw=black, very thick,
    text width=10em, minimum height=3em, text centered, on chain
  },
  line/.style    = {draw, thick, <-},
  element/.style = {
    tape, top color=white, bottom color=blue!50!black!60!,
    minimum width=8em, draw=blue!40!black!90, very thick,
    text width=10em, minimum height=3.5em, text centered, on chain
  },
  every join/.style = {->, thick, shorten >=1pt},
  decoration = {brace},
  tuborg/.style  = {decorate},
  tubnode/.style = {midway, right=2pt},
}

\makeatletter
\pgfdeclareshape{datastore}{
  \inheritsavedanchors[from=rectangle]
  \inheritanchorborder[from=rectangle]
  \inheritanchor[from=rectangle]{center}
  \inheritanchor[from=rectangle]{base}
  \inheritanchor[from=rectangle]{north}
  \inheritanchor[from=rectangle]{north east}
  \inheritanchor[from=rectangle]{east}
  \inheritanchor[from=rectangle]{south east}
  \inheritanchor[from=rectangle]{south}
  \inheritanchor[from=rectangle]{south west}
  \inheritanchor[from=rectangle]{west}
  \inheritanchor[from=rectangle]{north west}
  \backgroundpath{
    \southwest \pgf@xa=\pgf@x \pgf@ya=\pgf@y
    \northeast \pgf@xb=\pgf@x \pgf@yb=\pgf@y
    \pgfpathmoveto{\pgfpoint{\pgf@xa}{\pgf@ya}}
    \pgfpathlineto{\pgfpoint{\pgf@xb}{\pgf@ya}}
    \pgfpathmoveto{\pgfpoint{\pgf@xa}{\pgf@yb}}
    \pgfpathlineto{\pgfpoint{\pgf@xb}{\pgf@yb}}
  }
}
\makeatother

\tikzstyle{peers} = [
  draw, circle, violet,
  bottom color=\lav, top color=white,
  text=violet, minimum width=10pt
]
\tikzstyle{superpeers} = [
  draw, circle, burntorange,
  left color=\oran,
  text=violet, minimum width=20pt
]
\tikzstyle{legendsp} = [
  rectangle, draw, burntorange, rounded corners, thin,
  bottom color=\oran, top color=white,
  text=burntorange, minimum width=2.5cm
]
\tikzstyle{legendp} = [
  rectangle, draw, violet, rounded corners, thin,
  bottom color=\lav, top color=white,
  text=violet, minimum width=2.5cm
]
\tikzstyle{legend_general} = [
  rectangle, rounded corners, thin, burntorange,
  fill=white, draw, text=violet,
  minimum width=2.5cm, minimum height=0.8cm
]

\usepackage{pstricks}
\usepackage{sgame}
\usepackage{egameps}
\usepackage{dlfltxbcodetips}

\usepackage{fancyhdr}
\usepackage{enumerate}
\usepackage{comment}
\usepackage{lscape}
\usepackage{lineno}
\usepackage{soul}
\usepackage[titletoc,title]{appendix}

\usepackage[round]{natbib}
\usepackage[
  colorlinks = true,
  linkcolor  = red,
  urlcolor   = blue,
  citecolor  = blue
]{hyperref}
\usepackage[capitalize]{cleveref}

\newcommand{\E}{\mathbb{E}}

\makeatletter
\newcommand*\bigcdot{\mathpalette\bigcdot@{.5}}
\newcommand*\bigcdot@[2]{%
  \mathbin{\vcenter{\hbox{\scalebox{#2}{$\m@th#1\bullet$}}}}}
\makeatother

\newcommand{\disc}{\texttt{disc}}
\newcommand{\appl}{\texttt{appl}}

\newcommand{\appr}{\texttt{appr}}

\newcommand{\MKTCAP}{\texttt{mktcap}}
\newcommand{\phasei}{\texttt{phase I}}

\makeatletter
\newcommand\EightPtClose{\@setfontsize\EightPtClose\@viiipt{9}}
\newcommand\TenPtType{\@setfontsize\TenPtType\@xpt\@xiipt}
\def\notesize{\EightPtClose}
\newenvironment{figurenotes}[1][\vspace{1em}Note]{%
  \begin{minipage}[t]{\linewidth}\notesize{\itshape#1: }}{%
  \end{minipage}}
\makeatother

\begin{document}

\title{Valuing Pharmaceutical Drug Innovations\thanks{We thank Lam K. Bui, Will K. Nehrboss, Ishaan Dey and You Wang for outstanding research assistance, and graduate and undergraduate students at UVA enrolled in empirical industrial organization classes for helpful comments. We also thank Bart Hamilton, Christopher Adams, Paco Buera, Paul Mahoney, Derek Lemoine, Maxim Engers, Tim Simcoe, Nicolae Garleanu, Rebekah Dix, Rena M. Conti and the participants and discussants at the UVA, UMass-Amherst, SEA 2021, MaCCI 2022, ASHEcon 2022, EARIE 2022, SciencesPo, Indiana University, BU TPRI, NASM-ES 2023, MaCCI-EPoS: Frontiers in Empirical IO 2023, JHU, IIM-B, 2024 Health Economics Conference (Toulouse School of Economics), 2025 Conference on the Economics and Finance of Healthcare and Medicine (WashU Olin), IIOC 2026 and IP Day 2026 (BU Questrom). We acknowledge financial support from the Batten Research Grant Program Quantitative Collaborative and the Bankard Fund for Political Economy at UVA.}}

\author{Gaurab Aryal\thanks{ Department of Economics, Boston University, \href{mailto:aryalg@bu.edu}{aryalg@bu.edu}}\hspace{0.3in}
 Federico Ciliberto\thanks{ Department of Economics, University of Virginia, DIW and CEPR, \href{mailto:ciliberto@virginia.edu}{ ciliberto@virginia.edu}}\\ 
 Leland E. Farmer\thanks{ Department of Economics, University of Virginia, \href{mailto:lefarmer@virginia.edu}{ lefarmer@virginia.edu}}\hspace{0.3in}
 Ekaterina Khmelnitskaya\thanks{Sauder School of Business, University of British Columbia, \href{mailto:ekaterina.khmelnitskaya@sauder.ubc.ca}{ekaterina.khmelnitskaya@sauder.ubc.ca}}
}
\date{\today\\
}

\maketitle

\begin{abstract}
\begin{spacing}{1}
We propose a methodology to estimate the market value of pharmaceutical drugs. Our approach combines the event study method with a discounted cash flow model that infers drug values from stock market responses to drug development announcements. We estimate the average value of a drug developed by small firms (those below the 95th percentile of market capitalization) to be \$2.16 billion. At the preclinical stage, the risk-adjusted and present discounted average net value of drugs is \$50 million. Leveraging these estimates, we also determine the expected drug development cost at the start of the discovery stage to be \$38 million. We estimate values and costs for several therapeutic areas (e.g., neoplasm, infections) and explore applying these estimates to design policies that support drug development through drug buyouts and targeted preclinical interventions.
\end{spacing}
\end{abstract}

\maketitle

\section{Introduction}
 
Measuring the value of pharmaceutical drug innovations is essential for determining appropriate incentives for innovators, assessing the effectiveness of current policies, and guiding industry decision-making. Specific estimates of drug values are of particular interest because they often affect policy choices. However, research on drug valuation is limited, likely due to a lack of drug-specific R\&D expenditure data and the challenges of modeling complex, drawn-out drug development processes.
 
To address these issues, we develop a systematic two-step methodology to determine the market value of drugs by combining the \textit{event study} approach with a \textit{discounted cash flow} model. Using our approach, we estimate drug value from stock market reactions to drug development announcements across a large sample of drugs under development. While event study and discounted cash flow models are not new, we show that combining them enables us to estimate the market value of drugs using publicly available data.\footnote{For some applications of the event study method, see \cite{WhinstonCollins1992, Hwang2013, KoganPapanikolaouSeruStoffman2017, LangerLemoine2020, RubioTurnerWilliams2020} and \cite{SinghRocafortCaiSiahLo2022}.}
 
Our objective is to produce reliable estimates of the market value of drugs that are useful to firms and policymakers to support drug development, e.g., drug buyouts or cost-sharing, which we discuss later.\footnote{We note that throughout the paper we use the term ``value of a drug'' to mean the market value of the drug, and not the social value of a drug, which also includes the value of its health benefits to society.} As we discuss below, having reliable estimates of values will help us select drug development policies that balance societal benefits with costs while providing incentives for firms to invest in R\&D to address the pharmaceutical industry's ``grand challenge'' \citep{Pauletal2010}.

We apply our methodology to the Cortellis dataset from Clarivate, which covers more than 70,000 drug candidates. We focus on drugs developed by publicly traded companies in the US and supplement the data with daily US stock prices and market capitalization data from the Center for Research in Security Prices. First, we estimate the change in the firm's market value following a drug development announcement. To that end, we use the \emph{unrestricted market model} to estimate the expected \emph{cumulative abnormal returns} (CAR) associated with three types of announcements for which the announcement dates are genuine surprises--discovery (the start of preclinical research), discontinuation of discovery, and FDA approval--and apply the signal extraction approach of \cite{KoganPapanikolaouSeruStoffman2017} to filter out noise. Second, we rely on the semi-strong form of the \emph{efficient markets hypothesis} \citep{Fama1965, Samuelson1965, FamaFisherrJensenRoll1969}, which implies that the expected CAR measures the market's revised
estimate of the value created by the drug announcement.\footnote{We restrict attention to public firms because market values are unobservable for private firms. Appendix \ref{app:public_private} shows that the probability of success and durations are similar across public and private firms. Throughout the paper, we use ``the market," ``investors," and ``market participants" interchangeably as shorthand for the equilibrium price reaction implied by the semi-strong form of the \emph{efficient markets hypothesis}. Expressions such as ``the market expects," ``the market updates," or ``the market anticipates" should be read as referring to the joint behavior of competing traders whose orders aggregate into observed stock prices, not to a single strategic decision maker.}
 
To illustrate the intuition, consider a firm that announces FDA approval of a drug. The approval news resolves the uncertainty about the FDA's decision. Consequently, the change in the firm's market value after this single surprise news is the difference between the drug's full value (now that it has been approved) and its value just before the approval announcement, which equals the full value times the probability of approval conditional on submitting an application. Rearranging, we recover the value of an approved drug using the estimated CAR and the probability of success.

To value a drug at earlier stages, we go backward: we discount the value of an approved drug for the expected time to approval and the probability of success from that stage. We estimate the expected discount rate and the probability of success using a competing-risks model that accounts for both observed and unobserved heterogeneity. The observed heterogeneity includes firms' market size, therapeutic area (e.g., neoplasm vs. infections), and decade (to capture changes in science). Specifically, we estimate a stage-specific Gompertz proportional-hazards model with firm-level frailty to capture unobserved heterogeneity, using a duration sample of more than 13,000 drug-indication-stage spells.

In our analysis, we distinguish between large-cap firms (at or above the 95th percentile of market capitalization) and small-cap firms. This distinction matters because a drug can represent a much larger fraction of a small firm's value than of a large pharmaceutical firm, which can develop several drugs at the same time. The firm size affects, among other things, the probability of success and the time to success. Our preferred sample is drugs developed by small firms, because for these firms the drug in question is typically the \textit{focal} drug, allowing us to identify its value. For a large, multi-product firm, by contrast, a single drug announcement may reflect factors beyond the focal drug, including spillovers across its pipeline. Furthermore, there are only 16 large firms, so the large-firm estimates are noisier.

For small firms, we estimate the gross profit value of an approved drug to be \$2.16 billion. The present discounted and risk-adjusted value of expected gross profits at the discovery stage is \$88 million, and the average value of a drug at the discovery stage, net of present discounted development costs, is \$50 million. 

For large-firm drugs, we estimate \$20.8 billion in gross profit value, \$1.13 billion in risk-adjusted discovery-stage value, and \$824 million net of development costs, roughly an order of magnitude higher than the small-firm estimates. These large-firm estimates should be interpreted with caution. The gap relative to small firms is consistent with spillovers across large firms' pipelines, and also with them developing higher-value drugs on average.

A key advantage of our approach is that we recover the implied expected cost of drug development without observing R\&D expenditures. Because the market's valuation at discovery already embeds its assessment of future costs, the difference between the gross profit value (what an approved drug is worth, risk-adjusted back to the discovery stage) and the net value at discovery identifies total expected development costs. These costs are option-adjusted because they reflect that firms abandon unsuccessful candidates rather than committing all expenditures upfront. The adjustment is quantitatively important because we estimate that only 6.6\% and 8.5\% of drug candidates developed by small-cap and large-cap firms, respectively, ever receive FDA approval, so most candidates never incur late-stage costs. For small firms, the average option-adjusted development cost is \$38 million, but for large firms, it is \$305 million.\footnote{The sample of drugs differs across announcements because some drugs are discontinued, and some appear only in later announcements. So, we can estimate only the average stage-specific development costs. \label{footnote:sample}}

We assess internal consistency using a ``negative'' event: discontinuation 
announcements for drugs at the preclinical stage. The logic is that these 
announcements provide an independent estimate of the discovery-stage cost, 
which should be a small fraction of total costs because late-stage trials 
dominate development expenditures. For small firms, we estimate the preclinical cost at approximately \$19 million, 
around half of the total \$38 million development cost. This share is somewhat 
above existing estimates from \citet{DiMasi2016} and \citet{Sertkayaetal2024}, 
who both place preclinical costs at around 30\% of total expected drug 
development cost, though our 95\% confidence interval $[0, \$53\text{M}]$ 
encompasses the benchmark-implied magnitudes. For 
large firms, the consistency check fails, reinforcing our concern that 
single-drug announcements at multi-product firms also reflect pipeline 
spillovers. By providing market-based, risk-adjusted cost estimates from a representative sample of drugs, we complement the existing literature, which relies either on confidential surveys of a few firms \citep{DiMasi2016} or the accounting cost of a single trial \citep{SertkayaWongJessupBeleche2016}.

Disaggregating by therapeutic area reveals heterogeneity in both the level and the composition of drug value. For small-firm drugs, the gross profit at approval ranges from \$1.4 billion for cardiovascular drugs to \$7.1 billion for rare diseases and \$6.6 billion for neoplasms, likely reflecting the pricing power of oncology and orphan drugs. High approval profits, however, come with high development costs. For rare diseases and neoplasms, implied costs of \$389 million and \$208 million are roughly 85\% and 81\% of the present-discounted gross profit at discovery, leaving a net discovery value of only \$70 million and \$47 million, respectively. Infectious diseases follow a similar pattern: \$2.2 billion in approval profits, \$137 million in costs, and only 22\% of the present-discounted gross profit at discovery retained as net value. At the lower end, inflammatory and gastrointestinal drugs generate approval profits of \$1.4 and \$1.5 billion, respectively, with correspondingly modest implied costs of \$4 and \$3 million. Neurological drugs are an interesting intermediate case: \$2.8 billion in approval profits — third-highest in our sample — but only \$32 million in implied costs. For large firms, the ranking of implied costs across indications is broadly similar. While rare diseases, neoplasms, and infectious diseases have the highest development costs, the absolute levels are several times higher than those of smaller firms.

Finally, we use our estimates to inform the design of policies that support drug development \citep{Munos2009, PammolliMagazziniRiccaboni2011, Scannelletal2012}, focusing on drug buyouts at two endpoints of the development process: post-approval and discovery.\footnote{Although we focus on drug buyouts, our estimates can inform other policy tools such as cost-sharing agreements, where the government covers development costs in exchange for procurement commitments. However, to calibrate such agreements, we need stage-specific costs, which we do not use in all stages; we leave this application to future work.}

Under a drug buyout program, based on \cite{Kremer1998}'s patent buyouts, we envision the government purchasing a drug's manufacturing rights and making them public at a price equal to the drug's private value plus a markup, e.g., the social value.
A natural concern is that our estimates cannot 
simultaneously serve as valid measures of drug value \emph{and} as the 
basis for buyout payments. If firms anticipate universal buyouts, market 
reactions will reflect the expected payment rather than the underlying 
value of the drug, affecting our estimates, which is the critique of \cite{Lucas1976}. 
Our solution is that the government commits to implementing 
buyouts with a small (but exogenous) probability. 
Then the buyout is a surprise, and the observed market reactions can inform the value.

Furthermore, we can use our estimates to lower the costs of running this program than paying the 
social value of the drug as a markup. 
Instead, the policymaker can use the distribution of the estimated values to choose the 
take-it-or-leave-it price \citep{Myerson1981}: the government offers to buy at this fixed price, and the firm accepts if its private value exceeds it. 
We show that this approach lowers the expected government outlays 
relative to the social-value payment. 

A natural follow-up question is \emph{when} to apply the drug buyout: after the approval or at discovery. 
Post-approval buyouts are feasible but expensive. 
Discovery-stage buyouts are far 
cheaper than post-approval. Moreover, drug values increase monotonically across development stages, so the critical bottleneck lies at the discovery stage, where the viability of a drug candidate is the most uncertain. Support at this stage is more effective per dollar than after approval. 

However, an early-stage buyout raises a natural question: if the government acquires the drug at discovery, who funds and conducts the remaining development? We 
explore, among other instruments, the possibility of combining discovery-stage buyouts with an \emph{advanced market 
commitment} \citep{KremerGlennerster2004, KremerLevinSnyder2020PnP, 
KremerLevinSnyder2022}, under which the government pre-commits to purchase a fixed number of approved drugs at a pre-fixed price. This commitment serves as a prize for successful developers.

Our indication-specific estimates identify cardiovascular and inflammatory drugs as having the lowest approval-stage profits, which are also the areas that represent large unmet medical needs. These are 
the natural candidates for targeted buyout support, and our estimates 
provide the granularity needed to rank indications by the urgency and 
cost-effectiveness of intervention. This mapping from estimates 
to policy illustrates the value of our framework.

Our paper contributes to several strands of literature. First, it adds to the literature that estimates the value of innovations \citep[e.g.,][]{Griliches1981, Pakes1986, Austin1993, ChanLakonishokSougiannis2001, HallJaffeTrajtenberg2005, MunariOriani2011, AzoulayZivinLiSampat2018, McKeonPoegeSimcoe2022}. Unlike the extant literature on patents, we focus on products under development, particularly novel drugs, and propose a practical strategy to identify their commercial value. Our methodology can be applied to other regulated industries, such as medical devices, agrochemicals, and green technology sectors, where public announcements are available.
 
Second, our work complements \citet{SinghRocafortCaiSiahLo2022}, who apply event-study methodology to the stock market's reaction to clinical trial announcements in the pharmaceutical industry, but do not estimate drug values. Our broader empirical approach is closely related to \citet{KoganPapanikolaouSeruStoffman2017}, from whom we take the signal-extraction step with truncated-normal priors that filter noise out of observed CARs. We extend both lines of work by combining event-study estimates with a discounted cash flow model and a competing-risks duration model.
 
Further, by providing estimates of R\&D costs, this paper contributes to the small literature evaluating the cost of bringing new drugs to market \citep{DimasiHansenGrabowski2003, dimasi2004rd, DiMasi2016, DuboisKyle2016, SertkayaWongJessupBeleche2016, WoutersMcKeeLuyten2020, CBO2021, Schlanderetal2021, Sertkayaetal2024}, informing policy debates \citep{Frank2003} and regulations such as price interventions \citep{CongressReport2021}.
 
Lastly, our paper complements the literature that studies policies to spur pharmaceutical R\&D, such as patent buyouts \cite{Kremer1998} and transferable patents \cite{DuboisMoissonTirole2022}. Our drug buyout approach is closer to \cite{Kremer1998}, but our method of estimating drug values requires weaker assumptions and has practical advantages over the auctions proposed in \cite{Kremer1998}. We rely on stock prices that aggregate information dispersed among a large pool of self-interested investors \citep{Milgrom1981b}, providing a more accurate valuation than an auction. Furthermore, as drugs are often covered by multiple patents \citep{Gupta2021, McKeonPoegeSimcoe2022}, valuing a drug is preferable to combining patent values.

\section{Institutional Background\label{sec:institutional}}

\paragraph{Drug R\&D and Announcements.}
The R\&D process in the U.S.\ pharmaceutical industry consists of several distinct stages defined by the FDA. Figure \ref{fig:timing} is a schematic of the process. The first stage is the preclinical stage, which includes creating a new molecule (or a system of molecules) and testing it in the laboratory using \emph{in vitro} and \emph{in vivo} methods. We refer to this stage as either the discovery or preclinical stage.

If the preclinical research is successful, the firm can begin testing the drug candidate in humans through three phases of clinical trials. Phase I involves screening the drug for possible toxicity. In Phase II, firms test the drug's efficacy in a larger sample of individuals with the targeted diseases. Finally, Phase III involves double-blind studies to assess the drug's effectiveness in a large patient population.

 \begin{figure}[t!]
\begin{center}
\caption{ Schematic of Drug Development Process.}\label{fig:timing}
\begin{tikzpicture}
[font=\sffamily,
 every matrix/.style={ampersand replacement=\&,column sep=1.3cm,row sep=1cm},
 source/.style={draw,thick,rounded corners,fill=yellow!20,inner sep=.23cm},
 basicresearch/.style={draw,thick,rounded corners,fill=blue!10,inner sep=.23cm},
 process/.style={draw,thick,circle,fill=green!20},
 exitnode/.style={draw,thick,rounded corners,fill=red!90,inner sep=.23cm},
 sink/.style={source,fill=yellow!20},
 sinks/.style={source,fill=green!20},
 datastore/.style={draw,very thick,shape=datastore,inner sep=.23cm},
 dots/.style={gray,scale=2},
 to/.style={->,>=stealth',shorten >=1pt,semithick,font=\sffamily\footnotesize},
 every node/.style={align=center, scale=0.65}]

 \matrix{
 \node[basicresearch] (hisparcbox0) {Basic Research};\& \& \node[exitnode] (hisparcbox5) {exit};\\
 \node[source] (hisparcbox) {Discovery (Preclinical) Stage};
 \& \node[process] (daq) {Clinical Trials}; \& \node[sinks] (datastore00) {FDA Review }; \\

 \node[exitnode] (hisparcbox1) {exit}; \& \& \node[sinks] (hisparcbox9) {Market}; \\
 };

 \draw[to] (hisparcbox0) -- node[midway,above] {}(hisparcbox);

 \draw[to] (hisparcbox) -- node[midway,below] {success}(daq);
  
 \draw[to] (hisparcbox) -- node[midway,above] {failure} (hisparcbox1);  
 \draw[to] (datastore00) -- node[midway,right]{failure}(hisparcbox5);  
\draw[to] (datastore00) -- node[midway,right]{success}(hisparcbox9);  

 \draw[to] (daq) -- node[midway,below] {success}(datastore00);
 \draw[to] (daq) -- node[midway,left] {failure}(hisparcbox5);  
  
\end{tikzpicture}
\begin{figurenotes}This is a schematic representation of the drug development process, where the entries ``success'' and ``failure'' correspond to possible announcements by a firm.\end{figurenotes}
\end{center}
\end{figure}

After successful completion of clinical trials, the firm submits an FDA review application (an NDA or BLA). The FDA has a group of internal and external experts who review the results from clinical trials and the applicant firm's manufacturing capacity before deciding whether to approve the application. If the FDA approves the drug, the drug is launched in the market. We refer to this final milestone as FDA approval.

\begin{figure}[t!]
\begin{center}
\caption{\large \bf Stock Price and FDA Approval\label{fig:ChemoCentryx}}
\includegraphics[scale=0.25]{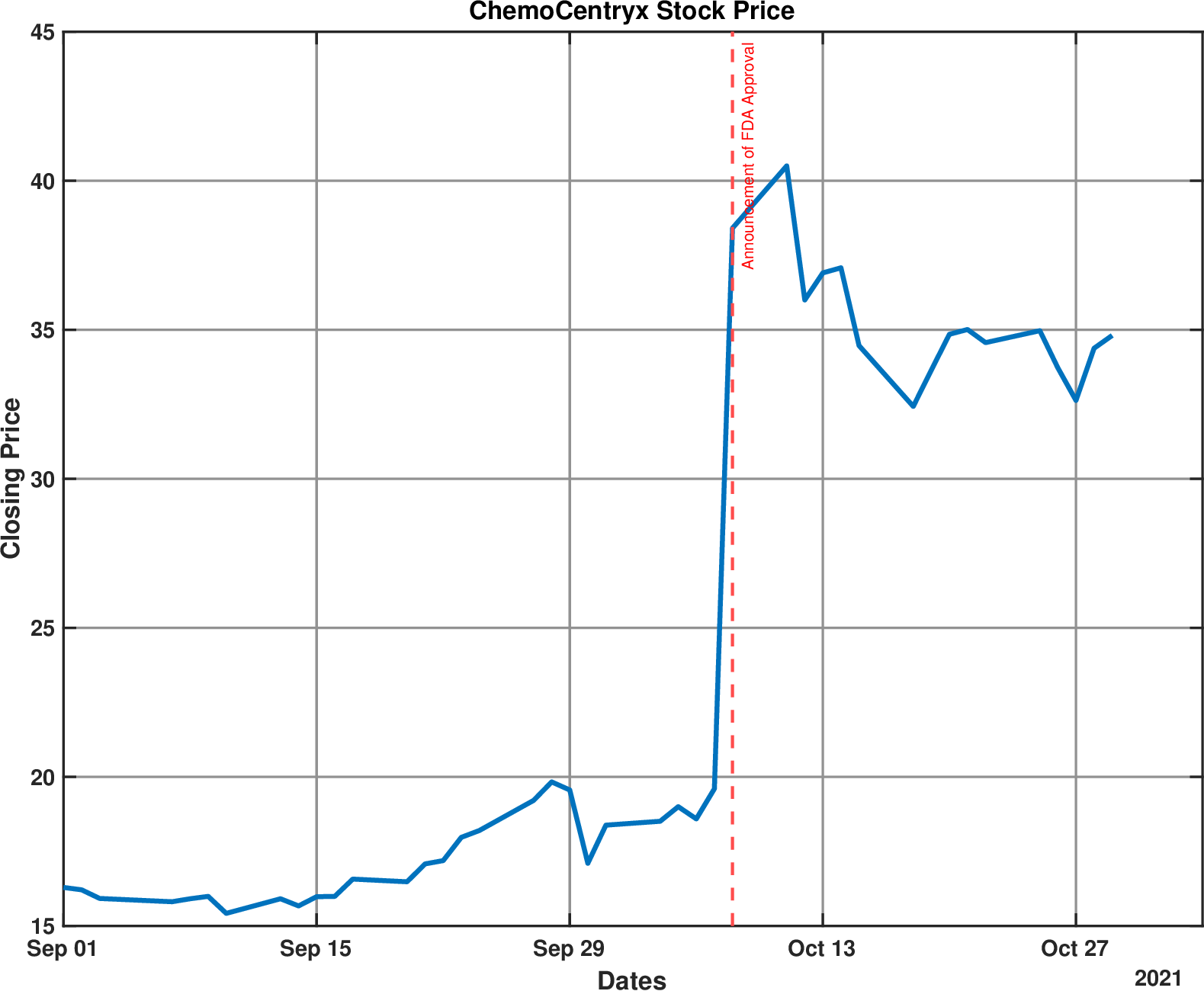}
\begin{figurenotes} Time series of daily stock prices for ChemoCentryx around the announcement date. On October 8, 2021, ChemoCentryx announced FDA approval for its vasculitis drug. 
\end{figurenotes}
\end{center}
\end{figure}

\paragraph{Example of an Announcement.}
Consider the case of ChemoCentryx, which was developing an ANCA-associated vasculitis therapy and had submitted an FDA application. On October 8, 2021, the FDA approved its application. As shown in Figure \ref{fig:ChemoCentryx}, the market responded positively to this news, and its stock price increased sharply.\footnote{Click \href{https://web.archive.org/web/20211008203634/https://ir.chemocentryx.com/news-releases/news-release-details/chemocentryx-announces-fda-approval-tavneostm-avacopan-anca}{here for ChemoCentryx's} (archived) announcement document. Last accessed November 14, 2023.} Assuming it was the only news that day from ChemoCentryx, we can attribute the increase in ChemoCentryx's stock price to the resolution of uncertainty surrounding FDA approval.

For our analysis, we focus primarily on discovery and FDA approval announcements (see Section \ref{sec:data} for our sample selection decision). In general, discovery announcements typically introduce a new drug candidate, such as ``Arrowhead Announces New Clinical Candidate ARC-AAT for Treatment of Alpha-1 Antitrypsin Deficiency-Associated Liver Disease.'' These announcements primarily originate in press releases, conference presentations reporting early or preclinical results, and corporate filings; they thus reflect the moment a firm chooses to disclose a candidate publicly. Approval announcements confirm FDA decisions, such as ``FDA Approves BELSOMRA (suvorexant) for the Treatment of Insomnia; Merck \& Co.\ Inc.,'' and similarly derive mostly from press releases issued by the firm or the FDA. We know the dates for both types of announcements, and we are confident in their accuracy.

In Table \ref{table:example}, we present examples of five successful drugs, including information about the dates of their discovery and FDA approval. These milestones also indicate the time it takes for a drug to reach the market from its initial discovery. 

\begin{table}[ht!]
\caption{ Examples of Timeline\label{table:example}}
\scalebox{0.9}{\begin{tabular}{llllll}
\toprule
Drug Name &
  \emph{caspofungin} &
  \begin{tabular}[c]{@{}l@{}}\emph{dimethyl}\\
   \emph{fumarate}\end{tabular} &
  \emph{evolocumab} &
  \begin{tabular}[c]{@{}l@{}}\emph{telotristat}\\ \emph{etiprate}\end{tabular} &
  \emph{ziprasidone} \\
\toprule
Firm                & Merck \& Co Inc   & Biogen Inc        & Amgen Inc       & \begin{tabular}[c]{@{}l@{}}Lexicon- \\ Pharma Inc\end{tabular} & Pfizer Inc      \\
Indication &Infectious &
  Neurological &
  Cardiovascular &
  Neoplasm &
  Neurological\\
Disease &
  Fungal Infection &
  \begin{tabular}[c]{@{}l@{}}Multiple \\ Sclerosis\end{tabular} &
  \begin{tabular}[c]{@{}l@{}}Hyper- \\ cholesterolemia\end{tabular} &
   \begin{tabular}[c]{@{}l@{}}Carcinoid \\ Syndrome\end{tabular} &
  \begin{tabular}[c]{@{}l@{}}Bipolar \\ Disorder\end{tabular}\\
Discovery           & Jun 12, 1996     & Nov 12, 2003 & Jun 16, 2009   & Feb 21, 2007  & Jan 1, 2002 \\
FDA Approval        & Jan 1, 2001   & Mar 27, 2013    & Jul 17, 2015   & Feb 28, 2017  & Aug 31, 2004 \\
\bottomrule  
\end{tabular}}
\begin{figurenotes}
This table displays key information for five successful drugs in our sample. For each drug, we report the drug name, the developing firm, the therapeutic indication, and the dates of discovery and FDA approval. 
\end{figurenotes} 
\end{table}

\subsection{Regulatory Environment}

Three regulatory paradigms govern firms' announcements about the success and failure of drug candidates throughout the development process. First, the Securities and Exchange Commission (SEC) requires public companies to disclose all material information to investors via the annual 10-K, quarterly 10-Q, and current 8-K forms, with \emph{Regulation Fair Disclosure} (2000) mandating timely disclosure of all material information. Second, the FDA controls what firms can announce about their drugs during development, requiring registration of clinical trials within 21 days of enrolling the first subject under the \emph{Food and Drug Administration Modernization Act} (1997) and disclosure of information about clinical trials and FDA application processes whenever relevant, ensuring announcements are not \emph{materially misleading}. Third, the \emph{Sarbanes-Oxley Act} (2002) allows the SEC to monitor firms' announcements about their FDA review process, and since 2004, the FDA has referred cases of false or misleading statements to the SEC.

These regulations incentivize firms to inform the market correctly and promptly. However, companies retain some discretion in determining what is considered ``material'' and ``not misleading.'' This ambiguity is more pronounced when clinical trial results are less prominent or when large-cap firms are developing multiple drugs. In such cases, firms may delay or ``bundle'' negative and positive announcements to soften the market reaction. Our analysis restricts attention to only those days on which firms make a single announcement.\footnote{Appendix \ref{app:single_announcement} discusses this restriction in detail, including the frequency of multi-announcements.}

\section{Data\label{sec:data}}

Our empirical analysis rests on two datasets. The first is the \textit{announcements dataset}, which links drug development milestones to stock market returns and is used to estimate cumulative abnormal returns (CAR) associated with pipeline events. The second is the \textit{duration dataset}, which tracks how long drugs spend in each development phase and whether they advance or fail, and is used to estimate stage-specific transition probabilities and durations via competing-risks hazard models. Both datasets draw on the same sources but use the data differently. Their construction is summarized below and described in Appendix \ref{app:data_details}.

The two datasets impose different precision requirements for milestone dates. CARs are estimated from narrow event windows, so the announcements dataset requires that each milestone date correspond to the date on which the market first learned the news. The duration dataset is less demanding because what matters is whether a drug advances or is discontinued at each stage, and approximate timing yields good estimates. Imprecision in intermediate milestone dates does not materially affect the estimation of transition probabilities.

\subsection{Data Sources}
\label{subsec:data_sources}

Our primary dataset on drug development comes from Cortellis, which is owned and managed by Clarivate Analytics. The database covers more than 70,000 drug candidates worldwide. For each development milestone, Cortellis records the announcement date, the drug name, the associated firm, and the target indication (and in some cases diseases), drawing on academic articles, patents, press releases, financial filings, earnings calls, conference presentations, and regulatory publications. Professional analysts working for Cortellis maintain the data.

We supplement Cortellis with clinical trial information from ClinicalTrials.gov. Cortellis provides ClinicalTrials.gov identifiers associated with drug and disease names, which we link to our data. In the announcements dataset, this link helps construct transitions across development stages. In the duration dataset, ClinicalTrials.gov provides trial start and completion dates that supplement the Cortellis data.

We obtain daily returns, prices, and shares outstanding for all biomedical and pharmaceutical companies publicly listed on U.S.\ stock exchanges from the Center for Research in Security Prices (CRSP). We match firm names between CRSP and Cortellis first using a Large Language Model, then manually validate each match.\footnote{Firms frequently change names or merge, so we use CRSP permanent identifiers (PERMNO) to ensure consistent matching, linking any historical name associated with a PERMNO to its counterpart in Cortellis. One limitation is that Cortellis does not track mergers and acquisitions. When an acquired firm ceases to trade independently, as was the case with Genentech after its acquisition by Roche, we cannot link its post-acquisition announcements to stock-market data.} We use monthly CPI data from the Federal Reserve Economic Data (FRED) to adjust for inflation.

\paragraph{Announcements dataset.} The announcements are drawn from Cortellis' milestone records spanning 1985--2019, covering drug development in Western markets (US, Canada, Europe, Australia, and New Zealand). For estimation, we focus on the US market from January 1, 2000--August 30, 2019, the period following the launch of ClinicalTrials.gov.

As mentioned earlier, we focus on two positive milestone types that mark the beginning and the end of the development process: \textit{discovery} (entry into preclinical development) and \textit{FDA approval}, or just \textit {Approval}. Discovery marks the first public disclosure of a drug candidate. FDA approval is a regulatory decision outside the firm's control. Importantly, both events are likely genuine surprises and provide clean information about the value.

We do not use intermediate milestones, e.g., those on clinical trials or the submission of the FDA application. Cortellis records the date on which a new phase begins, not the date on which results from the preceding phase were first disclosed. In practice, firms have discretion over when and how to report intermediate steps \citep{CunninghamEdererHodgsonWang2025}, and consequently, the dates for these intermediate stages recorded by Cortellis may not be the exact first dates on which the information was released. In such cases, the dates in our data will not correspond to the date on which uncertainty was resolved.

We also observe discontinuation announcements at various stages. The timing of these announcements is also imprecise, and clinical trial results are often released incrementally before a formal termination decision is made. 
A likely exception is the discontinuation of preclinical research. Just as with the start of the discovery stage, discontinuation announcements are \emph{not} tied to a preceding phase whose results could have ``leaked'' earlier. We therefore use discovery-stage discontinuation announcements for estimation.

The final announcements sample is constructed by applying sequential filters to our main data. These filters require that the announcement was relevant for the U.S., be the first announcement per drug-indication-event, restrict to single-firm announcements, and drop cases where a prior foreign disclosure preceded the U.S. announcement.\footnote{Multiple announcements can arise for the same drug-indication-event for two reasons. First, if a licensing or acquisition deal occurs after an initial announcement, the acquiring firm may be recorded as having independently made the same type of announcement for the same drug-indication. We address this multiplicity by screening for and dropping deal-related announcement titles. Second, when a drug is jointly developed, multiple firms may each be associated with the same announcement type for the same drug-indication on the same date. In both cases, we retain only the first announcement.}  Together, they address the main sources of noise in abnormal return estimation, namely limited U.S.\ valuation relevance, double-counting of milestones, and reduced informativeness following foreign pre-disclosure. The event study is further restricted to single-announcement firm-dates to ensure clean attribution of price reactions to individual events (see Appendix \ref{app:single_announcement}).

\paragraph{Duration dataset.}  The duration dataset tracks each drug-indication through the sequential stages of development (Discovery, Phase I, Phase II, Phase III, Application, and Approval), recording the time spent at each stage and the outcome, i.e., advance, discontinuation, or censoring. It is constructed from a broader set of Cortellis milestone records, restricted to U.S.\ development activity and supplemented with data from the ClinicalTrials.gov website. The two datasets are linked through a common firm-size measure, so that the transition probabilities entering the valuation framework are conditioned on the same firm characteristics that shape market reactions to pipeline news (see Appendix \ref{app:duration_pipeline}).

\subsection{Firm Heterogeneity \label{subsec:sample_characteristics}}
In our sample, firms differ in terms of market capitalization. We measure firm size using a within-day percentile rank of real market capitalization across all CRSP firms. 
To obtain a classification that is stable within each decade and that does not give disproportionate weight to firms with many announcements in a single year, we first compute the median percentile rank within each firm-year, and then take the median of those firm-year medians across all years the firm appears in a given decade. We define large firms as those at or above the 95th percentile of this decade-specific distribution, and the rest as small firms. In particular, a firm is classified as large in a decade if the median of its within-day market-capitalization percentile ranks, computed across its announcement dates in that
decade, is at or above 0.95. In our sample, the smallest firm classified as large had a median market capitalization of about \$3.3
billion in the 2000s and \$9.1 billion in the 2010s.

Heterogeneity between small and large firms is central to our approach. First, large firms may have greater expertise and resources for clinical development, which could raise their success rates and shorten development timelines \citep{MCockburnHenderson2001, KriegerLiPapanikolaou2022}. Second, large firms may also specialize in different therapeutic areas. Finally, because larger firms have higher market capitalizations, the same dollar value of a drug announcement translates into a smaller percentage change in the stock price, which affects the estimation and interpretation of CARs. So, we treat large and small firms separately.

Table \ref{tab:sample_descriptives} summarizes the estimation sample for announcements. The sample covers 601 unique firms and 4,232 unique drugs, yielding 5,301 single-announcement firm dates. Announcements are unevenly distributed across firms. The median firm makes 3 announcements, while the mean is 8.8, reflecting the concentration of development activities among a small number of large firms. 

\begin{table}[t!]
\begin{center}
\caption{ Sample Descriptives for Announcements}\label{tab:sample_descriptives}
\scalebox{0.935}{\begin{tabular}{lcc}
\toprule
& Count & Share\\
\midrule
\multicolumn{3}{l}{\textit{Panel A: Sample Coverage}} \\[3pt]
Unique firms & 601 & \\
Unique drugs & 4,232 & \\
Single-announcement firm-dates &  5,301 & \\
Mean announcements per firm &  8.8 & \\
Median announcements per firm &   3 & \\
\addlinespace
\multicolumn{3}{l}{\textit{Panel B: Distribution Over Time}} \\[3pt]
\quad 2000--2009 &  2,099 & (39.6\%) \\
\quad 2010--2019 &  3,202 & (60.4\%) \\
\addlinespace
\multicolumn{3}{l}{\textit{Panel C: Distribution by Announcement Type}} \\[3pt]
Discovery &  4,632 & (87.4\%) \\
FDA approval &    557 & (10.5\%) \\
Discontinued at discovery &    112 & ( 2.1\%) \\
\addlinespace
\multicolumn{3}{l}{\textit{Panel D: Distribution by Firm Size}} \\[3pt]
Large firms ($\geq$ 95th pctile market cap) &  1,830 & (34.5\%) \\
Small firms ($<$ 95th pctile market cap) &  3,471 & (65.5\%) \\
\addlinespace
& Mean & Median \\
\cmidrule(lr){2-3}
Market cap, large firms (\$B) & 137.3 & 115.9 \\
Market cap, small firms (\$B) &   4.8 &   0.5 \\
\addlinespace
\multicolumn{3}{l}{\textit{Panel E: Distribution of Firm Size by Decade}} \\[3pt]
& 2000s & 2010s+ \\
\cmidrule(lr){2-3}
Large firms &    935 &    895 \\
Small firms &  1,164 &  2,307 \\
\addlinespace
\multicolumn{3}{l}{\textit{Panel F: Indications}} \\[3pt]
Cardiovascular &    400 & ( 7.5\%) \\
Gastrointestinal &    642 & (12.1\%) \\
Immune disorders &    582 & (11.0\%) \\
Infectious diseases &    650 & (12.3\%) \\
Inflammatory &    654 & (12.3\%) \\
Neoplasm (cancer) &  1,670 & (31.5\%) \\
Neurological &    794 & (15.0\%) \\
Rare diseases &    646 & (12.2\%) \\
Other &  1,098 & (20.7\%) \\
\bottomrule
\end{tabular}}
\begin{figurenotes}
The estimation sample consists of  5,301 single-announcement firm dates 
across 601 unique firms spanning 4,232 unique drugs. 
Large firms are defined as those at or above the 95th percentile 
of the within-year market capitalization distribution (decade-specific median). 
Market capitalization figures are in real December 2020 dollars. 
Therapeutic areas are non-mutually exclusive: a drug may target 
multiple indications spanning different categories.
\end{figurenotes}
\end{center}
\end{table}

Discovery announcements account for 87.4\% of the sample, followed by FDA approvals at 10.5\% and discovery-stage discontinuations at 2.1\%. Discovery announcements dominate because most drugs that enter preclinical development do not reach Approval.

Large firms account for 1,830 observations (34.5\%). The size gap between the two groups is substantial. The mean market capitalization is \$137.3 billion (median \$115.9 billion) for large firms and \$4.8 billion (median \$0.5 billion) for small firms. The growth in sample size from the 2000s to the 2010s is concentrated among small firms, whose observations increase from 1,164 to 2,307, while large-firm observations decline modestly from 935 to 895.

The most common therapeutic area is neoplasms (which includes benign and malignant tumors), accounting for 31.5\% of observations, followed by neurological (15.0\%), infectious diseases (12.3\%), inflammatory (12.3\%), rare diseases (12.2\%), gastrointestinal (12.1\%), immune disorders (11.0\%), and cardiovascular (7.5\%). Because a drug may target multiple indications, an announcement can be associated with multiple indications, therapeutic areas are not mutually exclusive, and some indications can be associated with multiple therapy areas (e.g., ``brain cancer" is associated with Neoplasm and neurological disorder), the shares do not sum to 100. The ``Other" category is a catch-all category that includes indications with only a few observations. We treat this category separately, but do not report it.

\subsection{Development Histories 
\label{subsec:dev_histories}}

One of the defining features of drug development is that only a fraction of drug candidates get FDA approval to reach the market. Table \ref{tab:dev_histories} summarizes the \emph{duration dataset} along two dimensions: (raw) transition probabilities across development stages, which capture the likelihood of success, and the time spent at each stage, which captures the length of development. All statistics are computed from our sample of 13,766 firm-drug-indication-stage spells with non-missing firm size and non-outlier durations.

Panel A reports transition probabilities. On average, 8.6\% of drug candidates that enter discovery are ultimately approved.\footnote{A drug can be developed concurrently for multiple indications by more than one firm. We define an observation at the individual firm-drug-indication level.} Attrition is uneven across stages. The sharpest drop occurs at Phase II to Phase III, where only 34.5\% of drugs advance. The early clinical stages also exhibit substantial attrition: 71.4\% of discovery-stage drugs transition to Phase I, and 63.9\% of Phase I drugs transition to Phase II. In contrast, the transition from application to Approval is substantially higher, with a conditional probability of 92.9\%. The overall approval probability for newly discovered drugs is comparable to those in the literature, ranging from 8\% to 13.8\% \citep{Pauletal2010, Hay2014, Mullard2016, DiMasi2016, Wong2019}. The pattern of attrition also aligns with prior findings, with the transition to Phase III representing the most difficult threshold to cross.\footnote{Most previous studies use clinical trial data to estimate transition probabilities from Phase I to Approval; our corresponding estimate is 12.1\%. \cite{Pauletal2010} estimate a 69\% probability of transitioning from the preclinical stage to clinical development, which is close to our discovery-to-Phase I transition probability.}

\begin{table}[t!]
\begin{center}
\caption{ Development Histories}\label{tab:dev_histories}
\setlength{\tabcolsep}{4pt}
\begin{tabular}{l r r r r r r r r r r}
\toprule
\multicolumn{11}{l}{\textit{Panel A: Transition Probabilities} (\%)} \\[2pt]
& & \multicolumn{9}{c}{Conditional} \\
\cmidrule(lr){3-11}\\
& & & \multicolumn{4}{c}{By Firm Size} & \multicolumn{4}{c}{By Decade} \\
\cmidrule(lr){4-7}\cmidrule(lr){8-11}
& & & \multicolumn{2}{c}{Large} & \multicolumn{2}{c}{Small}
  & \multicolumn{2}{c}{2000--09} & \multicolumn{2}{c}{2010--19} \\
\cmidrule(lr){4-5}\cmidrule(lr){6-7}\cmidrule(lr){8-9}\cmidrule(lr){10-11}
Stage Transition
  & \multicolumn{1}{c}{Marginal}
  & \multicolumn{1}{c}{Overall}
  & \multicolumn{1}{c}{Prob.}
  & \multicolumn{1}{c}{N}
  & \multicolumn{1}{c}{Prob.}
  & \multicolumn{1}{c}{N}
  & \multicolumn{1}{c}{Prob.}
  & \multicolumn{1}{c}{N}
  & \multicolumn{1}{c}{Prob.}
  & \multicolumn{1}{c}{N} \\
\midrule
Discovery $\to$ Phase I      & 71.4 & 71.4 & 72.5 & 523 & 70.7 & 839 & 63.7 & 653 & 78.4 & 709 \\
Phase I $\to$ Phase II       & 45.6 & 63.9 & 61.4 & 567 & 67.0 & 466 & 63.2 & 532 & 64.7 & 501 \\
Phase II $\to$ Phase III     & 15.7 & 34.5 & 35.5 & 516 & 33.6 & 538 & 33.2 & 527 & 35.9 & 527 \\
Phase III $\to$ Application    &  9.3 & 59.0 & 63.5 & 249 & 54.9 & 275 & 54.6 & 260 & 63.3 & 264 \\
Application $\to$ Approval &  8.6 & 92.9 & 95.9 & 268 & 90.0 & 271 & 88.1 & 202 & 95.8 & 337 \\
\cmidrule(lr){1-11}
 Discovery $\to$ Approval  &        &  8.6 &  9.6 &     &  7.9 &     &  6.4 &     & 11.0 &     \\
\addlinespace[8pt]
\multicolumn{11}{l}{\textit{Panel B: Duration (months)}} \\[2pt]
  & \multicolumn{1}{c}{N}
  & \multicolumn{1}{c}{Mean}
  & \multicolumn{1}{c}{SD}
  & \multicolumn{1}{c}{P25}
  & \multicolumn{1}{c}{P50}
  & \multicolumn{1}{c}{P75}
  & \multicolumn{4}{c}{} \\
\midrule
\emph{Stage}--&&&&&&&\\
Discovery           & 6{,}438  & 36.4 & 27.2 & 16.9 & 29.9 & 51.1 & \multicolumn{4}{c}{} \\
Phase I             & 2{,}667  & 32.1 & 29.8 & 10.4 & 22.7 & 44.1 & \multicolumn{4}{c}{} \\
Phase II            & 3{,}101  & 36.5 & 26.8 & 16.2 & 29.0 & 49.8 & \multicolumn{4}{c}{} \\
Phase III           & 948      & 31.2 & 22.8 & 14.3 & 25.7 & 40.0 & \multicolumn{4}{c}{} \\
Application    & 612      & 13.3 & 16.7 & 5.9  & 9.4  & 12.0 & \multicolumn{4}{c}{} \\[2pt]
\emph{Outcome}--&&&&&&&\\
Success  & 2{,}806 & 15.4 & 12.3 & 5.9 & 11.8 & 22.5 & \multicolumn{4}{c}{} \\
Failure  & 1{,}706 & 25.8 & 17.9 & 12.2 & 22.2 & 35.3 & \multicolumn{4}{c}{} \\
Censored & 9{,}254 & 41.5 & 29.0 & 18.5 & 34.6 & 58.0 & \multicolumn{4}{c}{} \\
\bottomrule
\end{tabular}
\begin{figurenotes}
Panel A reports marginal probabilities (\% share of all discovery-stage projects
reaching each subsequent stage) and conditional transition probabilities
(share advancing conditional on entering a stage, excluding censored observations), across the entire sample and also split by firm size and decade.
Sample: 13,766 observations with non-missing firm size and non-outlier durations.
Panel B reports stage durations in months, by development stages and outcomes of the stages:
\textit{Success} means advance to next stage;
\textit{Failure} means discontinuation;
\textit{Censored} means no observed transition.
\end{figurenotes}
\end{center}
\end{table}

The firm size breakdown reveals differences concentrated in the late stages. Drugs developed by large-cap firms have a modestly higher cumulative approval rate (9.6\%) than those developed by smaller firms (7.9\%). The gap is most pronounced between Phase III and application (63.5\% vs.\ 54.9\%) and between application and Approval (95.9\% vs.\ 90.0\%). From Phase I to Phase II, however, the pattern reverses, with small firms showing a higher conditional probability (67.0\% vs.\ 61.4\%). 

This data feature may reflect two opposing forces. On the one hand, small firms appear more selective at the preclinical stage, potentially pursuing only the most promising hypotheses from the outset, consistent with their only modestly lower rate of advancing from discovery to Phase I. However, this mechanism is at odds with the later transition probabilities (from Phase II onwards) being smaller for small firms.
On the other hand, once a drug is in development, small firms may be less willing to terminate it even when interim results are discouraging, since discontinuation can hurt the firm \citep{GuedjScharfstein2004}. 

Comparing the decades, we see that overall raw approval rates nearly 
doubled from 6.4\% in the 2000s to 11.0\% in the 2010s, with 
improvements visible at every stage. This increase likely reflects a combination of better drug design, biomarker-guided development, and regulatory changes; it does not account for changes 
in firm composition, therapy areas, or censoring. Once we control for these factors in the Gompertz competing-risks 
model, the success rates are stable across decades, with a slight decline in the 2010s (Table \ref{tab:transition_heterogeneity_detail}).

Panel B reports stage durations in months. Discovery is the longest stage, with a median of 29.9 months, followed by Phase II (29.0 months), Phase III (25.7 months), and Phase I (22.7 months). Durations also differ by outcome. Successful transitions are faster than failures. The median duration to success is 11.8 months, compared with 22.2 months for failure, consistent with the idea that promising drugs move through stages more quickly. Drug spells are censored either because development is ongoing as of August 30, 2019, or because Cortellis assigns a ``No Development Reported'' (NDR) status to drug-indication pairs that have had no recorded activity for more than a year but were never formally discontinued, suggesting the project may have stalled without an official termination decision. Censored spells have the longest stage durations, with a median of 34.6 months, reflecting a combination of right-censoring and the delayed recording of inactivity by Cortellis.

We model stage durations using a Gompertz competing-risks hazard model with three risks (success, failure, and no development reported) and unobserved firm-level heterogeneity. The Gompertz specification keeps the model tractable while allowing the hazard for each risk to vary flexibly with time spent in stage, therapeutic area, and firm characteristics \citep{FineGray1999, HonoreMuney2006, AbbringVandenberg2007}. Full details on the estimation are in Appendix \ref{app:hazard_estimation}, and the estimation results are in Section \ref{sec:gompertz_results}.

\section{Drug Valuation\label{section:valuation}}

\subsection{Real Options Framework for Drug Development}

Before presenting our market-based valuation methodology, it is useful to frame drug development as a sequential investment problem with embedded real options. Drug development involves sequential stages, indexed by $\mathcal{K} = \{\disc, \text{phase1}, \text{phase2}, \text{phase3}, \appl, \appr\}$, an ordered set where stage $k\in\mathcal{K}$ depends on the successful completion of the previous stage.

For stage $k\in\mathcal{K}$, let $C_k$ denote the development cost, which we assume is sunk at the start of the stage, and let $V_k$ denote the \emph{net value} of a drug at the start of stage $k$ (after paying $C_k$). If a project discontinues at stage $k$, the firm incurs no subsequent development costs. Let $p_{k|k-1}$ be the probability of reaching stage $k$ from $k-1$, and let $\delta\in(0,1)$ be the discount factor.

Given approval value, $V_\appr$, the value at stage $k\in\mathcal{K}\backslash\{\appr\}$ is given recursively by 
\begin{equation}
V_{k} = \mathbb{E}(\delta^{\tau_{k \to k+1}}) \, p_{k+1|k} \, V_{k+1} - C_k, \label{eq:recursive_general}
\end{equation}
where $\mathbb{E}(\delta^{\tau_{k \to k+1}})$ is the expected discount factor and the expectation is taken over the time a drug stays in stage $k$, conditional on successfully reaching stage $k+1$. The discounting arises because a drug generates profits only after FDA approval, which takes time (see Table \ref{tab:dev_histories}, \emph{Panel B}), but this \emph{time to success} is a priori random.
Substituting recursively, the net value at discovery before any costs are incurred is
\begin{eqnarray}
V_\disc = \mathbb{E}(\delta^{\tau_{\disc \to \appr}}) \prod_{k\in\mathcal{K}} p_{k|k-1} \times V_\appr - \left(C_\disc+\sum_{k\in\mathcal{K}\backslash\{\disc\}} \mathbb{E}(\delta^{\tau_{\disc \to k}}) \prod_{\disc<\kappa \leq k} p_{\kappa|\kappa-1} \times C_k\right), \label{eq:V0_expanded}
\end{eqnarray}
where the last term in the parentheses is the expected present value of total development costs, which we denote as $\mathbb{E}_\disc^{Opt}(C)$. Each cost $C_k$ is weighted by the probability $\prod_{\disc<\kappa \leq k} p_{\kappa|\kappa-1}$ that the project survives to incur it, and discounted to discovery.\footnote{In practice, we estimate the Gompertz competing-risks model (see Appendix \ref{app:hazard_estimation}), which allows $p_{k|k-1}$ to evolve with the time a drug ``spends" in that stage. We also estimate the competing risks model for each of the five stages, so that we can determine $p_{k'|k}$ for any feasible stages $k$ and $k'$, not just adjacent ones.} The abandonment option is embedded in these probability weights, as late-stage costs receive low weight precisely because the drug may be abandoned early.

To appreciate the magnitude of the option value, it is helpful to compare $\mathbb{E}_\disc^{Opt}(C)$ with the cost under a \emph{passive} policy that commits to all stages upfront, regardless of intermediate failures, $\mathbb{E}_\disc^{Pa}(C):= C_\disc+\sum_{k\in\mathcal{K}\backslash\{\disc\}} \mathbb{E}(\delta^{\tau_{\disc \to k}}) \times C_k. $
The difference between the two is 
\begin{equation*}
\text{Option Value} = \mathbb{E}_\disc^{Pa}(C) - \mathbb{E}_\disc^{Opt}(C) = \sum_{k\in\mathcal{K}\backslash\{\disc\}} \mathbb{E}(\delta^{\tau_{\disc \to k}}) \left(1 - \prod_{\disc<\kappa \leq k} p_{\kappa|\kappa-1}\right) C_k. 
\end{equation*}
Because the success probabilities $p_{k|k-1}$ in drug development are small at each stage and late-stage clinical trials (Phases II and III) account for the bulk of development costs, this option value can be substantial. For example, a back-of-the-envelope calculation suggests that, with an overall discovery-to-approval success rate of roughly 7\% and late-stage trials accounting for approximately 80\% of total development costs, the abandonment option saves on the order of 74\% (= $93\%\times 80\%$) of the passive cost commitment. That is why pharmaceutical R\&D is viable despite low success rates, because the \emph{effective} cost is lower than the full cost.

This recursive structure also clarifies why net value tends to increase monotonically from stage to stage, a point with important implications for the timing of public policy interventions (see Section \ref{sec:policysection}). 
To see why, note that the recursion in (\ref{eq:recursive_general}) gives $V_{k+1} - V_k = V_{k+1}\times \bigl(1 - \mathbb{E}(\delta^{\tau_{k\to k+1}}) \, p_{k+1|k}\bigr) + C_k$. 
Since $C_k > 0$ and $\mathbb{E}(\delta^{\tau_{k\to k+1}}) \, p_{k+1|k} < 1$, the right-hand side is positive whenever $V_{k+1} > 0$, that is, whenever the project is worth continuing. Two forces drive this gap: the costs of stage $k$ become sunk and drop out of the forward-looking value, and the uncertainty about surviving stage $k$ resolves. Both raise the expected value of the project. Since firms rationally abandon projects with negative continuation value rather than advance them, $V_{k+1} > 0$ holds for projects that remain in active development, so net value increases monotonically across stages. This monotonic pattern indicates that the critical bottleneck for private investment lies at the earliest stages of development.

\subsection{Valuation at Approval}

Here, we determine the terminal condition for (\ref{eq:recursive_general}), namely the valuation of a drug at approval, $V_\appr$.
When a firm announces that the FDA has approved a drug under review, the firm's stock price should immediately increase following the announcement (e.g., Figure \ref{fig:ChemoCentryx}), and with it the firm's value. The size of this increase should equal the change in the \emph{expected} profits from selling the drug. Importantly, in this context, the only relevant change in the drug's status is the resolution of uncertainty about the FDA's decision. Therefore, conditional on knowing the transition probability $p_{\appr|\appl}$, we can use the change in the firm's market value after the announcement to recover the value of an approved drug.

To formalize this intuition, let $S_{k}\in\{0,1\}$ be a binary variable 
equal to one if the stage-$k$ announcement is positive and zero otherwise, 
so that $S_\appr=1$ if the FDA approves the drug. FDA approval represents 
the terminal stage, where the \emph{gross profit value} is defined as $\Pi_\appr := \mathbb{E}(\Pi \mid S_{\appr}=1)$, 
the present discounted value of profits from selling the approved drug, 
after all development costs are sunk. For example, if $\pi$ denotes the 
expected average yearly profit after approval and profits are constant 
over time, then $\Pi_\appr = \pi/(1-\delta)$.

Just before the approval announcement, the market expects the value of the drug to be $(\Pi_\appr \times p_{\appr|\appl})$. Immediately after the announcement, the uncertainty about the FDA's decision is resolved, and the market expects the value to be $\Pi_\appr$. It is reasonable to assume that by the time the FDA announces its approval, all other payoff-relevant information about the drug has already been disclosed to the market. When the only ``news'' that pertains to the firm is this approval decision, the change in the firm's market value, $\mathbb{E}\left(\text{CAR}_{\appr}\right) \times \MKTCAP$, equals the change in the market value of the drug, i.e.,
\begin{eqnarray}
\mathbb{E}\left(\text{CAR}_{\appr}\right) \times \MKTCAP &=& \Pi_\appr - \Pi_\appr \times p_{\appr|\appl} = \Pi_\appr \times (1-p_{\appr|\appl}). \label{eq:evs3}
\end{eqnarray}
We estimate $\mathbb{E}\left(\text{CAR}_{\appr}\right)$ and observe the firm's market capitalization, $\MKTCAP$. The transition probability $p_{\appr|\appl}$ is the probability that a drug eventually exits the application stage via a successful transition rather than through failure or censoring, and is recovered from the competing risks model as the cause-specific probability of the success outcome, integrated over the full duration distribution of the application stage.

While we distinguish between gross profit, $\Pi_k$, and net value, $V_k$, which is the gross profit minus the present value of all future development costs, Equation (\ref{eq:evs3}) identifies the gross profit value of a drug at approval, $V_\appr=\Pi_\appr$, because no future development costs remain. For all earlier stages, $\Pi_k > V_k$, as the latter accounts for development costs.

\subsection{Valuation at Discovery\label{subsection:disc_valuation}}

Having estimated the value of a drug at FDA approval, we now turn to estimating its value at the discovery stage. We use two complementary approaches. The first works backward through the recursive valuation framework to recover the \emph{gross} profit value at discovery, i.e., the value before development costs are accounted for. The second uses discovery announcements directly, in the same spirit as the approval event study, to recover the \emph{net} value at discovery, i.e., the value after accounting for the expected development costs.

\subsubsection{Gross Profit Value at Discovery}

The recursive framework in Equation (\ref{eq:recursive_general}) allows us to work backward from approval to discovery. Because we condition on a drug having already reached approval, development costs are sunk at that point and do not enter the valuation. The gross profit value at discovery, $\Pi_\disc$, is therefore obtained by appropriately discounting and risk-adjusting $\Pi_\appr$.

First, the cash flows must be discounted for the time elapsed between discovery and approval. Let $\mathbb{P}_\appr(\tau | S_k = 1)$ denote the probability distribution of time to success, that is, the probability that a drug is approved within the next $\tau$ years, conditional on being at stage $k$. The expected discount factor from discovery is 
$\mathbb{E}(\delta^{\tau_{\disc \to}}) = \sum_{\tau \geq 0} \delta^\tau \times \mathbb{P}_\appr(\tau | S_\disc = 1).$
Second, at discovery, there is additional uncertainty about whether the drug will ever get approved, captured by the probability
$p_{\appr|\disc}.$
The gross profit at discovery is given by 
\begin{eqnarray}
  &&\Pi_\disc 
  \equiv \mathbb{E}(\Pi | S_\disc = 1) = \left(\sum_{\tau \geq 0} \left(\sum_{t=\tau}^{\infty} \delta^t \pi_t \right) 
     \times \mathbb{P}_\appr(\tau | S_\disc = 1)\right) \times p_{\appr|\disc} \notag \\
     &=& \frac{\pi}{(1-\delta)}\times \left(\sum_{\tau \geq 0}  
    \delta^\tau \times \mathbb{P}_\appr(\tau | S_\disc = 1)\right) \times p_{\appr|\disc} = \Pi_\appr \times \mathbb{E}(\delta^{\tau_{\disc \to}}) 
     \times p_{\appr|\disc},\qquad \label{eq:evs1}
\end{eqnarray}
where the second equality relies on the assumption that the expected gross per-year profit after FDA approval is constant, i.e., $\pi_t=\pi$, and the last equality uses $\Pi_\appr = \frac{\pi}{(1-\delta)}$.\footnote{The constant yearly profit assumption affords us tractability by allowing us to separate the expected discount rate from present value of profit, without affecting the estimates of $\Pi_\appr$. For instance, we can relax this assumption by modeling profits to decline at a known parametric rate.} 

Thus $\Pi_\disc$ is the product of three identified terms: the present 
discounted value of profits from approval onward, the probability of 
reaching approval from discovery, and the expected discount factor. 
Implicitly, we assume that development announcements between discovery 
and approval affect the transition probabilities and the time to success, 
but not the drug's commercial value conditional on approval. This assumption would be violated if interim clinical results caused the market to revise its estimate of the drug's post-approval profitability. 
However, we do not 
use any intermediate milestone announcements because our identification relies 
solely on discovery and approval events. 
The preclinical stage is too early for such announcements, and, by the time an FDA 
application is submitted, the market has already incorporated the results 
of clinical trials into the firm's stock price. The approval announcement, therefore, resolves only the remaining regulatory uncertainty, leaving the 
drug's expected post-approval profitability unchanged.

\subsubsection{Net Value at Discovery}
When a firm announces that a drug candidate has entered preclinical 
development, the market updates the firm's stock price to reflect the 
expected net value of the drug. Before the announcement, suppose the market 
expects the drug to be discovered with probability $p_\disc$, so the 
anticipated value is $V_\disc \times p_\disc$. The announcement resolves 
this uncertainty, and the value becomes $V_\disc$. When there is only one 
piece of news that day, the change in the firm's market value equals the 
change in the value of the drug, identifying $V_\disc$, i.e., 
\begin{equation}
  \mathbb{E}(\text{CAR}_\disc) \times \MKTCAP 
  = V_\disc \times (1 - p_\disc). \label{eq:disc_consistent}
\end{equation}

The key challenge is that $p_\disc$ is not directly observable. Unlike
stage-to-stage transition probabilities, which we estimate from our data,
estimating $p_\disc$ would require knowing, at each point in time, how many
research programs are active. 
For small firms, their pre-discovery research is relatively less scrutinized
by the market than that of larger firms, making it hard for investors to
form precise expectations about $p_\disc$. Even if small firms communicate
their therapeutic focus, platform, and broad strategy, it is a reasonable starting point to
assume that the market may not anticipate the disclosure of a \emph{specific}
drug candidate.

We therefore set $p_\disc = 0$ as our baseline assumption. Under this
assumption, discovery announcements are fully unanticipated, and Equation
(\ref{eq:disc_consistent}) identifies
$V_\disc = \mathbb{E}(\text{CAR}_\disc) \times \MKTCAP$. Ours is a
conservative approach: our baseline estimate of $V_\disc$ is a \emph{lower
bound} on the true net value at discovery if $p_\disc > 0$. We validate this assumption in Section
\ref{subsection:consistency} for small firms, and explore sensitivity to it
in Appendix \ref{sec:sens_rho}.

\subsection{Development Cost at Discovery}\label{subsection:costs}

Having identified $\Pi_\disc$ from approval announcements and $V_\disc$ from discovery announcements, we can recover the option-adjusted present value of development costs by taking their difference. 
In particular, using Equation (\ref{eq:disc_consistent}) at $p_\disc = 0$ 
and (\ref{eq:V0_expanded}), the option-adjusted present value of 
development costs is identified as the difference between the gross 
profit value and the net value at discovery, i.e., $ \mathbb{E}_\disc^{Opt}(C) = \Pi_\disc - V_\disc 
  = \Big(\Pi_\appr \times \mathbb{E}(\delta^{\tau_{\disc\rightarrow}}) 
    \times p_{\appr|\disc}\Big)  - \Big(\mathbb{E}(\text{CAR}_{\disc}) \times \MKTCAP\Big).$ This approach requires no direct observation of R\&D expenditures: the 
market's valuation at discovery implicitly embeds its assessment of 
future costs, and our framework extracts that information as a residual.
This cost estimate has three properties worth emphasizing.

First, it is \emph{option-adjusted}. The probability weights in Equation (\ref{eq:V0_expanded}) ensure that late-stage costs 
enter only to the extent that the project survives to incur them. As 
discussed in Section \ref{section:valuation}, the difference between 
the option-adjusted cost and the passive 
cost, which commits to all stages 
regardless of intermediate outcomes, can be large when success probabilities are low and late-stage costs are high. 
$\mathbb{E}_\disc^{Opt}(C)$ is the relevant measure of development costs for 
firms facing sequential investment decisions with abandonment options.

Second, the estimate is recovered entirely from market data and 
estimated transition parameters, without imposing a parametric 
decomposition of how development costs, probabilities, and revenues co-vary across 
drugs. 

Third, the sign and magnitude of $\mathbb{E}_\disc^{Opt}(C)$ are 
both informative. A positive value is the standard case, where 
development costs reduce the drug's net value. A negative value may 
indicate that the discovery announcement conveys information beyond the 
focal drug, that the market partially anticipates the announcement 
($p_\disc > 0$), or that the assumed discount factor $\delta$ differs 
from the market's implicit rate. 

In terms of the magnitude, if $V_\disc$ is small relative to $\Pi_\disc$, development 
costs consume most of the gross profit value, and the net value at the true launch decision may well be negative for many drugs, especially 
after accounting for the unmodelled pre-discovery stage. In the 
notation of the recursive framework, the pre-discovery net value is 
$V_0 = \mathbb{E}(\delta^{\tau_{0\to\disc}}) \, p_\disc \, V_\disc - 
C_0 < V_\disc$, and even a modest $C_0$ could push $V_0$ below zero 
when $V_\disc$ is small. This has direct implications for the viability 
of drug development and the case for policy intervention, which we 
return to in Section \ref{sec:policysection}.

\subsection{Internal Consistency using Discovery Discontinuations\label{subsection:consistency}}

The identification strategy in the previous sections relies on market reactions to two \emph{positive} announcements: approval and discovery. We now show that market reactions to a third type of event, the \emph{discontinuation} of preclinical research, provide an independent consistency check. In particular, the discovery-stage cost $C_\disc$, identified from discontinuation of discovery announcements, should be a small fraction of the total costs $\mathbb{E}_\disc^{Opt}(C)$, as late-stage clinical trials dominate development expenditures. If this prediction is borne out in the data, it validates the model's internal structure across both positive and negative market events.

\paragraph{Pre-announcement market value.} As discussed in Section \ref{subsec:dev_histories}, drugs in discovery face three competing risks in our Gompertz specification: successful transition to Phase I, failure (formal discontinuation), and No Development Reported (NDR). The pre-announcement market value of the drug is determined by the law of total expectation over these outcomes. We adopt the convention that NDR and failure have the same valuation implications, in that both yield zero continuation value to the firm: a formally discontinued drug generates no further development or commercial cash flows, and an NDR drug is by definition one for which the firm has stopped recorded activity for an extended period. Hence, the market expects no further cash flows there either. The two outcomes nonetheless differ in their statistical properties: NDR is a silent fadeout with no associated announcement, while failure is an announced event with its own hazard structure. Pooling them into a two-risk model would impose the wrong hazard structure on drugs that fade out, even if it produced the same valuation, which is why we retain the three-risk Gompertz specification while treating NDR as economically equivalent to failure for valuation purposes.\footnote{This convention is consistent with the recursive valuation framework in Section \ref{section:valuation}, in which the transition probability $p_{k+1|k}$ is the cause-specific cumulative incidence of successful transition. Drugs that do not successfully transition, whether through failure or NDR, contribute zero to the expected continuation value.} Under this convention, the pre-announcement market value of a drug in discovery is $
    \text{Pre-announcement value} = V_{\phasei} \times p_{\phasei|\disc}$,  
where $p_{\phasei|\disc}$ is the cause-specific cumulative incidence of successful transition to Phase I, recovered from the Gompertz competing-risks model.

\paragraph{Identification of $C_\disc$.} The discontinuation announcement resolves the uncertainty about the drug's future, with the post-announcement value equal to zero. The change in market value following discontinuation therefore identifies $\mathbb{E}(\text{CAR}_{\text{drop}}) \times \MKTCAP_{\text{drop}} = -V_{\phasei} \, p_{\phasei|\disc}$. 
Solving the recursion $V_\disc = \mathbb{E}(\delta^{\tau_{\disc \to \phasei}}) \, p_{\phasei|\disc} \, V_{\phasei} - C_\disc$ for $V_{\phasei}$ and using it in the above equation and rearranging identifies the discovery-stage cost as
\begin{equation}
    C_\disc = -\,\mathbb{E}(\text{CAR}_{\text{drop}}) \times \MKTCAP_{\text{drop}} \times \mathbb{E}(\delta^{\tau_{\disc \to \phasei}}) - V_\disc.
    \label{eq:C_disc}
\end{equation}
Two features of this expression are worth noting. First, the success probability $p_{\phasei|\disc}$ does not appear explicitly: it cancels because both the pre-announcement market value and the recursion that defines $V_{\phasei}$ depend on it identically. This makes $C_\disc$ robust to the precise estimate of the success-cause hazard. Second, the only nuisance parameter is the expected discount factor, which is identified from the duration model.

\paragraph{The consistency check.} This estimate of $C_\disc$ should be
smaller than the total development cost $\mathbb{E}_\disc^{Opt}(C)$, because
late-stage trials dominate total expenditures. The estimate also provides a
diagnostic for the $p_\disc = 0$ assumption used in identifying $V_\disc$.
If the true $p_\disc > 0$, our estimate of $V_\disc$ from
\eqref{eq:disc_consistent} would be downward biased, and $C_\disc$ defined
in \eqref{eq:C_disc} would be upward biased. Broad consistency between these
two cases will support our approach and the assumption that discovery
announcements are largely unanticipated, at least for small firms.\footnote{Later, in Appendix \ref{app:eventstudy}, we find that the event-study is consistent with discovery announcements being largely unanticipated, and within the range covered by the
sensitivity analysis in Appendix \ref{sec:sens_rho}.} Our estimate of
$C_\disc$ is necessarily noisy, as it is identified from average market
reactions rather than drug-level costs and from a relatively small number of
discontinuation events. The consistency check is therefore best read as a magnitude
diagnostic, asking if $C_\disc$ is in a plausible range relative to
total cost and to external benchmarks.

\section{Estimation and Valuation Results\label{sec:estimation}}

This section presents all empirical results. We begin with the expected cumulative abnormal returns (Section \ref{sec:car_delta}), then turn to transition probabilities and durations estimated from competing-risks hazard models (Section \ref{sec:gompertz_results}), present the estimates of drug valuations and implied development costs (Section \ref{sec:valuation_results}), and conclude with comparing the estimated values with sales implied revenue for external validity (Section \ref{subsec:sales_validation}).

\subsection{Expected Cumulative Abnormal Returns\label{sec:car_delta}}

Stock returns around announcement dates are noisy. On any given trading day, a firm's return reflects not only the market's reaction to a drug-development announcement but also broader market movements and unrelated information flows. Because drug values depend on these returns, we must account for both the market return and this trading noise. Below, we explain the steps involved in estimating the expected CAR for each announcement.

\paragraph{Step 1: Cumulative Abnormal Returns.}
Let $i$ index firms and $t$ index announcement dates. For each firm-announcement pair $(i,t)$, we compute a daily abnormal return as the difference between the firm's realized return ($R_{it}$) and the contemporaneous value-weighted market return on day $t$ ($R_{t}$), i.e., $AR_{it} := R_{it} - R_{t}$.\footnote{As a robustness check, we also compute abnormal returns using the Fama--French five-factor model, where factor loadings are estimated over a 200-day rolling window ending ten days before each event. The results are similar, and throughout the paper, we report only the market-adjusted specification.} We add these daily abnormal returns over a three-day window $[0,+2]$ centered on the announcement date to obtain the \emph{cumulative abnormal return}, $\text{CAR}_{it} = \sum_{\tau=0}^{2} AR_{it+\tau}$. If the announcement date falls on a non-trading day, $t$ denotes the first subsequent trading day. The three-day window is meant to account for delayed price discovery because the market sometimes takes one or two trading days to fully incorporate information from pharmaceutical announcements, particularly for smaller firms with lower analyst coverage.

\paragraph{Step 2: Signal Extraction.}
The cumulative abnormal return $\text{CAR}_{it}$ contains both relevant and irrelevant (i.e., noise) information. Following \cite{KoganPapanikolaouSeruStoffman2017}, we model the observed return for firm $i$ at announcement date $t$ as $\text{CAR}_{it} = w_{it} + \varepsilon_{it}$, where $w_{it}\sim G_w(\cdot; \theta_w)$ is the true announcement-related CAR value and $\varepsilon_{it} \sim \mathcal{N}(0, \sigma_{\varepsilon,it}^2)$ is market noise. Our object of interest is the announcement-level posterior mean $\mathbb{E}[w_{it} \mid \text{CAR}_{it}]$.

Economic reasoning disciplines the support of $w_{it}$. For positive milestones such as drug discovery and FDA approval, the announcement should only add value, so we model $w_{it}$ as drawn from a truncated normal distribution constrained to be non-negative, i.e., $G_w(\cdot; \theta_w)=\mathcal{N}^+(0, \sigma^2_{w,it})$. For discontinuation announcements, termination can only convey negative information, so we constrain $w_{it} \leq 0$, i.e., $G_w(\cdot; \theta_w)=\mathcal{N}^-(0, \sigma^2_{w,it})$.

Under these truncated-normal priors, the posterior mean has a closed-form expression that depends on the signal-to-noise ratio $\sigma_{w,it}^2 / (\sigma_{w,it}^2 + \sigma_{\varepsilon,it}^2)$, which we allow to vary both by announcement type and by firm size and estimate using MLE. The \cite{KoganPapanikolaouSeruStoffman2017} framework is well-suited to this setting because it provides a tractable Bayesian estimator that shrinks the noisy observed return toward zero in proportion to how much of the total variance is noise, while respecting the sign constraints. Full details about the estimates are reported in Appendix \ref{section:fama} and Table \ref{tab:signal_extraction}.

\paragraph{Step 3: Expected CARs by Announcement Type.}
The resulting posterior mean is the \emph{expected CAR}, $\mathbb{E}(\text{CAR}_{it}) 
\equiv \mathbb{E}\left[w_{it} \mid \text{CAR}_{it}\right]$. Averaging these 
announcement-level posteriors within each type summarizes the typical magnitude of 
each event. As expected, Discovery and FDA approval announcements both generate 
positive mean expected CARs, with approval the larger of the two, consistent with 
the greater commercial certainty that regulatory approval conveys once the drug has 
cleared the highest hurdle, while discontinuation at the discovery stage generates a 
negative mean. Appendix \ref{subsec:raw_vs_expected} reports these means with 
standard errors alongside the corresponding raw CARs, and Appendix 
\ref{app:eventstudy} presents event-study plots of the underlying returns around 
each announcement type. These averages are descriptive; the valuation 
exercise uses the announcement-level posteriors themselves, averaged within cells 
defined by firm characteristics (Section \ref{sec:valuation_results}).

\subsection{Transition Probabilities and Discount Factors\label{sec:gompertz_results}}

The valuation exercise requires, for each stage of drug development, two additional inputs: the probability that a drug advances to the next stage, and the expected time spent there, which determines the discount factor applied to future cash flows. We estimate both from a duration dataset of drug-indication-stage observations using Gompertz competing-risks hazard models with firm-specific unobserved heterogeneity (or frailty). Appendix \ref{app:hazard_estimation} provides the details, including the model estimates and sensitivity analysis.

At each stage, a drug can either advance to the next stage (success), be discontinued (failure), or exit without a reported outcome, designated as NDR (no development reported), which we treat distinctly from failure. The Gompertz specification allows the baseline hazard to vary over time within a given stage, capturing the empirical reality that a drug's probability of advancing or failing changes as it ages within a given phase. Covariates include firm-size percentile, therapeutic area fixed effects, and a decade indicator. Gamma shared frailty at the firm level accounts for unobserved heterogeneity across firms.

We estimate transition probabilities independently at each stage rather than conditioning on complete pipeline histories. Most drug-indication pairs are observed in only one or two stages, and relatively few traverse the full pipeline from Discovery through approval. Conditioning on complete chains would rest on a small number of observations and would introduce selection bias by excluding drugs that exit the pipeline at intermediate stages. By pooling all drugs observed at a given stage, regardless of their prior history, we substantially increase the effective sample size at each transition. The estimation sample contains 13,766 drug-indication-stage observations with non-missing firm size and non-outlier durations. 

Table \ref{tab:gompertz_inputs} reports the predicted conditional transition probabilities and expected discount factors from the Gompertz models, separately for large and small firms. Reading across the pipeline, conditional success probabilities for 
large-firm drugs are lowest in the middle of the pipeline and rise 
sharply through the late stages: from 36.5\% at Phase II to Phase III, 
to 61.3\% at Phase III to Application, and 93.7\% at the application 
stage. The size advantage is similar to the raw data (Table \ref{tab:dev_histories}) and consistent across stages, except for Phase I to Phase II, where small firms have a modestly higher conditional probability (0.610 versus 0.588).

\begin{table}[t!]
\begin{center}
\caption{Transition Probabilities and Expected Discount Factors}
\label{tab:gompertz_inputs}
\medskip
\begin{tabular}{lcccc}
\toprule
& \multicolumn{2}{c}{Conditional} &
  \multicolumn{2}{c}{$\mathbb{E}(\delta^{\tau_{k\to k+1}})$} \\
\cmidrule(lr){2-3} \cmidrule(lr){4-5}
Transition & Large & Small & Large & Small \\
\midrule
Discovery $\to$ Phase I       & 0.667 & 0.619 & 0.9213 & 0.9185 \\
Phase I $\to$ Phase II        & 0.588 & 0.610 & 0.9262 & 0.9213 \\
Phase II $\to$ Phase III      & 0.365 & 0.363 & 0.8928 & 0.8858 \\
Phase III $\to$ Application   & 0.613 & 0.507 & 0.8769 & 0.8683 \\
Application $\to$ Approval    & 0.937 & 0.902 & 0.9533 & 0.9493 \\
\addlinespace
Discovery $\to$ Approval ($p_{\appr|\disc}$)     & 0.085 & 0.066 &        &        \\
\bottomrule
\end{tabular}

\begin{figurenotes}
Predicted conditional transition probabilities ($p_{k+1\mid k}$) and 
expected discount factors ($\mathbb{E}(\delta^{\tau_{k\to k+1}})$) from 
Gompertz competing-risks hazard models with Gamma shared frailty, 
evaluated at population-averaged predictions separately for large firms 
and small firms. Expected discount factors are computed at the baseline 
$\delta = 0.95$. Covariates include size percentile, therapeutic 
area dummies, and decade indicators.
\end{figurenotes}
\end{center}
\end{table}

In the Gompertz success hazard, the firm size coefficient is positive and statistically significant at Discovery ($0.642$, p-value $= 0.004$), Phase III ($0.806$, p-value $= 0.041$), and Application ($0.698$, p-value $= 0.054$), which are predominantly the stages where regulatory expertise and commercial infrastructure matter most.

The cumulative effect of these stage-by-stage differences compounds into a meaningful gap in unconditional approval rates. Chaining the five group-mean transition probabilities ($0.667 \times 0.588 \times 0.365 \times 0.613 \times 0.937 \approx 0.082$) gives an approximate sense of the unconditional rate; the observation-level chained estimate, which preserves cross-stage covariance, is $0.085$.\footnote{For each drug-indication spell, we predict the conditional success probability at all five stages using the spell's covariates and the stage-specific Gompertz estimates, multiply them to obtain a drug-level unconditional approval probability, and then average within size groups. This approach preserves the positive cross-stage covariance induced by drug and firm characteristics.} The analogous calculation for small firms gives 0.066. These estimates suggest that out of every 100 drug candidates that enter Discovery, around 9 large-firm drugs and 7 small-firm drugs will reach the market.

The expected discount factors $\mathbb{E}(\delta^{\tau_{k\to k+1}})$ in 
Table \ref{tab:gompertz_inputs} translate each transition probability into 
present-value terms by accounting for the time a drug spends at each stage. 
Throughout the paper, we use a flat baseline discount factor of 
$\delta = 0.95$, corresponding to an annual discount rate of approximately 
5.3\%. As a robustness check (Table \ref{tab:sensitivity_scenarios}), we 
also report results using a decade-specific rate, $\delta_{rf+5}$, which 
combines the average ten-year real interest rate in each decade 
(approximately 0.74\% per year over the full sample) with the 
5 percentage-point pharmaceutical risk premium estimated by 
\cite{KoijenPhilipsonUhlig2016}, yielding $\delta_{rf+5} \approx 0.931$ 
on average. The resulting expected discount factors at the baseline 
range from 0.868 at Phase III to 0.953 at Application, and are similar 
across firm sizes.

\paragraph{Heterogeneity in Transition Probabilities.}
Table \ref{tab:transition_heterogeneity_detail} presents Gompertz-predicted transition probabilities across three additional dimensions of heterogeneity. Panel A confirms the firm-size pattern from Table \ref{tab:gompertz_inputs}. Large firms have higher conditional success rates at four of five stages, with the gap most pronounced at Phase III and FDA Application. The overall probability of success across all firms is 7.4\%.

\begin{table}[t!]
\caption{ Heterogeneity in Transition Probabilities}
\label{tab:transition_heterogeneity_detail}
\medskip
\hspace{-0.4in}\begin{tabular}{lccccccc}
\toprule
& \multicolumn{5}{c}{Conditional} & &  \\
\cmidrule(lr){2-6}
& Disc$\to$P1 & P1$\to$P2 & P2$\to$P3 & P3$\to$Appl & Appl$\to$Appr & &
  $p_{\appr|\disc}$ \\
\midrule
\emph{Panel A: Firm Size}&&&&&&&\\
Small firms  & 0.619 & 0.610 & 0.363 & 0.507 & 0.902 & & 0.066 \\
Large firms  & 0.667 & 0.588 & 0.365 & 0.613 & 0.937 & & 0.085 \\
All firms    & 0.615 & 0.604 & 0.355 & 0.592 & 0.925 & & 0.074 \\[6pt]

\emph{Panel B: Therapeutic Area}&&&&&&&\\
Cardiovascular   & 0.550 & 0.468 & 0.301 & 0.579 & 0.862 & & 0.040 \\
Gastrointestinal & 0.586 & 0.665 & 0.335 & 0.586 & 0.886 & & 0.073 \\
Immune disorders & 0.687 & 0.709 & 0.405 & 0.667 & 0.940 & & 0.122 \\
Infectious       & 0.617 & 0.581 & 0.518 & 0.809 & 0.956 & & 0.140 \\
Inflammatory     & 0.615 & 0.653 & 0.396 & 0.595 & 0.928 & & 0.083 \\
Neoplasm       & 0.735 & 0.629 & 0.321 & 0.508 & 0.986 & & 0.073 \\
Neurological     & 0.501 & 0.595 & 0.299 & 0.595 & 0.929 & & 0.049 \\
Rare diseases    & 0.701 & 0.729 & 0.398 & 0.570 & 0.975 & & 0.115 \\[6pt]
\emph{Panel C: Decades}&&&&&&&\\

2000  & 0.587 & 0.642 & 0.377 & 0.576 & 0.897 & & 0.077 \\
2010 & 0.630 & 0.584 & 0.343 & 0.602 & 0.941 & & 0.072 \\

\bottomrule
\end{tabular}
\begin{figurenotes}
All entries are predicted transition probabilities from Gompertz
competing-risks hazard models with Gamma shared frailty, evaluated at
population-averaged predictions. Panel A conditions on firm size (market
capitalization percentile); Panel B on therapeutic area; Panel C on decade of
stage entry. Unconditional probability of approval, $p_{\appr|\disc}$, is
computed by chaining predicted probabilities across all five stages, then
averaging within groups. See Appendix \ref{app:hazard_estimation} for
estimation details. Each panel aggregates across the dimensions not reported;
firm-size-specific probabilities by indication are reported in
Table \ref{tab:disease_heterogeneity}.
\end{figurenotes}
\end{table}

Panel B documents substantial variation in success rates across therapeutic areas. Infectious diseases have the highest unconditional success rate at 14.0\%, driven by strong late-stage performance (Phase III success rate of 80.9\% and application success rate of 95.6\%). Clear clinical endpoints and the regulatory priority often accorded to anti-infectives are consistent with this outcome. Immune disorders and rare diseases also exhibit high success rates, at 12.2\% and 11.5\%, respectively. For rare diseases, favorable regulatory pathways, such as orphan drug designation and accelerated approval, are consistent with the higher late-stage success rates. At the other end, cardiovascular and neurological diseases have the lowest unconditional success rates, at 4.0\% and 4.9\%. These estimates are comparable with previous estimates \citep{Hay2014, Wong2019} that also document low success rates for these therapeutic areas.\footnote{Our estimates suggest slightly higher success probabilities for Neoplasm than other studies for Oncology, possibly because Neoplasm includes non-malignant tumors.}

Panel C examines whether transition probabilities have changed over time. The unconditional approval rates are similar across decades: 7.7\% in the 2000s and 7.2\% in the 2010s. The composition of success, however, has shifted. The 2010s show higher rates at Discovery and at the late stages of Phase III and FDA Application, but lower rates at Phase I and Phase II, which again is consistent with  \cite{Wong2019}.

\paragraph{Public and Private Firms.} Our valuation framework requires stock market 
data, so private firms are necessarily excluded from the valuation step. The 
transition dynamics can be estimated for all firms; however, the framework could 
be applied directly to private firms if their valuations were observable. In 
Appendix \ref{app:public_private} we compare Gompertz estimates using pooled data with those on the publicly traded subsample and find that restricting to publicly traded firms does not adversely affect our 
valuation exercise, and our estimates should be broadly informative about drug 
development in the wider population of firms.

\subsection{Estimated Drug Values and Development Costs}
\label{sec:valuation_results}

We now combine the expected CARs from Section \ref{sec:car_delta} with the transition probabilities and discount factors from Section \ref{sec:gompertz_results} to compute drug valuations at the discovery stage. All results use the competing-risk specification for NDR exits, $\delta = 0.95$, and $p_{\disc} = 0$ unless otherwise noted. Sensitivity to both parameters is examined in Section \ref{sec:sens_rho}.

Before turning to the estimates, we describe how we construct $\Pi_{\appr}$, 
$\Pi_{\disc}$, and $V_{\disc}$. We group announcements into cells defined by firm 
size, alone or interacted with decade or therapeutic indication. Within each cell, 
we average the announcement-level change in the firm's value, defined as the product of the expected CAR from 
Section \ref{sec:car_delta} and the firm's real market capitalization on the 
announcement date, together with the Gompertz-implied transition probabilities and 
expected discount factors, each evaluated at the drug's own covariates. We then 
apply the appropriate formulae \eqref{eq:evs3}, \eqref{eq:evs1} and \eqref{eq:disc_consistent} to these cell means.
Because indications are not mutually exclusive, an announcement is in every 
indication cell to which it belongs.

\paragraph{Baseline Estimates.}
Table \ref{tab:by_size} presents the main results, separately for 
small and large firms. For small-firm drugs, we estimate gross profits 
at approval of $\Pi_{\appr} = \$2.16$ billion, and the corresponding 
present value of expected gross profits at Discovery is $\Pi_{\disc} = 
\$88$ million. The net value of a small-firm discovery, inferred from 
the discovery announcements, is $V_{\disc} = \$50$ million. Because 
the samples underlying $\Pi_{\disc}$ and $V_{\disc}$ are drawn from 
different announcement types and need not contain the same drugs (c.f.  
footnote \ref{footnote:sample}), we cannot recover development costs 
at the individual drug level. Instead, we take the difference of the 
two sample means, which identifies the average total development cost 
as $\mathbb{E}^{Opt}_{\disc}(C) = \Pi_{\disc} - V_{\disc} = \$38$ 
million. For large-firm drugs, the corresponding estimates are 
substantially larger: $\Pi_{\appr} = \$20.8$ billion, $\Pi_{\disc} = 
\$1.1$ billion, $V_{\disc} = \$824$ million, and implied development 
costs of \$305 million. In both cases, expected gross profits exceed the 
discovery valuation, consistent with the market pricing in 
positive development costs.

\begin{table}[t!]
\caption{Average Drug Values and Development Costs, by Firm Size}
\label{tab:by_size}
\hspace{-0.4in}\begin{threeparttable}
\begin{tabular}{lcccccc}
\toprule
Firm Type & Market Cap & $p_{\appr|\disc}$ & $\Pi_{\appr}$ & $\Pi_{\disc}$ &
      $V_{\disc}$ & $ \mathbb{E}_\disc^{Opt}(C)$ \\
\midrule
Small & 4,189 & 0.066 & 2,157 & 88 & 50 & 38 \\
 & & {[0.049, 0.083]} &
  {[848, 3,466]} & {[27, 149]} &
    {[34, 66]} & {[0, 91]} \\[8pt] 
Large & 136,456 & 0.085 & 20,837 & 1,129 & 824 & 305 \\
 & & {[0.059, 0.110]} &
  {[11,959, 29,715]} & {[526, 1,732]} &
    {[529, 1,120]} & {[0, 869]} \\   
\bottomrule
\end{tabular}
\end{threeparttable}
\begin{figurenotes}
Large firms are the top 5\% by market capitalization.
Market Cap is the average market capitalization at Discovery. $p_{\appr|\disc}$
is the unconditional probability of FDA approval conditional on Discovery,
computed by observation-level chaining across all five stages. $\Pi_{\appr}$
is gross profit at approval; $\Pi_{\disc}$ is its present value at
discovery. $V_{\disc}$ is the market's valuation of the Discovery inferred
from the expected dollar CAR. The implied development cost is
$ \mathbb{E}^{Opt}_{\disc}(C) = \Pi_{\disc} - V_{\disc}$. Baseline parameters:
$\delta = 0.95$, $p_{\disc} = 0$, NDR as a competing risk, and all values are in Millions of US Dollars. Confidence
intervals combine cluster-robust standard errors (clustered by firm) for the cell-level dollar CARs with the delta method for Gompertz-derived quantities. The interval for $\mathbb{E}_\disc^{Opt}(C)$ is the one-sided \cite{CoxShi2023} interval. 
\end{figurenotes}
\end{table}

However, the ratio of large-firm to small-firm $\Pi_{\appr}$ is approximately 9.7 to 1, reflecting the larger market capitalizations and dollar CARs of large-firm approvals. The corresponding ratio for $V_{\disc}$ is even larger, at approximately 16 to 1. If large-firm announcements were informative only about the focal drug, the two ratios should be similar, as they would both be reflecting the same size differential. We read this pattern as consistent with large-firm discovery announcements conveying information beyond the focal drug's expected profits, potentially reflecting the value of the firm's broader research platform or pipeline \citep{NiederreiterRiccaboni2022}. The discontinuation-based diagnostic in Section \ref{sec:identification_disc} is consistent with the pipeline interpretation, but also with large firms developing higher-value drugs.

The table also shows the 95\% confidence intervals, which combine cluster-robust standard errors (clustered by firm) for the cell-level dollar CARs with the delta method for Gompertz-derived quantities, evaluating analytical gradients of cell-average predictions at the point estimates and propagating through the Gompertz variance-covariance matrix.\footnote{The delta method is preferred here because the discovery-to-Phase I transition has a strongly negative Gompertz shape parameter ($\hat{\gamma} = -0.38$) and substantial frailty ($\hat{\theta} = 0.65$), which creates a near-cure-fraction parameterization. In simulation-based inference, small perturbations often drive the predicted Stage 1 success probability to zero, leading to degenerate confidence intervals. The delta method avoids this problem by evaluating the gradient at the point estimate, where the local curvature is well-behaved.} Because the development cost is the difference between two estimated quantities, $\mathbb{E}_\disc^{Opt}(C) = \Pi_{\disc} - V_{\disc}$, the model implies a non-negativity constraint, and the standard procedure is inapplicable because the parameter (cost) can be at the boundary \citep{Andrews2000}. So, we construct one-sided confidence intervals for the cost following the procedure in \cite{CoxShi2023}.

\begin{figure}[t!]
\centering
\caption{CDF and PDFs of Values and Profits, by Firm type\label{fig:valuation_ecdf}}
\includegraphics[scale=0.4]{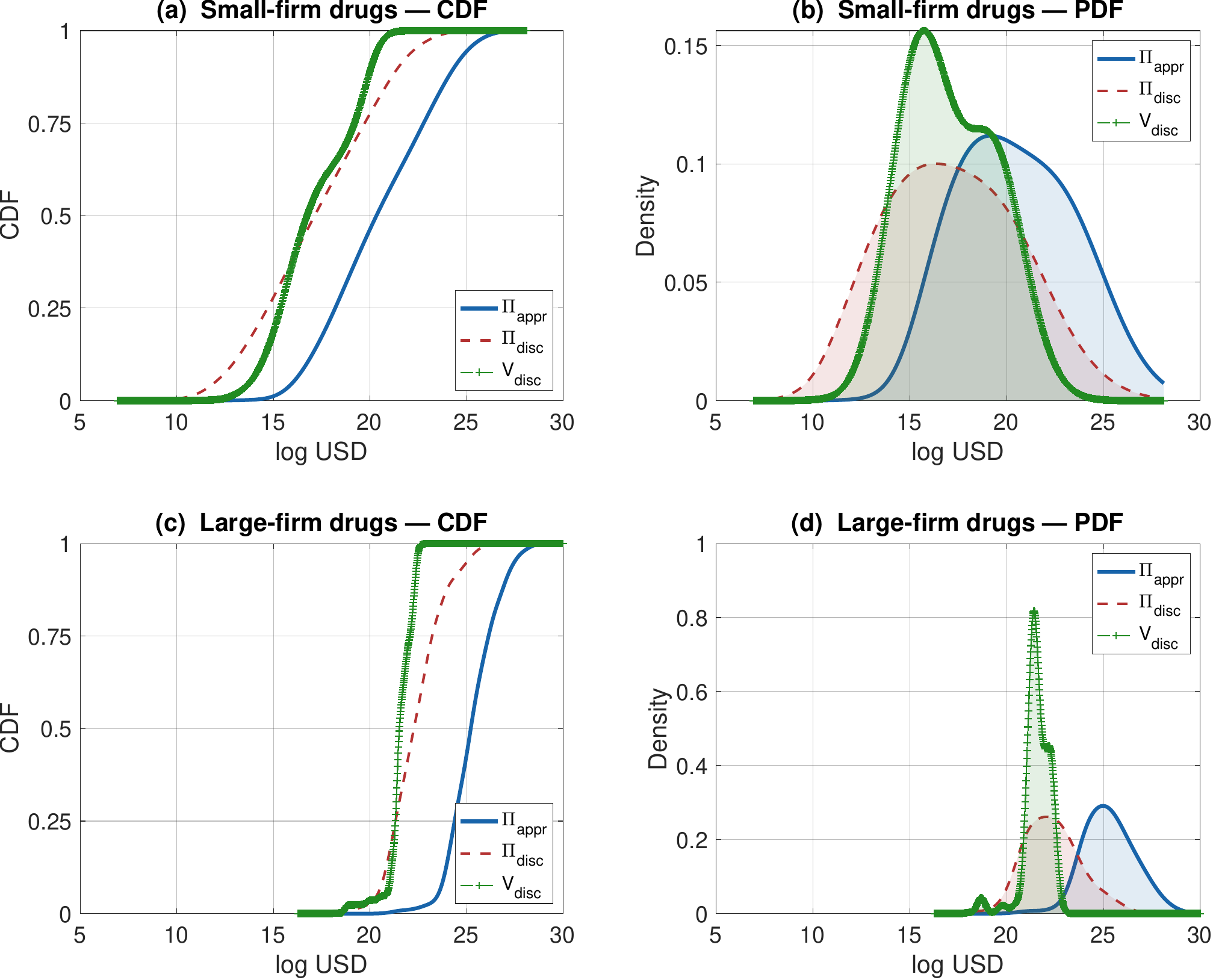}
\begin{figurenotes}
Gaussian kernel density estimates of drug-level valuations separately for small firms (panels a--b) and large firms  (panels c--d). The left panels (a, c) show cumulative distribution functions (CDFs) and the right panels (b, d) show probability density functions (PDFs) of $\hat{\Pi}_\appr$ (blue), $\Pi_{\disc}$ (red), and $V_\disc$ (green). All valuations are expressed in logarithmic USD. Each density is computed over its natural sample of announcements: $\Pi_{\appr}$ and $\Pi_{\disc}$ over approval announcements; $V_{\disc}$ over discovery announcements. Multi-indication drugs contribute once per top indication.
\end{figurenotes}
\end{figure}

Figure \ref{fig:valuation_ecdf} plots CDFs and PDFs of the log of values and gross profit, by firm size. The patterns reinforce the conclusion from Table \ref{tab:by_size} that firm size is the primary determinant of the market's reaction to a discovery announcement. In Appendix \ref{app:additional}, we examine the sensitivity of these estimates to several modeling decisions.

\paragraph{Heterogeneity across Diseases.}
Next, we estimate the mean values and development costs by firm size 
and indication. Because $\Pi_\disc$ and $V_\disc$ are estimated from 
different samples of drugs, rather than taking raw within-cell averages, 
we compute each quantity using cell-specific inputs. That is, the 
Gompertz-predicted transition probabilities and discount factors are 
averaged within each size--indication cell, and the expected dollar CARs 
are averaged within the same cells.\footnote{Specifically, for each 
size--indication cell we average the observation-level Gompertz implied 
$p_{\appr|\disc}$ and $\mathbb{E}(\delta^{\tau})$, and separately average 
the dollar CARs at Discovery and approval.}

\begin{table}[t!!]
\centering
\caption{Drug Values and Development Costs by Therapeutic Area}
\label{tab:disease_heterogeneity}
\begin{threeparttable}
\begin{tabular}{lcccccc}
\toprule
Therapeutic Area & $p_{\appr|\disc}$ & $\Pi_{\appr}$ & $\Pi_{\disc}$ & $V_{\disc}$ & $\mathbb{E}_\disc^{Opt}(C)$ & 95\% CI for $\mathbb{E}_\disc^{Opt}(C)$ \\
\midrule
\multicolumn{6}{l}{\textit{Panel A: Small Firms}} \\
\addlinespace
Cardiovascular   &  0.034 &     1,417 &        30 &        47 & $\dagger$ &  \\
Gastrointestinal &  0.064 &     1,496 &        61 &        58 &         3 & {[0,         59]} \\
Immune           & 0.111 &     2,099 &       145 &        64 &        81 & {[0,        229]} \\
Infectious       & 0.126 &     2,156 &       175 &        38 &       137 & {[0,        337]} \\
Inflammatory     &  0.068 &     1,408 &        61 &        58 &         4 & {[0,         74]} \\
Neoplasm         &  0.065 &     6,576 &       256 &        47 &       208 & {[0,        547]} \\
Neurological     &  0.046 &     2,757 &        80 &        47 &        32 & {[0,        128]} \\
Rare             & 0.105 &     7,096 &       459 &        70 &       389 & {[0,      1,143]} \\
\midrule
\addlinespace[6pt]
\multicolumn{6}{l}{\textit{Panel B: Large Firms}} \\
\addlinespace
Cardiovascular   &  0.048 &    13,245 &       408 &     1,240 & $\dagger$ &  \\
Gastrointestinal &  0.085 &    14,808 &       829 &       754 &        74 & {[0,        825]} \\
Immune           & 0.133 &    15,196 &     1,296 &       507 &       789 & {[0,      1,894]} \\
Infectious       & 0.159 &    31,893 &     3,387 &     1,152 &     2,235 & {[0,      5,789]} \\
Inflammatory     &  0.096 &    16,489 &     1,048 &       741 &       308 & {[0,      1,238]} \\
Neoplasm         &  0.085 &    81,584 &     4,244 &       529 &     3,715 & {[0,      8,968]} \\
Neurological     &  0.053 &    16,789 &       580 &       804 & $\dagger$ &  \\
Rare             & 0.136 &   116,732 &     9,886 &       193 &     9,693 & {[0,     26,043]} \\
\bottomrule
\end{tabular}
\end{threeparttable}
\begin{figurenotes}
Valuations and development costs (in millions of USD), by therapeutic area, 
separately for large firms (top 5\% by market capitalization) and small firms. 
$p_{\appr|\disc}$ is the unconditional approval probability from the hazard estimation. $\Pi_{\appr}$ is gross profit at approval; 
$\Pi_{\disc}$ is its present value at discovery; $V_{\disc}$ is the market's 
net valuation at discovery; $\mathbb{E}_\disc^{Opt}(C)$ is the development 
cost. Baseline: $\delta = 0.95$, $p_{\disc} = 0$, NDR as competing risk. 
CIs are one-sided $[0, \hat{\mathbb{E}_\disc^{Opt}(C)} + z_{0.95}\hat{\sigma}]$ 
following \cite{CoxShi2023}. Multi-indication drugs appear once per 
indication. $\dagger$ indicates a negative implied cost ($V_{\disc} > 
\Pi_{\disc}$), suggesting that discovery announcements in these areas 
convey value beyond the focal drug, such as pipeline or platform 
information. Implied development cost is computed as $\mathbb{E}_\disc^{Opt}(C) 
= \Pi_\disc - V_\disc$ from the unrounded inputs; the rounded values shown 
in adjacent columns may not exactly reproduce it. $p_{\appr|\disc}$ is 
firm-size specific within each indication and may differ from the 
indication-level probabilities in Table \ref{tab:transition_heterogeneity_detail} Panel B, 
which pool across firm sizes.
\end{figurenotes}
\end{table}

Table \ref{tab:disease_heterogeneity} reports the findings. We focus 
first on Panel A, which contains our preferred small-firm estimates. 
Two features stand out. The first is the wide dispersion in gross 
profit values across indications. At one end, rare diseases and 
neoplasms command approval-stage profits of \$7.1 billion and \$6.6 
billion, respectively, likely driven by the pricing power of oncology drugs 
and the monopoly rents that orphan drug designation confers. Infectious 
diseases also generate substantial approval profits (\$2.2 billion), 
likely driven by chronic antivirals (e.g., HIV treatments). For example, Gilead's HIV treatment, Biktarvy, sales were reported at \$14.3 billion in 2025 \citep{Gilead2025}. Neurological and immune drugs sit in the middle of the distribution, at \$2.8 billion and \$2.1 billion, respectively. At the lower end, gastrointestinal, inflammatory, and cardiovascular drugs generate approval profits in a tighter range of \$1.4 to \$1.5 billion, with correspondingly lower implied development costs.

The second feature is the stability of $V_{\disc}$ across indications. 
Despite the large variation in $\Pi_{\appr}$, the net discovery-stage 
value ranges narrowly from \$38 million for infectious diseases to \$70 million for rare diseases, with most cells clustering between \$47 and \$64 million. This stability arises because high approval profits 
are offset by high development costs: for rare diseases, neoplasms, and infectious diseases, implied costs of \$389, \$208, and \$137 million absorb most of 
$\Pi_{\disc}$, leaving net values close to the cross-indication average. Neurological drugs reveal a striking exception, however: with the third-highest approval profit in our sample (\$2.8 billion) but an implied cost of just \$32 million, neurological discoveries appear to deliver substantial commercial value at low marginal expense. The ranking of implied costs across indications, with rare diseases, neoplasms, and infectious diseases at the top, and inflammatory and gastrointestinal drugs at the bottom, is consistent with what has been reported in the literature.\footnote{For example, \cite{SertkayaWongJessupBeleche2016} report broadly similar findings using accounting cost estimates derived from expenses associated with individual clinical trials, by phase and therapeutic area. Their classification omits inflammatory conditions but includes immunomodulation, dermatology, and gastrointestinal categories, which substantially overlap. However,  unlike them, our estimates incorporate costs associated with the initiation of human trials and expenditures not directly tied to trial execution (e.g., manufacturing, trial design, regulatory oversight). They are risk-adjusted with discounting of later-stage costs, particularly in therapeutic areas where progression to late stages is less likely, such as neurology.} We note that we cannot estimate the cost for Cardiovascular drugs, which are marked by $\dagger$ in the table.\footnote{In practice, the implied average cost is negative, which suggests a small ``true'' cost. Negative cost estimates imply that the stock market reaction to discovery announcements is larger than can be rationalized by the corresponding expected discounted approval profits. As we mentioned in the introduction, one possible explanation is that such discoveries contain information relevant to other drugs developed by the firm.}

Panel B shows that large firms follow a broadly similar ranking across indications: rare diseases, neoplasms, and infectious diseases again account for the highest implied costs, at \$9.7 billion, \$3.7 billion, and \$2.2 billion, respectively. For cardiovascular and neurological drugs, we cannot estimate costs, suggesting these are more platform-intensive therapeutic areas, where a single discovery announcement is more likely to convey information beyond the focal drug. The absolute levels for large firms are several times higher throughout, with rare diseases reaching \$116.7 billion in approval profit and neoplasms reaching \$81.6 billion. These eye-popping levels reinforce the contamination argument that large-firm valuations reflect firm-level spillovers rather than the value of a single drug — no individual oncology drug, no matter how successful, plausibly carries an approval-stage value of \$80 billion. We note that all indication-level estimates are imprecise, with confidence intervals that include zero for most cells, so these comparisons should be interpreted with appropriate caution.

\paragraph{Heterogeneity by Decade.}
\begin{table}[t!]
\centering
\caption{Drug Valuations by Firm Size and Decade}
\label{tab:bootstrap_decade}
\begin{threeparttable}
\begin{tabular}{llrc}
\toprule
Firm Type & Decade & $\Pi_{\appr}$ & $\mathbb{E}_\disc^{Opt}(C)$ \\
\midrule
Small & 2000s  &  1,484  & 26 \\
      &        &         & {[0,\ \ \ 67]} \\[4pt]
 & 2010s &  2,590  & 47 \\
      &        &         & {[0,\ \ \ 122]} \\
      \addlinespace[8pt]
      
      Large & 2000s  & 18,616  & $\dagger$ \\
      &        &         & {[0,\ \ \ 205]} \\[4pt]
 & 2010s & 24,397 & 1,254 \\
      &        &         & {[0,\ 1,978]} \\
\bottomrule
\end{tabular}
\end{threeparttable}
\begin{figurenotes}
All dollar values are in millions. Large firms are the
top 5\% by market capitalization. $\Pi_{\appr}$ is gross profit at approval,
computed using observation-level chaining of predicted transition
probabilities across all five development stages. Implied cost is
$\mathbb{E}_\disc^{Opt}(C)$, evaluated at baseline
parameters $\delta = 0.95$, $p_{\disc} = 0$, with NDR treated as a
competing risk. Confidence intervals combine cluster-robust standard errors for the cell-level dollar CARs with the delta method for Gompertz-derived quantities; the interval for the cost is the one-sided \cite{CoxShi2023} interval. $\dagger$ indicates a negative implied cost ($V_{\disc} > \Pi_{\disc}$).
\end{figurenotes}
\end{table}

Table \ref{tab:bootstrap_decade} disaggregates the baseline results by 
firm size and decade. For small firms, implied development costs are 
modest in both periods, at \$26 million in the 2000s and \$47 million 
in the 2010s, even as approval-stage profits rise substantially from 
\$1.5 billion to \$2.6 billion. The cost increase is small in absolute 
terms relative to the profit increase, so most of the additional 
approval-stage profitability accrues to firm value rather than being 
absorbed by higher costs.\footnote{The increase in expected profitability of drugs approved in the 2010s relative to the 2000s may reflect a shift toward specialty, oncology, and rare-disease therapies, which tend to command higher prices per patient. One example is ivacaftor (Kalydeco), developed by Vertex Pharmaceuticals and approved for cystic fibrosis in 2012, with annual sales reaching approximately \$1 billion by 2019 \citep{Vertex2021}.} As 
with the indication-level estimates, the confidence intervals include zero and should be interpreted 
accordingly.
For large firms, we report the corresponding estimates for completeness: 
in the 2000s the implied cost is negative ($V_{\disc} > \Pi_{\disc}$, 
marked $\dagger$), while in the 2010s it is \$1.3 billion. As discussed earlier, large-firm valuations are likely 
inflated by firm-level information spillovers, so these figures should 
not be taken as literal estimates of per-drug development costs.

 \subsubsection{Discovery-Stage Costs}
\label{sec:identification_disc}

As discussed in Section \ref{subsection:consistency}, discontinuations of 
preclinical research enable us to identify the costs of preclinical research, 
which we refer to as discovery-stage costs. They also provide an internal 
consistency check on the model because the discovery-stage cost $C_{\disc}$ 
recovered from equation (\ref{eq:C_disc}) should be small relative to total 
implied development costs $\mathbb{E}_\disc^{Opt}(C)$.

Table \ref{tab:C_disc} reports the results. For small firms, the estimated 
discovery-stage cost is \$18.7 million, or about 49\% of the total implied development cost of \$38 
million. For large firms, $C_{\disc}$ is negative, which is consistent with 
spillovers across drugs. Our estimate of $C_{\disc}$ is also therefore consistent with the baseline 
assumption that for small firms discovery announcements are surprises, i.e., 
$p_{\disc} = 0$. Together with the parameter stability documented in Appendix \ref{sec:sens_rho}, Table 
\ref{tab:sensitivity_scenarios}, this strengthens the case for treating the 
small-firm estimates as the preferred benchmark.

\begin{table}[t!]
\centering
\caption{Discovery-Stage Development Costs}
\label{tab:C_disc}
\begin{threeparttable}
\begin{tabular}{lccc}
\toprule
Firm Type & $V_{\disc}$ & $\mathbb{E}_\disc^{Opt}(C)$ & $C_{\disc}$ \\
\midrule
Small & 50            & 38         & 18.7        \\
           & {[34,\ 66]}   & {[0,\ 91]} & {[0,\ 53]}  \\
\addlinespace
Large & 824               & 305         & $\dagger$    \\
           & {[529,\ 1,120]}   & {[0,\ 869]} & {[0,\ 253]}  \\
\bottomrule
\end{tabular}
\end{threeparttable}
\vspace{4pt}
\begin{figurenotes}
All dollar values are in millions. Confidence intervals are
Cox and Shi (95\%) CIs and appear in brackets below each estimate.
$C_{\disc}$ is the estimate of discovery-stage costs identified from
equation (\ref{eq:C_disc}), using the dollar CARs at Discovery and
discontinuation and the cell-level expected discount factor
$\mathbb{E}(\delta^{\tau_{\disc \to \phasei}})$. $\mathbb{E}^{Opt}_{\disc}(C)$
is the total development cost from Table \ref{tab:by_size}. Both are
evaluated at baseline parameters $\delta = 0.95$ and $p_{\disc} = 0$.
$\dagger$ indicates a negative point estimate.
\end{figurenotes}
\end{table}

In an important paper, \cite{DiMasi2016} estimates the cost is \$2.6 billion per 
approved drug. Their figure includes the cost of every abandoned drug candidate, 
which they observe from the survey, whereas we estimate the expected (average) 
cost at the start of R\&D for a single candidate. Using their transition 
probabilities, average phase durations, and stage-specific costs, we compute the 
discounted expected cost of clinical development that is comparable to our 
estimates.\footnote{\cite{DiMasi2016} do not report the probability of 
transitioning from the discovery stage (referred to as compound synthesis in 
their paper) to clinical development. We assume that compounds entering trials 
are synthesized during the lead optimization stage and use the corresponding 
transition probability from \cite{Pauletal2010}. For discounting, we apply the 
same factor as in our baseline specification, 0.95.} This yields an expected 
cost of clinical development at the time of Discovery of approximately \$71.7 
million (in 2020 USD). While \cite{DiMasi2016} costs are higher than other 
studies \citep{WoutersMcKeeLuyten2020}, including our estimate of \$38 million, 
it lies inside our 95\% confidence interval of $[0, \$91]$ million (Table 
\ref{tab:by_size}). 
The same holds for the discovery-stage share.  \citet{Sertkayaetal2024} estimates the cost of the preclinical phase using the preclinical/clinical ratio from \citet{DiMasi2016} to be approximately \$12.2 million (in 2020 USD), which is close to our estimate of \$18.7 million (Table \ref{tab:C_disc}) and lies inside the 95\% confidence interval of [0, \$53] million. Further, \cite{Pauletal2010} implies a comparable cost of roughly \$17 to \$18 million 
(in 2020 USD).\footnote{We assume our discovery announcements occur at the lead optimization stage, so the comparable
object is the cost of advancing a candidate from lead optimization through preclinical
development. \cite{Pauletal2010} (Figure 2) report cost per development project by stage
(in 2008 USD): lead optimization \$10M and preclinical \$5M. The probability of transitioning from lead optimization is reported to be 0.85, and the duration of the stage is 2 years. Taking the unadjusted sum gives \$15M, or \$18.2M in Dec-2020 USD.
Weighting the preclinical outlay by the probability of reaching it and discounting it two
years at $\delta = 0.95$ gives $10 + 0.85 \times 0.95^{2} \times 5 = \$13.8$M, or  
\$16.7M (in 2020 USD).}

Furthermore, in a survey of 45 estimates, 
\citet{Schlanderetal2021} report that costs per \emph{approved} drug range from 
\$161 million to \$4.54 billion (in 2019 USD), with most estimates below \$1 
billion. While we estimate the expected cost per candidate \emph{entering 
Discovery}, converting it into per-approval terms using $p_{\appr|\disc} = 0.066$ 
(Table \ref{tab:transition_heterogeneity_detail}) gives 
$\$38\text{M}/0.066 \approx \$576$ million (in 2020 USD) or \$569 million (in 2019 USD) in Discovery-date dollars, which falls 
inside that range. 

We also note that there may be at least two reasons why our estimates differ slightly from the 
literature. First, our cost is option-adjusted: each stage cost enters weighted by 
the probability that the project survives to incur it, so the late-stage 
expenditures that dominate published totals receive small weight. 
Second, our sample is small-cap public firms, with almost one third of 
announcements about Neoplasm drugs, whereas \citet{DiMasi2016} restrict attention 
to \emph{self-originated} compounds from 10 multinational pharmaceutical firms, and \citet{Schlanderetal2021} note that most published estimates are also drawn from large pharmaceutical firms. So, our estimates are best interpreted as the market's 
expectation of what a small firm will spend, not the industry-wide 
cost of bringing a compound to market.

\subsection{External Validation using Drug Sales \label{subsec:sales_validation}}

As an external check on our estimate of the gross profit at approval, $\Pi_{\appr}$,
we compare it against the present value of actual drug sales. If our estimates are reasonable, they should be broadly
consistent with what approved drugs actually earn in the market. Importantly,
this comparison is only feasible for a subset of drugs for which sales data are
available, so it constitutes a partial but informative validity check rather
than a full-sample test. We also compare our cost estimates with the estimates from the literature. 

Sales data come from the Cortellis worldwide sales database, which covers
a subset of 1,041 approved drugs over the period 1980--2019.
These drugs are more likely to be commercially
significant. For each drug in the Cortellis
sample, we compute average annual sales across all years with available data
(the median coverage is six years), summing across marketing firms when multiple
companies sell a drug. Of these, 307 can also be matched to an approval
announcement in our event sample and are used for the comparisons that follow.
Panel A of Table \ref{tab:sales_summary} reports the distribution of average
annual worldwide sales per drug for that subsample, separately by firm size.
 
\begin{table}[t!]
\centering
\caption{Drug Sales and Sales-Based Profit, by Firm Size}
\label{tab:sales_summary}
\label{tab:sales_comparison}
\begin{threeparttable}
\small
\setlength{\tabcolsep}{5pt}
\begin{tabular}{l ccc ccc}
\toprule
& \multicolumn{3}{c}{Small firms} & \multicolumn{3}{c}{Large firms} \\
\cmidrule(lr){2-4}\cmidrule(lr){5-7}
Margin & 40\% & 45\% & 50\% & 40\% & 45\% & 50\% \\
\midrule
\multicolumn{7}{l}{\textit{Panel A: Average annual worldwide sales, margin-invariant}} \\
\addlinespace
Mean drug & \multicolumn{3}{c}{447} & \multicolumn{3}{c}{1,264} \\
Median drug & \multicolumn{3}{c}{113} & \multicolumn{3}{c}{816} \\
p75 drug & \multicolumn{3}{c}{368} & \multicolumn{3}{c}{1,482} \\
p90 drug & \multicolumn{3}{c}{1,094} & \multicolumn{3}{c}{2,672} \\
p95 drug & \multicolumn{3}{c}{2,710} & \multicolumn{3}{c}{3,691} \\
\addlinespace
\multicolumn{7}{l}{\textit{Panel B: Sales-based \(\Pi_{\appr}\), perpetuity (\(\delta = 0.95\), factor = 20)}} \\
\addlinespace
Mean drug & 3,572 & 4,019 & 4,465 & 10,111 & 11,375 & 12,639 \\
Median drug & 907 & 1,021 & 1,134 & 6,526 & 7,341 & 8,157 \\
p90 drug & 8,754 & 9,848 & 10,942 & 21,379 & 24,052 & 26,724 \\
p95 drug & 21,683 & 24,394 & 27,104 & 29,524 & 33,215 & 36,905 \\
\midrule
\(N\) drugs & \multicolumn{3}{c}{185} & \multicolumn{3}{c}{122} \\
\bottomrule
\end{tabular}
\end{threeparttable}
\begin{figurenotes}
The sample is the 307 drugs in the Cortellis worldwide sales database for which
an approval announcement is also observed, of which 185 are developed by small firms and 122 by large firms. Panel A reports worldwide sales in 2020 million US dollars using the CPIAUCSL series. 
Here, each drug is one observation, averaged across all years with data and summed across marketing
firms. Sales do not depend on the margin, but for readability, Panel A is shown under the 45\%-margin column. Panel B reports sales-based
\(\Pi_{\appr} = \text{margin} \times \text{avg annual sales} \times
\text{capitalization factor}\), with perpetuity factor \(1/(1-\delta)\).
\end{figurenotes}
\end{table}

To compare sales with $\Pi_{\appr}$, which is a present value at the time of approval, we convert average annual sales into a stock. We multiply average annual sales by a gross profit margin and capitalize
either as a perpetuity or a finite annuity, reflecting different
assumptions about patent life. We define the gross profit margin as
revenues net of cost of goods sold and selling, general, and
administrative expenses, but \emph{before} R\&D, approximately 45\% for
the typical branded pharmaceutical product. When relevant, we also consider a plausible range of
margin between 40\% and 50\%.\footnote{This margin sits between two more familiar benchmarks. It is narrower than the headline gross margin of roughly 76\% reported for large pharmaceutical firms, which nets only cost of goods sold \citep{ledley2020}, and broader than operating or EBITDA margins, which already deduct R\&D. R\&D is added back here because it is precisely what we seek to recover from the cost residual; manufacturing, distribution, marketing, and sales-force expenses are netted out. The 45\% benchmark follows the drug-level cash-flow literature where \citet{grabowski2002returns} report a lifecycle average of 45\% (with 40--50\% sensitivity bounds) for new drugs introduced in the 1990s, an assumption maintained in subsequent work \citep{dimasi2004rd,giaccotto2005}. Firm-level Compustat evidence is consistent, e.g., \citet{ledley2020} imply a pre-R\&D residual of roughly 48\% for large pharmaceutical firms over 2000--2018, and \citet{scherer2001} finds 40--45\% for the industry aggregate over 1962--1996.} 

A further approximation concerns the allocation of sales across drugs and firms.
When multiple firms market the same compound, we sum reported sales
across marketing partners and treat the total as the cash flow
attributable to the drug. 
However, we do not adjust licensing fees,
royalty splits, or co-promotion arrangements that determine how that cash flow is divided. The resulting drug-level sales figures should therefore be read as approximate measures of the commercial scale of
each compound rather than precise per-firm revenues.
Panel B of Table \ref{tab:sales_comparison} reports the resulting sales-based
$\Pi_{\appr}$ under alternative margin assumptions.
 Appendix \ref{app:additional}, Table \ref{tab:sales_capitalization} reports
the corresponding figures under alternative capitalization assumptions.

\paragraph{Sales by firm size.}
Our estimates of $\Pi_{\appr}$ differ across the two groups, so it is useful
to ask how much of that difference reflects large firms bringing higher-value
drugs to market. Large-firm drugs sell more. Their mean real annual sales are
\$1,264 million against \$447 million for small-firm drugs, a ratio of 2.83.
 
Because $\Pi_{\appr}$ is constructed as a cell mean, averaging $\mathbb{E}\left(\text{CAR}_{\appr}\right) \times \MKTCAP$  across approval announcements within each size group, the
comparable sales statistic is the mean of the same group's own sales
distribution. At a 45\% margin under the perpetuity assumption, 
average sales for small firms imply a value of \$4.019 billion, which is comparable with our estimate of \$2.157
billion. For large firms, the implied value is \$11.375 billion, against our estimate of
\$20.837 billion. Therefore, larger firms on average are likely to develop more valuable drugs than smaller firms.

Our estimates are also consistent with \citet{Duboisetal2015}, who
estimate the elasticity of pharmaceutical innovation to the market size of
approximately 0.23 and infer that roughly \$2.5 billion in additional expected revenue is required to support the invention of one additional
new chemical entity. Applying our 45\% gross profit margin, this
translates into a marginal profit of about \$1.13 billion per new
drug (or more precisely a new chemical entity (NCE)). Our
estimate of $\Pi_{\appr}=\$2.157$ billion for small firms lies above
this benchmark. The difference could be because 
\citet{Duboisetal2015} focus on NCEs and report marginal, not average, profit. 
  
\section{Designing Policies to Support Drug Development\label{sec:policysection}}

We now consider how our estimates can be used in designing policies to support drug development. 
We explore a set of policy instruments related to \emph{drug buyouts}, in which the government purchases the 
manufacturing rights to a drug and places it in the public domain. We consider 
two stages at which such support could be implemented: (i) after FDA approval and 
(ii) at the discovery stage. Although the current system is not designed for drug 
buyouts, our analysis is motivated by considerations similar to those in patent 
buyouts \citep{Kremer1998} and transferable patents \cite{DuboisMoissonTirole2022}.

The objective of the government in using drug buyout after approval is to improve access to approved drugs by reducing monopoly-induced deadweight loss, i.e., \emph{ex post} efficiency. 
We also explore tradeoffs the government will face in trying to lower the fiscal cost of policy, taking approval as given and choosing a Myersonian reserve price. Throughout, we use small firm estimates in Tables \ref{tab:by_size} and \ref{tab:disease_heterogeneity}.

\subsection{Drug Buyout After Approval\label{sec:buyout_approval}}

We first consider the case of the government buying out a drug after FDA approval. By then, all uncertainties associated with the drug's development would have been resolved, and all development costs would have been sunk. All that remains is to determine the buyout price. 

A key practical concern for a policy intervention based on our approach is  \cite{Lucas1976}'s critique. To illustrate the problem, consider an example calibrated to our small-firm estimates. The expected CAR associated with FDA approval is 3.23\% (Table \ref{tab:expected_vs_actual}, column 2), and the probability of approval given application $p_{\appr|\appl}$ is $0.9$ (Table \ref{tab:gompertz_inputs}). Suppose that the market capitalization of a hypothetical firm developing the drug is \$4 billion. Then the private value of the drug is $\Pi_{\appr}=(\text{3.23\% of \$4 billion})/(1-0.9)= \$1.3$ billion. So, the government wants to buy the drug for \$1.3 billion plus a markup based on the social value of the drug.
The social value of drugs tends to be larger than the market value \citep{ChabotGoetghbeurGregorie2004, Schrag2004, Yinetal2012, HowardBachBerndtConti2015, ContiGruber2020}, and so, for this example, suppose the social value of the drug is $W=\$5$ billion, and it is common knowledge. Further, suppose the government plans to pay $W$ for the drug.\footnote{Paying more than $\Pi_{\appr}$ is a transfer that is not a requirement of \emph{ex post} efficiency, and any strengthening of \emph{ex ante} incentives is incidental. We think it is reasonable for a democratic government to share the social value it creates with the innovator. Whether it shares full $W$ or part of it is not pinned down by our framework, so we begin with $W$ as a benchmark and later study how far the outlay can be reduced.} 
 
If the government announces that it intends to implement the drug buyout policy \emph{before} the FDA approval, then after the announcement of FDA approval, competing traders will react as if the value of the drug is \$5 billion (its social value) and not \$1.3 billion (private value), because they expect the government to pay its social value. Insofar as the market expects the payout to exceed the private value, we cannot use our approach to estimate the ``true'' market value of drugs \emph{and} use those estimates to implement drug buyout policies.
 
To address this Lucas' critique, we start by envisioning that the government announces its intention to implement a drug buyout with a small \textit{exogenous} probability $\varphi \in (0,1)$ and pay the drug's social value $W$. Then, the expected value of the drug is $\Pi_\appr^*:= (1-\varphi) \times \Pi_{\appr} + \varphi \times W$. Then, Equation (\ref{eq:evs3}) identifies $\Pi_\appr^*$, which in turn identifies the market value as $
\Pi_{\appr} = \frac{\Pi_\appr^* - \varphi \times W}{1 - \varphi}.$ Thus, by committing to implementing the drug buyout program with probability $0<\varphi<1$, the government can estimate the market value of drugs and run the program.  

So far, the drug buyout idea is to offer the full social value $W$ as an incentive for the drug-developing firm. It brings the drug into the public domain and hence provides ex post more efficient access to the drug through lower drug prices. In practice, however, the government may also face fiscal constraints, and paying $W$ outright can be too expensive. In this case, it can use our estimated value distribution to set a take-it-or-leave-it offer $\Pi_0 \leq W$, which lowers the cost of the buyout. The tradeoff is that the firm may not always accept, but the price can be chosen optimally \citep{Myerson1981} to balance the two forces, as we show below.

Let $F_{\Pi_\appr}(\cdot)$ denote the distribution of the $\Pi_\appr$ (e.g., Figure \ref{fig:valuation_ecdf} shows the CDFs and PDFs of log of $\Pi_\appr$). The expected savings from offering $\Pi_0$ instead of $W$ is $ \varphi \times (W - \Pi_0) \times F_{\Pi_\appr}(\Pi_0)$, 
where $F_{\Pi_\appr}(\Pi_0)$ is the probability that the drug's market value is smaller than the offered $\Pi_0$ and the firm accepts the offer. The government can choose $\Pi_0$ \emph{before} the drug's value is realized to maximize its savings (relative to $W$), and the solution is the optimal (reserve) price $\Pi_0^*$ of \cite{Myerson1981}, which is determined by
$
\Pi_0^* + \frac{F_{\Pi_\appr}(\Pi_0^*)}{f_{\Pi_\appr}(\Pi_0^*)} = W,
$
where the left-hand side is the marginal cost of offering $\Pi_0^*$ and the right-hand side is the marginal benefit. Because $\Pi_0^*<W$, the government can use our estimates to lower costs.

We implement this idea using our estimates. The exercise proceeds as follows. We estimate a Weibull distribution for $\Pi_{\appr}$ and use the estimated $\hat{F}_{\Pi_{\appr}}(\cdot)$ and its density $\hat{f}_{\Pi_{\appr}}(\cdot)$ in Myerson's optimality condition to determine $\Pi_0^*$. We do not observe the social value $W$, so we parameterize it as $W = \tau \times \text{median}(\Pi_{\appr})$ and consider $\tau$ from 1 to 10, spanning scenarios in which the government values a drug at anywhere from one to ten times its typical market value. For each value of $W$, after determining the optimal reserve price $\Pi_0^*$, we compute the expected savings $(W - \Pi_0^*) F_v(\Pi_0^*)$. 
 
\begin{figure}[t!!]
\caption{Drug Buyout Policy: Optimal Reserve Price Simulation\label{fig:buyout_sim}}
\centering
\includegraphics[scale=0.3]{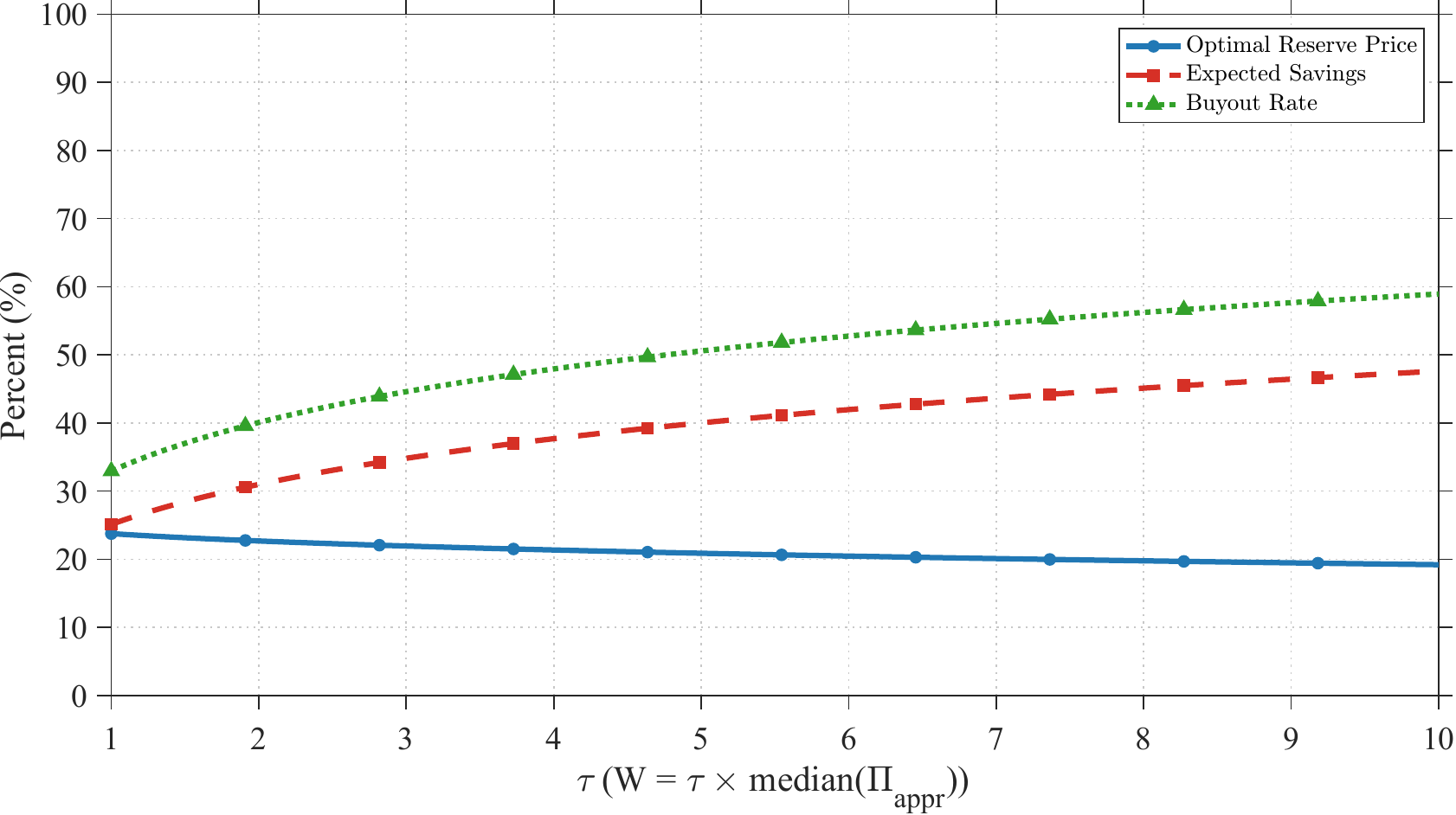}
\begin{figurenotes}
Figure showing the three outcomes of drug buyout simulations (expressed as a percentage of the social value $W$) for drugs developed by small firms. 
Here $\tau$ indexes the assumed social value of a drug relative
to its market value: $W = \tau \times \text{median}(\Pi_{\appr})$. The \textit{optimal
reserve price} is $\Pi_0^*/W \times 100$, where $\Pi_0^*$ is the
solution to the Myerson optimality condition $\Pi_0^* +
F_v(\Pi_0^*)/f_v(\Pi_0^*) = W$, with $F_v(\cdot)$ and $f_v(\cdot)$
denoting the Weibull CDF and PDF fitted to the distribution of
$\Pi_{\appr}$. The \textit{expected savings} series plots
$[(W - \Pi_0^*) \cdot F_v(\Pi_0^*)]/W \times 100$, which measures the
expected reduction in government expenditure per drug relative to paying
full social value $W$ to all firms. The \textit{buyout rate} series
plots $F_v(\Pi_0^*) \times 100$, the fraction of drugs whose market
value falls below $\Pi_0^*$, and these are therefore purchased under the
optimal mechanism.
The Weibull distribution is estimated separately for
each group by maximum likelihood. The estimated median approval-stage
drug values are $\text{median}(\Pi_{\appr}) = \$1.19$ billion, so that $\tau = 2$, for example, corresponds to
$W = \$2.38$ billion.
\end{figurenotes}
\end{figure}
 
 The results are presented in Figure \ref{fig:buyout_sim}, where, for comparison, we express everything as a percentage of the social value 
$W$. The central finding is that the government can achieve substantial savings relative to paying the full social value $W$ across the entire range of $\tau$, and that the savings increase monotonically with $W$. For small 
firms, the optimal reserve price declines from approximately 24\% of $W$ 
at $\tau = 1$ to around 19\% at $\tau = 10$. The corresponding expected 
savings rise from roughly 25\% of $W$ at $\tau = 1$ to 48\% at 
$\tau = 10$, with buyout rates ranging from approximately 33\% to 59\% 
across values of $\tau$. Put differently, even when social value equals 
the median market value, the government needs only to offer roughly 
one-fifth of $W$ to elicit voluntary participation from firms whose 
drugs are worth less than that threshold.  
Although we do not have an estimate for the social value of drugs, these results suggest that the drug buyout program need not be fiscally prohibitive.

A natural consequence of setting $\Pi_0^* < W$, however, is that not every firm will accept the offer. The buyout rate $F_{\Pi_{\appr}}(\Pi_0^*)$ in Figure \ref{fig:buyout_sim} suggests that at $\tau = 1$, roughly one-third of drugs are purchased, rising to just under three-fifths at $\tau = 10$. This means that a non-trivial share of approved drugs remains in private hands under the optimal mechanism. One might worry that this undermines the program's goal of expanding access. We argue that this concern is mitigated once the program is understood as a repeated, evolving institution rather than a one-off intervention. Because $\Pi_0^*$ and $\varphi$ are publicly announced prior to each approval cycle, each round in which the program operates generates a new observation of $\Pi_{\appr}$, which can be used to update $\hat{F}_{\Pi_{\appr}}(\cdot)$ and recalibrate $\Pi_0^*$ for the next round.
 
Formally, from Equation (\ref{eq:evs3}), the market value of the current drug is identified from the observed $\mathbb{E}(\text{CAR}_{\appr}) \times \MKTCAP$ in either case, namely whether $\Pi_{\appr} \leq \Pi_0^*$ or $\Pi_{\appr} > \Pi_0^*$:
\begin{eqnarray*}
\mathbb{E}\left(\text{CAR}_{{\appr}}\right) \times {\MKTCAP}=\begin{cases}
\Big((1-\varphi)\times \Pi_{\appr}+\varphi \times  \Pi_0^*\Big)\times (1-p_{\appr|\appl}),& \text{if} \quad\!\!\Pi_{\appr} \leq \Pi_0^*\\
\Pi_{\appr}\times (1-p_{\appr|\appl}), & \text{if}\quad\!\! \Pi_{\appr} > \Pi_0^*\end{cases},
\end{eqnarray*}
where everything is known except $\Pi_{\appr}$, which allows the government to recover $\Pi_{\appr}$ and update the distribution iteratively. Over time, as $\hat{F}_{\Pi_{\appr}}(\cdot)$ becomes more precisely estimated, the optimal reserve price converges to its true value.

\subsection{Drug Buyout at Discovery and Early-Stage Support}
\label{sec:early_buyout}

We now consider support at Discovery. The tradeoffs
differ from support after approval, because by then clinical uncertainty is
resolved and R\&D costs are sunk, so the government needs only to set a price, but at Discovery, both lie ahead. 
We still consider the primary objective to improve access to approved drugs without reducing the rate of innovation. 
Thus the policy instruments we discuss below have two components: a buyout at Discovery, which 
delivers access, paired with entry-stage policies to support follow-on development.
Table \ref{tab:policy_taxonomy} summarizes the instruments, the margin each
acts on, and the estimate that calibrates it. The
indication-level estimates are imprecise, so the groupings above should be read
as illustrating how the estimates would be used rather than as a ranking of
indications.

\begin{table}[t!]
\centering
\caption{Early-Stage Policy Instruments\label{tab:policy_taxonomy}}
\begin{threeparttable}
\small
\begin{tabular}{@{}p{3.1cm}p{3.1cm}p{3.1cm}p{3.1cm}p{3.3cm}@{}}
\toprule
Instrument & Margin & Information & Indications & Inputs \\
\midrule
Buyout and development contract & Ownership and pricing of the approved drug & Costs (contract); approval only (tournament) & All indications & $\left\{V_{\disc}, \mathbb{E}^{Opt}_{\disc}(C)\right\}$ \\
\addlinespace
Cost sharing & Development cost & R\&D costs & High-cost: rare, neoplasm, infectious, immune & $\left\{\mathbb{E}^{Opt}_{\disc}(C)\right\}$ \\
\addlinespace
Discovery grant & Pre-discovery cost $C_0$ & Only entry & Low-cost: cardiovascular, gastro., inflammatory, neurological & $\left\{V_{\disc}\right\}$ \\
\addlinespace
Advance market commitment & Approval-stage prize & Neither costs nor effort & All. & $\left\{\Pi_{\appr}, p_{\appr|\disc}, \mathbb{E}(\delta^{\tau})\right\}$ \\
\bottomrule
\end{tabular}
\end{threeparttable}
\begin{figurenotes}
Indication groups from small-firm estimates in Table
\ref{tab:disease_heterogeneity}: implied development costs are \$81 to \$389
million for the high-cost group and \$0 to \$32 million for the low-cost group.
For the buyout, the accompanying incentive mechanism is listed in decreasing
order of what the government must observe. Calibrating magnitudes are given in
the text.
\end{figurenotes}
\end{table}

\paragraph{A buyout at Discovery.} Rather than purchasing a drug after
approval, the government may wish to purchase it at Discovery, so that the
compound enters the public domain before development rather than after. 
For instance, the government would offer to pay \$50 million to buy
out the drug (the value of $V_{\disc}$ in Table \ref{tab:by_size}), and a
development contract worth \$38 million to cover the option-adjusted expected
cost, with disease-specific calibrations from Table
\ref{tab:disease_heterogeneity}. The buyout delivers access, and the contract
preserves the innovation incentive that the purchase would otherwise remove. 
The buyout is designed to leave innovation incentives unchanged: 
by purchasing at $V_{\disc}$, it does not affect the entry
margin because it is a transfer to the firm of the present value of
the profits it would have earned on its own. 
As we discussed in Section \ref{sec:buyout_approval}, the buyout should be implemented with
probability $\varphi \in (0,1]$ to address the Lucas critique, though for ease
of exposition we set $\varphi = 1$.

A challenge with a Discovery-stage buyout is that once the drug is public, monopoly profits no longer
accrue to the first firm to obtain approval, eroding the incentives to develop the drug. 
Furthermore, development effort is non-contractible, and tools available to the government to restore incentives depend on its information. The most informationally demanding tools are the \emph{incentive contracts} where firms report their R\&D costs as development proceeds, and the government reimburses expenditure while
preserving effort incentives \citep{LaffontTirole1986}, trading rent extraction
against efficiency \citep{Rogerson1994}. Milestone payments
tied to clinical progress are a practical implementation, linking
compensation to observable outcomes at each stage rather than requiring
continuous monitoring. We do not have estimates of costs by clinical phase, so
milestone payments require additional data. 

When costs are unobserved, the government can instead use a \emph{research tournament} that awards a prize to the first firm to obtain FDA approval, inducing effort through competition \citep{Taylor1995, FullertonMcAfee1999, CheGale2003}. The drug is already public, so the prize must also cover development costs, for which our \$38
million estimate is a lower bound, as no firm would participate for less in
expectation. In general, contracts should also tolerate early
failure, to encourage exploration of high-variance research directions
\citep{Holmstrom1989, Manso2011}. The least informationally demanding
instrument is the advance market commitment, which we discuss below.

\paragraph{Policies to induce entry.} A government aiming to support innovation
would have to use different instruments. But which drugs need support?
 Equation (\ref{eq:recursive_general}) suggests that net values
rise across development stages, so the tightest constraint falls before
Discovery: a firm initiates discovery if $V_0 = \mathbb{E}(\delta^{t_0 \to
t_1}) p_{1|0} V_{\disc} - C_0 \geq 0$, where $C_0$ is the pre-discovery
research cost.\footnote{Estimates of $C_0$ are limited. The most widely cited project-level benchmark \citep{Pauletal2010} implies an out-of-pocket cost of approximately \$6.1 million (in 2020 USD) to successfully reach the lead optimization stage, which is where we assume our discovery announcements begin.} Thus, our estimate of $V_{\disc}$ is an upper bound for $V_0$, and it is
notably stable across indications, ranging from \$38 to \$70 million (Table
\ref{tab:disease_heterogeneity}). If the entry constraint binds, marginal projects have small $V_0$, so a policy raises innovation if it raises $V_{\disc}$ or lowers $C_0$.

\paragraph{Cost sharing and development subsidies.} Consider first a subsidy
that reimburses a fraction $\lambda\in(0,1)$ of expected development cost. Because
$V_{\disc} = \Pi_{\disc} - \mathbb{E}^{Opt}_{\disc}(C)$, such a subsidy raises
the net value of a discovery by $\lambda\times \mathbb{E}^{Opt}_{\disc}(C)$ and hence
relaxes the $V_0 \geq 0$ condition in direct proportion to the development cost
it offsets. So cost sharing is most effective for \emph{high-cost} indications. Table \ref{tab:disease_heterogeneity} Panel A shows that the development costs are
concentrated in a small number of therapeutic areas: \$389 million for rare
diseases, \$208 million for neoplasms, \$137 million for infectious diseases, and \$81 million for immune conditions. For them, a subsidy of even a modest fraction of expected cost is
large relative to a net discovery value of \$38 to \$70 million, and would relax the entry constraint. 

\paragraph{Discovery grants.} In the remaining indications, cardiovascular,
gastrointestinal, inflammatory, and to a lesser extent neurological, with
implied costs of less than \$32 million, cost sharing may not be effective because the development cost is already low, so subsidizing it will barely change $V_0$. We refer to these as \emph{low-cost} indications. 
For such drugs, a discovery grant to defray $C_0$ is a better instrument. 
To see that, note the entry condition is $\mathbb{E}(\delta^{t_0 \to t_1}) p_{1|0} V_{\disc} \geq C_0$, so a grant will offset the cost $C_0$, inducing entry at the margin. Because these drugs are not currently being
developed, such grants help start new R\&D activities without distorting the existing pipeline.

\paragraph{Advance market commitments.} Finally, an advance market commitment
(AMC), under which the government commits to purchase a specified quantity at a
pre-fixed price \citep{KremerGlennerster2004, KremerLevinSnyder2020PnP,
KremerLevinSnyder2022}, can raise innovation incentives in every indication and
requires neither cost observation nor procurement. The commitment functions as
a guaranteed market that restores the profit motive the public-domain status
removes. Our $\Pi_{\appr}$ estimates anchor the commitment in two steps. First,
they pin down the floor: an AMC that commits to pay $\Pi_{\appr}$ at approval
replicates the profit the firm would have earned under exclusivity and
therefore sustains, without raising, the current rate of innovation. For instance, for inflammatory and cardiovascular drugs the floor is \$1.4 billion and for rare diseases \$7.1 billion (see Table \ref{tab:disease_heterogeneity}).

Second, they price the markup. Because we are interested in drugs whose social value $W$ exceeds $\Pi_{\appr}$,
a planner pursuing the ex ante objective would set the commitment above that
floor, and the increment translates into entry incentives at a rate our
estimates identify. In particular, because $\Pi_{\disc} = \mathbb{E}(\delta^{\tau})
p_{\appr|\disc} \Pi_{\appr}$, an additional dollar committed at approval raises
discovery-stage value by $\Pi_{\disc} / \Pi_{\appr}$, which can be determined using our estimates in Table \ref{tab:disease_heterogeneity}. That pass-through ranges from
$2.1\%$ for cardiovascular drugs to $9.2\%$ for immune conditions.
Equivalently, raising $\Pi_{\disc}$ by \$10 million requires a commitment above
the floor of roughly \$109 million for immune conditions but \$469 million for
cardiovascular drugs.
The markup should decrease with the probability of success, so AMC is the most expensive for drugs that have the lowest probability of success and vice versa.
For example, developing drugs for neurological conditions such as Alzheimer's, Parkinson's,
and ALS is difficult \citep{Bespalov2016} with $p_{\appr|\disc} = 0.046$ (see Table \ref{tab:transition_heterogeneity_detail} Panel B). We do not estimate the social value of drugs, so we cannot say for which indications the gap between social and private value is largest. Determining the optimal number of entrants and the total commitment
size requires a portfolio-level model that we leave for future work.

In summary, the four instruments we discussed above act on four different margins: who owns the
approved drug, the cost of development, the cost of discovery, and the size of
the approval prize. The key lesson is that no single instrument acts on all of
them, and our estimates of value and costs help calibrate them. Finding the appropriate mechanism
is beyond our scope, but our estimates provide inputs necessary to calibrate those policies, including the size of the prize, the development cost at discovery, and
the approval probability. 

\section{Conclusion}\label{section:conclusion}

In this paper, we develop a tractable framework for valuing pharmaceutical innovations by combining event-study methods with a discounted cash flow model. Using market responses to drug development announcements, we estimate drug values and development costs.

We find that the value of an approved drug developed by small firms is \$2.16 billion. Discounted back to the discovery stage, the market's net valuation of a discovery is \$50 million, and the option-adjusted development cost evaluated at Discovery is \$38 million. 

We illustrate how these estimates can inform policy design. A government seeking to expand access to innovations through drug buyouts, whether at approval or at Discovery, can use our framework to compute the cost of placing drugs in the public domain. The same estimates can be used to evaluate early-stage subsidies.

Our results point to several directions for future research. First, the contrast between large- and small-firm estimates is a puzzle. Our results are consistent with large-firm announcements conveying information beyond the focal drug, potentially reflecting platform and pipeline spillovers. However, we cannot rule out that large firms develop higher-value drugs, and disentangling these explanations is an open question. Modeling spillovers explicitly, for instance by examining how a discovery announcement affects the market valuation of related drugs in the same firm's pipeline, would help separate the focal drug's value from broader firm-level signals. Then one could repeat our estimation across the full distribution of firm size, for example by size quintiles, and examine the value–size relationship.

Second, competitive dynamics are absent from the current framework. When two firms are developing drugs for the same indication, the market reaction to one firm's announcement likely incorporates information about the other firm's prospects.  
Third, our analysis focuses on milestone announcements because intermediate trial dates do not consistently coincide with the resolution of uncertainty. Linking intermediate milestones to the dates of results disclosure would permit estimation of stage-specific value and R\&D costs, giving a detailed picture of the valuation as drugs progress through the development pipeline.

\bibliographystyle{aer}
\bibliography{stock_pharma}
\begin{appendix}
\begin{center}
\noindent\textbf{\huge{Appendix}}
\end{center}

\singlespacing
\renewcommand{\theequation}{\thesection.\arabic{equation}}
\renewcommand{\thefigure}{\thesection.\arabic{figure}}
\renewcommand{\thetable}{\thesection.\arabic{table}}
\counterwithin*{equation}{section}
\counterwithin*{figure}{section}
\counterwithin*{table}{section}

\section{Data Construction\label{app:data_details}}

In this section, we describe the two datasets used in our empirical analysis. Both datasets draw on the Cortellis database of drug development milestones, but we use them differently. The first is the \textit{announcements dataset}, which links pipeline events to daily stock returns for publicly traded firms. We use it to estimate the cumulative abnormal returns (CARs) associated with each milestone. The second is the \textit{duration dataset}, which tracks how long each drug remains at each development stage and records whether it advances, is discontinued, or is removed from the record; in the latter case, we classify it as \emph{no development reported} (NDR). We use this dataset to estimate stage-specific transition probabilities and expected development times via competing-risks Gompertz hazard models.

The two datasets are built from overlapping Cortellis records, but we treat those records differently. In the announcements dataset, milestone dates are events that may trigger stock price reactions. In the duration dataset, the same dates are used to construct spell lengths and to classify outcomes. We link the two datasets through a common firm-size measure, which we construct in the announcements pipeline and merge into the duration pipeline. This ensures that the transition probabilities entering our valuation framework are conditioned on the same firm characteristics that shape the market's reaction to pipeline news. Figure \ref{fig:data_architecture} presents a schematic of the two pipelines and shows how they feed into the final valuation.

\begin{figure}[htbp]
\centering
\caption{Schematic of the Data Structure and Estimation Framework\label{fig:data_architecture}}

\begin{tikzpicture}[node distance=0.35cm and 0.4cm]


\node[anhdr] (anhdr)
  {\textbf{Announcements pipeline}
   \hfill \textit{\small Unit: firm-drug-date-event}};

\node[annode, below=0.3cm of anhdr, xshift=-4.5cm] (cortellis1)
  {\textbf{Cortellis}\\milestones};

\node[annode, right=0.4cm of cortellis1] (geofilter)
  {\textbf{Geographic \&}\\event-type filters};

\node[annode, right=0.4cm of geofilter] (crsp)
  {\textbf{Merge CRSP}\\returns};

\node[annode, below=0.35cm of cortellis1, xshift=4.5cm] (signal)
  {\textbf{Signal extraction}};

\node[annode, right=0.4cm of signal] (ecar)
  {\textbf{Expected CARs}};

\node[bynode, below=0.3cm of signal, xshift=0cm] (byprod)
  {\textit{Byproducts:}\quad
   firm $\times$ year size percentile
   $\;\cdot\;$
   firm $\times$ decade market cap};

\draw[arr] (cortellis1) -- (geofilter);
\draw[arr] (geofilter)  -- (crsp);
\draw[arr] (crsp)       |- ($(crsp.south)!0.5!(signal.north)$)
                        -| (signal);
\draw[arr] (signal)     -- (ecar);


\node[bridgenode, below=0.55cm of byprod] (bridge)
  {\textit{Firm size flows from announcements to duration pipeline}};
\vspace{0.2in}
\draw[fatarr] (byprod)  -- (bridge);


\node[durhdr, below=2.1cm of byprod] (durhdr)
  {\textbf{Duration pipeline}
   \hfill \textit{\small Unit: drug-indication-stage}};
\draw[fatarr] (bridge.south) -- (durhdr.north);

\node[durnode, below=0.3cm of durhdr, xshift=-4.5cm] (cortellis2)
  {\textbf{Cortellis}\\+ ClinicalTrials.gov};

\node[durnode, right=0.4cm of cortellis2] (usfilter)
  {\textbf{US-only filter}};

\node[durnode, right=0.4cm of usfilter] (phase)
  {\textbf{Phase correction}};

\node[durnode, below=0.35cm of cortellis2, xshift=4.5cm] (reshape)
  {\textbf{Reshape}\\drug-indication-stage};

\node[durnode, right=0.4cm of reshape] (outcome)
  {\textbf{Classify:} success\\  failure  NDR};

\node[durnode, right=0.4cm of outcome] (duration)
  {\textbf{Durations}\\+ merge firm size};
\vspace{0.1in}
\node[durnode, below=0.35cm of reshape, xshift=0cm] (gompertz)
  {\textbf{Gompertz hazard models}\\competing risks $\cdot$ frailty};

\node[durnode, right=0.4cm of gompertz] (transprob)
  {\textbf{Transition probabilities}\\Expected discount factors};

\draw[arr] (cortellis2) -- (usfilter);
\draw[arr] (usfilter)   -- (phase);
\draw[arr] (phase) |- ($(phase.south)!0.5!(reshape.north)$) -| (reshape);
\draw[arr] (reshape)    -- (outcome);
\draw[arr] (outcome)    -- (duration);
\draw[arr] (duration) |- ($(duration.south)!0.5!(gompertz.north)$) -| (gompertz);
\draw[arr] (gompertz)   -- (transprob);


\node[bridgenode, below=0.55cm of transprob, xshift=-4.5cm] (bridge2)
  {\textit{Transition probs + CARs combine in valuation}};

\draw[fatarr] (transprob.south) |- (bridge2.east);
\draw[fatarr] (bridge2.south)   -- ++(0,-0.55);


\node[valhdr, below=2.2cm of transprob, xshift=-4.5cm] (valhdr)
  {\textbf{Valuation}
   \hfill \textit{\small Unit: firm $\times$ size}};

\node[valnode, below=0.3cm of valhdr, xshift=-4.5cm] (inputs)
  {\textbf{Expected CARs}\\+ transition probs};

\node[valnode, right=0.4cm of inputs] (delta)
  {\textbf{Expected Discount factors}};

\node[valnode, right=0.4cm of delta] (pipevalue)
  {\textbf{Value}};

\draw[arr] (inputs)   -- (delta);
\draw[arr] (delta)    -- (pipevalue);

\end{tikzpicture}
\begin{figurenotes}
The announcements pipeline processes milestone records from Cortellis into expected CARs and firm-size measures. The duration pipeline combines Cortellis and ClinicalTrials.gov records to estimate stage-specific transition probabilities via competing-risks hazard models. The valuation step combines both sets of estimates to recover drug values and implied development costs by firm size.
\end{figurenotes}
\end{figure}

\subsection{Data Sources\label{app:data_sources}}

Both datasets draw on four sources, which we describe next.

\paragraph{Cortellis.} Our primary source of drug development information is Cortellis, which is owned and managed by Clarivate Analytics. The database covers more than 70,000 drug candidates worldwide. For each development milestone, Cortellis records the announcement date, the drug name, the associated firm, and the target disease indication. Professional analysts maintain the database using academic articles, patents, press releases, financial filings, earnings calls, conference presentations, and regulatory publications. As noted above, we use these records in two different ways. In the announcements dataset, each milestone is a dated event. In the duration dataset, we use the same dates to measure how long a drug spends at each development stage and to determine if and when that spell ends.

\paragraph{ClinicalTrials.gov.} ClinicalTrials.gov supplements Cortellis in two ways. First, Cortellis provides crosswalks linking its drug-indication entries to ClinicalTrials.gov trial identifiers. We use these links in the duration dataset to correct ambiguous phase designations and to obtain trial start and primary completion dates that fill gaps in the Cortellis timeline. Second, the merged data help us identify the stage at which a given drug-indication was terminated. ClinicalTrials.gov is particularly valuable for identifying the NDR designation, because we can use its trial completion dates to quantify the lag between when a drug actually stopped developing and when Cortellis records its exit.

\paragraph{CRSP.} We obtain daily returns, prices, and shares outstanding for all biomedical and pharmaceutical companies publicly listed on U.S.\ stock exchanges from the Center for Research in Security Prices (CRSP). We match firm names between CRSP and Cortellis using a two-step procedure. First, we generate draft matches using a large language model. Second, we manually validate each match. Firms rename, merge, or are acquired over our sample period, so we rely on CRSP permanent firm identifiers (PERMNOs) rather than names alone to ensure consistent firm identification. A Cortellis firm is considered matched if any historical name associated with its PERMNO corresponds to a name in the Cortellis data.

\paragraph{FRED.} We use monthly CPI data from the Federal Reserve Economic Data (FRED) database to convert nominal market capitalizations to real values when constructing our firm-size measure.

\subsection{Shared Processing Pipeline\label{app:data_processing}}

Before constructing the announcements and duration datasets, we pass the raw Cortellis records through a shared four-stage processing pipeline. Each stage addresses a specific issue in the raw data.

\paragraph{Stage 1: Import and Cleaning.} The first stage brings the raw data into a consistent, analysis-ready form. We standardize country and territory names to ensure consistent geography-related filters. We harmonize firm names across time to account for corporate rebranding and mergers. We correct date-formatting errors introduced during data entry. We map development statuses to numeric identifiers for subsequent classification steps. We resolve duplicate drug records that arise when Cortellis stores both current and historical versions of the same entry. We remove academic institutions, non-profits, and government entities because these are not commercially motivated firms whose stock market reactions we are interested in measuring. Finally, we apply a set of case-by-case corrections for known data entry errors in the status, date, or country fields, identified during manual review of the underlying source documents.

\paragraph{Stage 2: Clinical Trial Linkage.} The second stage links ClinicalTrials.gov records to their corresponding Cortellis drug entries and source documents. We proceed in two passes. In the first pass, we parse multi-drug intervention strings from ClinicalTrials.gov and match trial intervention names against Cortellis drug names within each therapeutic indication. In the second pass, we match remaining unlinked trials based on shared sponsor or collaborator firm names. We also link drug records to extracted source document filenames using both exact and fuzzy name matching, and we merge the development history with source identifier metadata so that each status change is associated with its underlying source document. This linkage is what enables the date validation and phase correction carried out in Stage 3.

\paragraph{Stage 3: Date Validation and CRSP Matching.} The third stage has two purposes: ensuring that event dates are reliable and connecting each firm to its stock market record. On the date side, we cross-reference source dates against ClinicalTrials.gov registry dates and research-assistant-cleaned dates for non-standard source types, and we retain only those dates that can be independently confirmed. On the CRSP side, we match firm names in Cortellis to CRSP stock identifiers using a combination of original manual matches, additional research-assistant-verified matches, and LLM-assisted matches with manual verification. In this stage, we also merge clinical trial dates from both Cortellis and ClinicalTrials.gov, restrict the sample to Western countries, and compute geographic coverage indicators for each event.

\paragraph{Stage 4: Event Classification and Final Assembly.} In the fourth stage, we construct the event-level dataset used in estimation. We classify each source announcement using both a machine learning classifier and manual review into categories such as approvals, submissions, product launches, deals, and other news types. We exclude announcements identified as mergers, acquisitions, or licensing deals because these reflect strategic corporate events rather than scientific progress. When multiple status changes are recorded for the same drug-indication-firm on the same date, we resolve duplicates by retaining the status most consistent with the source classification. We clean U.S. regulatory events separately, including FDA application, approval, and launch, with particular attention to correctly distinguishing approval dates from submission dates. For discontinued projects, we determine the stage of development by determining the highest stage reached prior to the termination event, using the development history at the drug-indication-firm, drug-indication, and drug levels. We then merge the final dataset with CRSP identifiers and real market capitalization data to estimate cumulative abnormal returns (CARs).

\subsection{Announcements Dataset\label{app:announcements_pipeline}}

The announcements dataset links drug development milestones to daily stock returns for publicly traded pharmaceutical and biomedical firms. Each observation is at the firm-drug-indication-date level. The dataset serves two purposes in our analysis. First, it is the source of the CAR estimates that enter the valuation formula. Second, it produces the firm-size measure that is passed to the duration pipeline. We describe each in turn.

\subsubsection{Sample Construction\label{app:sample_construction}}

We build the announcements sample through a sequence of filters. We start from a broad initial set of Cortellis records and progressively apply restrictions designed to ensure that each retained observation represents a single, cleanly attributed, informationally distinct U.S.-relevant drug development event.

\paragraph{Starting sample and event types.} Each observation covers one of four milestone categories: Discovery, FDA Application (also known as pre-registration), FDA Approval (or registration), and Discontinuation. We retain only announcements associated with the United States, the Americas, or worldwide geographic designations. Announcements with these geographic tags constitute approximately 75\% of the raw milestone data. The remaining 25\% come predominantly from Asia (76\% of those excluded), Central and Latin America and the Caribbean (12\%), the Middle East (8\%), and Africa (4\%). We exclude these regions because development conducted solely in non-U.S.\ and non-European markets is unlikely to form the basis of a marketing application to U.S.\ or European regulators. Therefore, announcements from these regions provide limited information for valuing U.S.-listed firms.

\paragraph{Removing deal-related announcements.} For approximately 88\% of raw milestone announcements, Cortellis reports the title of the underlying source document. We examined these titles and found that announcements classified as Discovery, Application, Approval, or Discontinuation are sometimes press releases announcing corporate transactions, such as licensing agreements or acquisitions, rather than genuine pipeline progress. This happens because Cortellis assigns milestone tags based on the drug's status change for a given firm. For example, if a firm licenses a drug that is already registered, Cortellis records a registration event for that firm even though the announcement describes a deal rather than a regulatory decision. We identify and remove deal-related announcements using a bag-of-words classifier applied to source document titles, followed by extensive manual verification. After this step and the geographic restriction, the dataset contains 21,576 firm-drug-date-event observations spanning 1985 to 2019.

\paragraph{Five content filters.} Starting from 21,576 observations, we apply five sequential filters.

\begin{enumerate}

\item \textit{U.S.\ component required.} We retain only announcements that include a U.S.\ component in their country or region field. Drug development milestones announced exclusively in non-U.S.\ markets are unlikely to generate meaningful stock price reactions for U.S.-listed firms.

\item \textit{Exclude small non-U.S.\ regions.} We exclude announcements from small countries or regional groupings that lack a U.S., E.U., or worldwide component. Milestones tagged only to small markets have limited implications for U.S.-listed firm valuations.

\item \textit{First announcement per drug-indication-event.} For each drug-indication pair and event category, we retain only the first (chronologically) announcement. This prevents double-counting when the same milestone is recorded in multiple countries or by multiple firms on different dates.

\item \textit{Single-firm announcements only.} We restrict to announcements attributable to exactly one firm. Approximately 15\% of observations involve multiple firms announcing on the same drug-indication-date, and we drop these observations to ensure unambiguous attribution of the market reaction.

\item \textit{No prior foreign announcement.} We drop U.S.\ announcements for which an earlier announcement of the same drug-indication-event has already occurred in another country. If a milestone was first disclosed abroad, the subsequent U.S.\ announcement carries reduced informational content because the market has likely already processed the news.

\end{enumerate}

After applying all five content filters and restricting to the post-2000 period, the announcements dataset contains 8,440 observations. Table \ref{tab:mid_filters} documents the cumulative effect of each filter.

\begin{table}[t!!]
\centering
\begin{threeparttable}
\caption{Announcements Data: Sequential Filter Application}
\label{tab:mid_filters}
\small
\begin{tabular}{@{}lrc@{}}
\toprule
Step & Observations & Dropped \\
\midrule
Full Announcements Dataset (starting point) & 21,576 & --- \\
After Filter 1: U.S.\ component required & 16,209 & 5,367 \\
After Filter 2: Exclude small non-U.S.\ regions & 16,209 & 0 \\
After Filter 3: First announcement per drug-indication-event
  & 14,817 & 1,392 \\
After Filter 4: Single-firm announcements only & 12,480 & 2,337 \\
After Filter 5: No prior foreign announcement & 12,480 & 0 \\
After year restriction (2000+) & 10,211 & 2,269 \\
After near-duplicate removal and indication collapse
  & 8,440 & 1,771 \\
\bottomrule
\end{tabular}
\begin{figurenotes}Sequential application of filters to construct the announcements dataset. Each row shows the number of observations remaining after the corresponding filter. In the final row, 12 announcements occurring within two weeks of another announcement for the same firm-drug are removed (reducing firm-drug-indication-date observations from 10,211 to 10,199), and we then collapse the remaining observations across indications to yield 8,440 unique firm-drug-date observations. The final estimation sample is further restricted to 5,301 single-announcement firm-dates with available expected CARs across 601 firms.
\end{figurenotes}
\end{threeparttable}
\end{table}

\paragraph{Restriction to estimation event types.} The 8,440-observation dataset contains all four milestone categories. In our expected-CAR estimation, we retain only two positive milestone types, Discovery and Approval, plus discontinuations at the discovery stage. We exclude FDA application announcements because submitting a regulatory application conveys less new information than the final approval decision. We exclude intermediate-phase discontinuations because the market's response to termination depends heavily on the drug's phase, which would require separate modeling. These restrictions remove 1,641 observations: 700 application announcements, 285 Phase I discontinuations, 431 Phase II discontinuations, 201 Phase III discontinuations, and 24 late-stage discontinuations. The signal extraction is then further restricted to single-announcement firm-dates, meaning dates on which exactly one event occurs for a given firm. This yields the final estimation sample of 5,301 observations. The 112 retained discovery-stage discontinuations are included in the variance decomposition as a third announcement type, with their own estimated signal-to-noise ratio. These observations receive expected CARs from a truncated normal distribution constrained to be non-positive, reflecting the fact that terminations can convey only non-positive information about firm value.

\subsubsection{Firm Size Classification\label{app:size_pipeline_details}}

We construct the firm-size measure in the announcements pipeline and then pass it to the duration pipeline. For each announcement date, we rank the announcing firm's real market capitalization against all CRSP firms with non-missing market capitalization on that date. The percentile is computed as $(\mathrm{rank} - 0.5) / N_{\mathrm{firms}}$, yielding a time-varying, event-level measure scaled from 0 to 1. Of the 14,113 unique firm-date pairs in the announcements data, 1,396 (9.9\%) fall on non-trading days, including 603 weekends, 26 major holidays, and 78 other market holidays. We map these to the most recent prior trading day by stepping back up to ten calendar days, which recovers all but one observation. We then obtain the firm-level size classification used in estimation by first taking the median within each firm-year, then the median across firm-years within each decade. This produces a time-invariant, decade-level measure of firm size that is stable enough to condition the hazard model estimates while still capturing the long-run differences in firm scale that are the object of interest in the valuation exercise.

\subsubsection{Single versus Multiple Announcement Dates\label{app:single_announcement}}

On approximately 20\% of firm dates in our sample, firms make more than one drug development announcement on the same day. To isolate stock price reactions to individual events, we restrict to dates with exactly one announcement per firm, because compound announcements make it impossible to attribute the observed return to any single pipeline event. Therefore, our main analyses use single-announcement firm-dates. Table \ref{tab:n_per_day} reports summary statistics for daily announcement counts by type. The median number of announcements per firm-date is 1, but the distribution has a right tail, with a maximum of 23.

\begin{table}[t!]
\begin{center}
\caption{Summary Statistics for Daily Announcements}
\label{tab:n_per_day}
\small
\begin{tabular}{lccccc}
\toprule
Announcement Type & N & Mean & Median & Max & Std.\ Dev. \\
\midrule
Discovery                   & 5,205 & 1.36 & 1 & 19 & 1.02 \\
Approval & 580   & 1.30 & 1 & 23 & 1.21 \\
Discontinued at discovery   & 145   & 1.92 & 1 & 17 & 2.13 \\
All          & 5,917 & 1.37 & 1 & 23 & 1.09 \\
\bottomrule
\end{tabular}
\begin{figurenotes}
Summary statistics for the number of daily announcements by type. The unit of observation is a firm date. Each row restricts to firm-dates with at least one announcement of the given type and reports the distribution of that type's announcement count across those dates. The ``All types combined'' row covers all 5,917 firm-dates in the sample, of which 1,222 (20.7\%) have two or more announcements; the estimation sample is restricted to the remaining single-announcement firm-dates.
\end{figurenotes}
\end{center}
\end{table}

\subsection{Duration Dataset\label{app:duration_pipeline}}

The duration dataset tracks each drug-indication combination through the development pipeline. It records how long each drug spends at each stage and how that stage ends. Constructing this dataset requires us to solve several problems that do not arise in the announcements pipeline. First, stage transitions must be inferred from discrete milestone records and not from what is recorded directly by Cortellis. Second, phase designations are sometimes ambiguous or missing. Third, the same drug may be developed simultaneously by multiple firms. Fourth, many observations are censored because development is still ongoing at the end of our sample window. We address each of these problems through a six-step construction pipeline. Table \ref{tab:duration_pipeline} summarizes the observation counts at each step.

\begin{table}[t!!]
\centering
\begin{threeparttable}
\caption{Duration Data Construction Pipeline}
\label{tab:duration_pipeline}
\small
\begin{tabular}{@{}cllrr@{}}
\toprule
Step &  & Description & Observations & Dropped \\
\midrule
 & & Cortellis milestone records (Western countries) & 689,404 &  \\
 & & \quad After removing duplicate records & 688,434 & 970 \\
 & & \quad After US-only filter & 445,088 & 243,346 \\
1 &  & \quad After CT.gov merge, phase correction, quality
  filters & 146,132 & 298,956 \\
2 &  & Add parent indication hierarchy & 146,132 & 0 \\
3 &  & Reshape; parent inheritance; single-firm restriction
  & 67,354 & 78,778 \\
4 &  & Calculate durations and classify outcomes & 67,354 & 0 \\
5 &  & Drop skipped transitions (retain sequential only)
  & 64,102 & 3,252 \\
6 &  & Summary statistics and quality checks & 64,102 &  \\
\bottomrule
\end{tabular}
\begin{figurenotes} Sequential application of filters to construct the duration dataset. In Step 1, we remove 970 duplicate records, followed by the US-only filter, which removes 35.3\% of the remaining records. The subsequent drops reflect the removal of non-start and non-completion date types. These ambiguous phase records could not be corrected using CT.gov (2,069), non-commercial entities (19,604), records with unknown firm names (24), placeholder and pre-1985 dates (3,790), and missing dates (15,339). In Step 3, we reshape the data from event-level to spell-level (one row per drug-indication-stage), apply parent-indication inheritance (recovering 5,136 records), and restrict to single-firm observations (86.8\% of records retained). In Step 5, we drop 3,252 observations (4.8\%) involving non-sequential stage transitions.
\end{figurenotes}
\end{threeparttable}
\end{table}

\paragraph{Step 1: Base Data and U.S.-Only Filter.} We begin with 689,404 Cortellis milestone records restricted to Western countries, supplemented by merged ClinicalTrials.gov trials from the same regions. The first task is to standardize the data and restrict it to U.S.\ development. We standardize date formats, imputing the first of the month for the 108,991 records that provide only month and year. We resolve 970 duplicate records where Cortellis stores both a current and a historical version of the same entry, keeping the current record, which yields 688,434 records. We then filter to U.S.\ development, dropping 243,346 non-U.S.\ records (35.3\%) and retaining 445,088. The U.S.-only restriction serves three purposes. First, it ensures consistency with ClinicalTrials.gov, which is a U.S.-based registry. Second, it avoids the complications that arise when combining development timelines from different regulatory jurisdictions. Third, it produces duration estimates that reflect a single regulatory environment, which makes the resulting transition probabilities interpretable as probabilities specific to the U.S.\ drug approval pathway.

Next, we process ClinicalTrials.gov records. The 445,088 retained records include 274,275 with information in ClinicalTrials.gov and 170,813 with information only from Cortellis. ClinicalTrials.gov provides multiple dates for every trial. We retain only start dates (40,097 records) and primary completion dates (38,658 records), and we discard 195,520 records associated with other date types, such as submission dates, posting dates, and full study completion dates. We prefer primary completion dates, defined as the last patient's last visit for the primary outcome measure, over full study completion dates. This is because the former more precisely approximates the point at which the clinical work of a given phase concludes. When multiple trials exist for the same drug-indication-firm-status combination, we retain the earliest primary completion date, as this most closely approximates when initial clinical evidence for that phase became available. After reshaping, we obtain 22,705 unique drug-indication-firm-status combinations with a ClinicalTrials.gov start date and 21,948 with a primary completion date. Of these, 9,427 U.S.-based and 6,741 non-U.S.-based ClinicalTrials.gov combinations have no corresponding Cortellis record. We subsequently remove these records using the quality filters described below, so the final output of Step 1 contains only Cortellis-based observations, augmented with ClinicalTrials.gov dates where available.

Approximately 3,067 records in this step carry an ambiguous phase designation, coded as ``Phase not specified,'' ``Phase Not Applicable,'' or ``Clinical'' without a phase number. For 998 of these, we resolve the ambiguity using ClinicalTrials.gov: we assign the most advanced CT.gov phase for the same drug-indication-firm combination that began before the ambiguous record's date. This yields 722 assignments to Phase 3, 183 to Phase 2, and 93 to Phase 1. We drop the remaining 2,069 records whose phase cannot be determined. We then remove 18,391 records from academic institutions and non-profit organizations, 1,189 records from U.S.\ government entities, and 24 records with unknown firm names, because these are not commercially motivated firms whose behavior we can link to stock market data. We apply a crosswalk of 16 known firm name changes to standardize the firm identifiers. We manually correct known data errors in Alzheimer's disease records. Finally, we drop 3,544 records with placeholder dates (coded as 01/01/1200 for unknown dates), 246 records with pre-1985 dates, and 15,339 records with missing dates. The output of Step 1 is 146,132 observations covering 36,320 unique drugs, 81,177 unique drug-indication-firm projects, and 8,200 unique firms. Of these, 45.6\% are from publicly traded companies, 11.5\% from private firms, and 42.9\% are from unclassified companies.

\paragraph{Step 2: Parent Indication Hierarchy.} Cortellis organizes therapeutic indications into a hierarchy in which specific diseases (such as ``non-small cell lung cancer'') are nested under broader categories (such as ``lung cancer'' and ``cancer''). A drug may be developed for a specific indication but have its development history recorded partly at the parent level. To recover this information, we load the Cortellis indication hierarchy, which contains 7,342 rows covering parent paths up to 11 levels deep, and reshape it to identify all ancestor indications for each disease. Because 5,389 rows involve indications with more than one path through the hierarchy, we take the union of all unique ancestors, which yields 13,872 unique indication-parent pairs. We add these parent columns to the base data in preparation for the inheritance procedure in Step 3.

\paragraph{Step 3: Reshape, Inheritance, and Single-Firm Restriction.} In this step, we transform the event-level records into the spell-level structure required by the hazard model, with one observation per drug-indication-stage combination. We also apply three further operations.

The first operation handles the 5,263 records classified as ``Outlicensed.'' For 4,832 of these, we assign the most advanced prior stage: 1,986 are assigned to Discovery, 539 to Phase 1, 1,018 to Phase 2, 300 to Phase 3, 94 to Application, and 895 to Approval. We drop the remaining 431 records with no identifiable prior stage.

The second operation is parent indication inheritance. For each drug-indication-firm combination, we inherit the stage and outcome records from parent indications in the Cortellis disease hierarchy, provided the same firm develops both the child and the parent indication, and the child indication does not already have the relevant stage recorded. This many-to-many join across up to 24 parent columns generates inherited candidate records. After deduplication (keeping the earlier date when both child and parent have the same stage), we retain 7,218 inherited records. We then perform timeline quality checks, removing 932 drug-indication pairs where stage entries appear after approval (146 cases), after termination (767 cases), or after both (19 cases; terminations and approvals may be associated with different firms), as well as individual records where outcomes precede stage starts (35 approval records, 607 termination records, and 400 NDR records) or where termination precedes approval (43 cases).

The third operation restricts to drug-indication-stage combinations attributable to a single firm. Of 77,630 unique combinations, 67,354 (86.8\%) involve exactly one firm. The remaining 10,276 (13.2\%) involve two or more firms (9,343 with two, 842 with three, and 91 with four or more), and we drop these observations. The output contains 67,354 records. By stage, the distribution is: Discovery 40,283 (59.8\%), Phase 1 10,081 (15.0\%), Phase 2 11,082 (16.5\%), Phase 3 3,818 (5.7\%), and FDA Application 2,090 (3.1\%). Of these records, 5,136 (7.6\%) were inherited from parent indications.

\paragraph{Step 4: Duration Calculation and Outcome Classification.} For each drug-indication-stage spell, we determine what happens next. We identify the earliest subsequent stage the drug reaches, allowing for stage-skipping, such as a drug moving directly from Discovery to Phase 2. Among the 13,644 successful transitions, 76.2\% follow the normal sequential path with no stages skipped, 19.2\% skip one stage, 3.4\% skip two, 0.8\% skip three, and 0.3\% skip four. We validate all timelines, setting to missing the 1,835 cases where the destination stage precedes the current stage's start date. We then assign each observation one of four outcomes: \textit{success} (13,644 observations, 20.3\%), meaning the drug advances to any later stage or reaches approval; \textit{failure} (6,893, 10.2\%), meaning the drug is formally discontinued, suspended, or withdrawn; \textit{censored NDR} (22,226, 33.0\%), meaning Cortellis records a ``No Development Reported'' status with no subsequent activity; or \textit{censored ongoing} (24,591, 36.5\%), meaning the drug is still active at the August 30, 2019 data download date. We measure duration as the elapsed time, in months, from the stage start to the outcome date.

\paragraph{Step 5: Sequential Transitions Only.} We restrict the sample to sequential stage transitions, that is, Discovery to Phase 1, Phase 1 to Phase 2, and so on, and we drop 3,252 observations (4.8\%) where a drug skips one or more stages. The most common skipped transition is Discovery directly to Phase 2 (2,097 cases, 64.5\% of all skipped transitions), followed by Discovery to Phase 3 (397, 12.2\%), Phase 1 to Phase 3 (273, 8.4\%), and Phase 2 to Application (144, 4.4\%). By originating stage, 2,643 of the 3,252 skipped transitions (81.3\%) begin at Discovery, 336 at Phase 1, 166 at Phase 2, and 107 at Phase 3. The reason for excluding these observations is that all 3,252 skipped transitions resulted in successful outcomes, with no failures and no censored cases. This degenerate outcome distribution makes it impossible to estimate meaningful cause-specific hazard rates, because there is no outcome variation from which to identify failure and NDR hazards. The skipped transitions also differ substantially in duration from sequential transitions, with a median of 18.0 months compared to 31.6 months for the retained sequential sample. This suggests they reflect genuinely different development pathways, such as accelerated programs, combination trials, or cases where Cortellis did not capture intermediate stages.

\paragraph{Final Dataset.} After all steps, the duration dataset contains 64,102 observations covering 31,100 unique drugs, 53,495 drug-indication projects, 1,769 therapeutic indications, and 6,826 firms, observed from March 1, 1985 through August 29, 2019. The stage distribution is: Discovery 37,640 (58.7\%), Phase 1 9,745 (15.2\%), Phase 2 10,916 (17.0\%), Phase 3 3,711 (5.8\%), and Application 2,090 (3.3\%). The great majority of drug-indication projects are observed at only one stage (84.5\%). Further, 11.9\% span two stages, 3.0\% span three, 0.4\% span four, and only 81 projects (0.15\%) traverse all five stages from Discovery through Application. Of the 53,495 projects in the dataset, 21,014 (39.3\%) reach Phase 1, 13,992 (26.2\%) reach Phase 2, 4,598 (8.6\%) reach Phase 3, 2,090 (3.9\%) reach Application, and 1,600 (3.0\%) achieve approval. For the 64,090 observations with non-missing company-type entries, 44.4\% are from publicly traded firms, 13.1\% from private firms, and 42.5\% are unclassified. Duration data are available for 44,178 observations (68.9\%). Coverage is lower for earlier stages, where right-censoring is more prevalent, rising from 62.0\% in Discovery to 88.9\% in Application. In terms of time coverage, the sample is concentrated in recent decades: 28.6\% of observations have stage start dates from 2015 to 2019, 25.8\% in 2010 to 2014, and 19.7\% in 2005 to 2009.

\subsubsection{Outcome Distribution and Censoring}
\label{app:outcome_distribution}

Understanding the distribution of outcomes is important for appreciating both what the hazard model estimates and what the censoring patterns imply for identification. We address three questions in turn. First, what do the outcomes look like across stages? Second, what does the NDR category mean, and why does it require special treatment? Third, how do the raw-count success rates compare to industry benchmarks?

\paragraph{Outcomes by stage.} Censored observations, combining NDR and ongoing spells, account for 73.0\% of the sequential sample (46,817 observations). This high share reflects the long-term nature of drug development and the substantial fraction of projects whose status remains unresolved at the end of our sample window. The observed success rate, defined as successes as a share of successes plus failures, varies considerably across stages: 58.9\% in Discovery, 68.3\% in Phase 1, 38.7\% in Phase 2, 61.3\% in Phase 3, and 91.4\% in Application. Phase 2 has the lowest success rate. This is consistent with the well-documented Phase 2 attrition problem in clinical development, where the translation from early-stage biological activity to demonstrated efficacy in larger patient populations is most likely to fail. Among the 6,893 failure observations, 89.3\% are classified as discontinued, 10.6\% as suspended, and 0.1\% as withdrawn. Mean durations, using the best available measure and preferring ClinicalTrials.gov dates over Cortellis dates when both are present for sequential transitions, range from 20.8 months in Application to 46.6 months in Discovery, with Phase 1 at 32.1 months, Phase 2 at 36.8 months, and Phase 3 at 33.5 months. Successful transitions have substantially shorter durations than failures or censored spells: a mean of 23.8 months compared to 37.5 months for failures and 47.4 months for censored observations, a pattern that holds across all five stages.

\paragraph{The NDR category and its dating problem.} Of the 46,817 censored observations, 22,226 (34.7\% of all observations) are classified as no development reported (NDR). This means Cortellis records no further development activity for the drug-indication-firm combination without recording a formal termination. NDR is not randomly distributed across stages. It accounts for 44.6\% of Discovery observations but only 5.3\% of Application observations. Within the censored category, the share that is NDR rather than truly ongoing also varies by stage, from 54.0\% in Discovery to 31.9\% in Phase 1 and 38.3\% in Phase 2. The median time to NDR is approximately 42 months across stages. This is longer than the median time to success (16 months) or failure (30 months), a pattern that initially appears puzzling but has a straightforward explanation.

For the 1,136 NDR cases (5.1\% of all NDR observations) with linked ClinicalTrials.gov data, we can directly measure the lag between when development actually stopped and when Cortellis recorded the cessation. The ClinicalTrials.gov primary completion date precedes the Cortellis NDR date by a median of 848 days, or approximately 2.3 years. In 96.6\% of these 1,136 cases (1,097 observations), the ClinicalTrials.gov trial completion date falls before the Cortellis NDR date. The ClinicalTrials.gov-based duration for these cases has a median of 546 days (approximately 18 months), while the Cortellis-based duration has a median of 1,637 days (approximately 54 months), a median difference of 901 days. This evidence strongly supports the interpretation that the Cortellis NDR date reflects the point at which Clarivate analysts noticed that no new information had been filed, rather than the point at which development activity actually ceased. Therefore, durations computed using the dates from Cortellis for NDR spells substantially overstate the true time at risk.

Our baseline specification treats NDR as a third competing risk rather than as simple censoring or as a failure event. This choice reflects the fact that NDR is neither a neutral exit from the data (as true censoring would imply) nor a clean failure signal. It occupies an intermediate position in which development has effectively stopped, but no formal termination has been recorded. Treating it as a distinct risk type allows the model to capture the systematic patterns in when and where NDR occurs without forcing it into either the success or the failure category. Appendix \ref{app:hazard_estimation} reports robustness checks treating NDR as censored, as a failure, or excluding it entirely. Treating NDR as a failure substantially changes the observed success rates: from 58.9\% to 16.5\% in Discovery, from 68.3\% to 43.4\% in Phase 1, from 38.7\% to 21.8\% in Phase 2, from 61.3\% to 47.8\% in Phase 3, and from 91.4\% to 85.9\% in Application.

\paragraph{Validation against industry benchmarks.} As a final check on the plausibility of the duration data, we compute cumulative transition probabilities from raw success-rate counts, excluding censored observations from the denominator. Multiplying the stage-specific success rates from Phase 1 through FDA Application gives $p_{\appr|\phasei} = 0.683 \times 0.387 \times 0.613 \times 0.914 = 0.148$. This falls within the commonly cited industry benchmark range of 10-15\%. The cumulative probability from Discovery through approval is 8.7\%. These raw-count estimates serve as a reference point for evaluating the hazard model predictions, and the close alignment between them and the industry benchmarks provides reassurance that the duration dataset accurately captures the underlying biology and regulatory economics of pharmaceutical development.

\section{Estimating Abnormal Returns\label{section:fama}}

In this section, we present the methodological details underlying the expected cumulative abnormal return (ECAR) estimates reported in the main text. We follow the signal extraction framework of \cite{KoganPapanikolaouSeruStoffman2017} (hereafter KPSS), adapting it to pharmaceutical announcements. As explained in Section \ref{sec:car_delta}, we first determine the CAR for firm $i$ at time $t$. Then, following KPSS, we decompose the CAR into two parts $\text{CAR}_{it} = w_{it} + \varepsilon_{it},$ where $w_{it}$ is the true announcement value as a fraction of the firm's market capitalization and $\varepsilon_{it} \sim \mathcal{N}(0, \sigma^2_{\varepsilon,it})$ is noise. Our object of interest is the posterior mean $\mathbb{E}\text{CAR}_{it} \equiv \mathbb{E}[w_{it} \mid \text{CAR}_{it}]$.

Positive milestones (such as drug discovery and FDA approval) should only add value to a program. So, we model $w_{it} \sim \mathcal{N}^+(0, \sigma^2_{w,it})$, a Gaussian distribution truncated at zero from below. Discontinuation announcements can only convey negative information. So, we model $w_{it} \sim \mathcal{N}^-(0, \sigma^2_{w,it})$, truncated at zero from above. Under these assumptions, the expected CAR for positive announcements is
\begin{equation}
   \mathbb{E}\text{CAR}_{it} = \eta_{it}\,\text{CAR}_{it}
    + \sqrt{\eta_{it}}\,\sigma_{\varepsilon,it}
    \frac{\phi\!\left(-\sqrt{\eta_{it}}\,\dfrac{\text{CAR}_{it}}
    {\sigma_{\varepsilon,it}}\right)}
    {1 - \Phi\!\left(-\sqrt{\eta_{it}}\,\dfrac{\text{CAR}_{it}}
    {\sigma_{\varepsilon,it}}\right)},
    \label{eq:ecar_positive}
\end{equation}
where $\phi$ and $\Phi$ are the standard normal PDF and CDF, and $
    \eta_{it} = \frac{\sigma^2_{w,it}}{\sigma^2_{w,it} +
    \sigma^2_{\varepsilon,it}}$ is the signal-to-noise ratio. The formula for discontinuation announcements is similar to equation \eqref{eq:ecar_positive}, producing non-positive posterior means. Equation \eqref{eq:ecar_positive} is increasing in $\text{CAR}_{it}$: when $\eta_{it}$ is close to one, the expected value tracks the observed return closely; when it is close to zero, the estimate is shrunk toward the prior mean, with the truncation ensuring the correct sign.

To implement equation \eqref{eq:ecar_positive}, we must estimate $\eta_{it}$, which requires estimates of both $\sigma^2_{w,it}$ and $\sigma^2_{\varepsilon,it}$. Estimating these separately for every firm-year cell is infeasible. Following KPSS, we assume that the signal-to-noise ratio is constant within estimation groups defined by firm size, decade, and announcement type, while allowing the noise variance to vary across firm-years. Recall that we classify firms as large or small based on median market capitalization, using the 95th percentile as the cutoff. We further split by decade (2000s and 2010s), which yields four estimation groups.

\paragraph{Variance regression.} Return volatility is higher on announcement days than on non-event days. We exploit this to identify the signal variance. In particular, we regress log-squared CARs on announcement-type indicators:
\begin{equation}
    \log(\text{CAR}_{it}^2) = \sum_{\kappa=1}^{\mathbb{K}} \gamma_\kappa\, X_{it}^{(\kappa)}
    + \alpha_i + \alpha_t + \tilde{u}_{it},
    \label{eq:var_decomp}
\end{equation}
where $X_{it}^{(\kappa)}$ is an indicator for announcement type $\kappa$, and $\alpha_i$ and $\alpha_t$ are firm and year fixed effects, absorbed via iterative demeaning. We winsorize the dependent variable at the 0.5th and 99.5th percentiles before taking logs. Because event-day variance cannot be lower than non-event-day variance, we impose $\gamma_\kappa \geq 0$ via constrained least squares.

\paragraph{Signal-to-noise ratio.} The coefficient $\gamma_\kappa$ measures the log increase in variance on announcement days of type $\kappa$ relative to non-event days. The implied signal-to-noise ratio is $\hat{\eta}_\kappa = 1 - e^{-\hat{\gamma}_\kappa}$.

\paragraph{Noise variance.} For each firm-year cell $(i,t)$, we recover the noise variance by adjusting the total empirical variance for the event-day signal contribution:
\begin{equation}
    \hat{\sigma}^2_{\varepsilon,it} =
    \frac{\mathbb{W} \times \hat{\sigma}^2_{\text{total},it}}
    {1 + \mathbb{W} \times \sum_{\kappa=1}^{\mathbb{K}} d_{it,\kappa}\,(e^{\hat{\gamma}_\kappa} - 1)},
    \label{eq:noise_var}
\end{equation}
where $\mathbb{W} = 3$ is the event window length, $\hat{\sigma}^2_{\text{total},it}$ is the mean squared daily abnormal return in firm $i$'s year $t$, and $d_{it,\kappa}$ is the fraction of trading days in that cell with a type-$\kappa$ announcement. The denominator removes the signal contribution on event days, which provides a cleaner estimate of the noise floor. This correction is negligible when $\gamma_\kappa$ is small, but it is material here, given the substantially larger gammas for pharmaceutical announcements.

\paragraph{Parameter estimates.} Table \ref{tab:signal_extraction} reports the estimated $\hat{\gamma}_\kappa$ (Panel A), the implied signal-to-noise ratios $\hat{\eta}_\kappa$ (Panel B), and the distribution of noise variances across firm-year cells (Panel C). We find that approval events carry the most signal in both size groups and both decades. This is consistent with FDA approval being the most definitive and price-relevant announcement in the pharmaceutical pipeline. Second, the signal content of discontinuation announcements varies markedly across groups. For small firms in the 2010s, discontinuation announcements carry the highest signal-to-noise ratio ($\hat{\eta} = 0.81$), whereas for large firms in the 2000s, they are uninformative ($\hat{\eta} \approx 0.00$). This reflects the fact that a single drug termination is far more consequential for smaller firms. Third, the decade split reveals meaningful time variation: approval-event gammas roughly double from the 2000s to the 2010s for small firms ($0.62 \to 0.93$), consistent with increasing market attentiveness to regulatory milestones over time.

\begin{table}[t!]
\begin{center}
\caption{Signal Extraction Parameters}\label{tab:signal_extraction}
\medskip
\begin{tabular}{lcccc}
\toprule
& Discovery & Approval & Discontinuation & \\
\midrule
\multicolumn{5}{l}{\textit{Panel A: Estimated $\hat{\gamma}_\kappa$ (log-variance increase on event days)}} \\[4pt]
\quad 2000s small firms & 0.2319 & 0.6159 & 0.6277 & \\
\quad 2000s large firms & 0.2572 & 0.2639 & 0.0000 & \\
\quad 2010s small firms & 0.2093 & 0.9292 & 1.6661 & \\
\quad 2010s large firms & 0.0001 & 0.2738 & 0.5256 & \\[8pt]
\multicolumn{5}{l}{\textit{Panel B: Implied signal-to-noise ratio $\hat{\eta}_\kappa = 1 - e^{-\hat{\gamma}_\kappa}$}} \\[4pt]
\quad 2000s small firms & 0.2070 & 0.4599 & 0.4661 & \\
\quad 2000s large firms & 0.2268 & 0.2319 & 0.0000 & \\
\quad 2010s small firms & 0.1888 & 0.6051 & 0.8109 & \\
\quad 2010s large firms & 0.0001 & 0.2395 & 0.4088 & \\[8pt]
\multicolumn{5}{l}{\textit{Panel C: Noise variance $\hat{\sigma}^2_{\varepsilon}$ across firm-year cells}} \\[4pt]
\quad 2000s small firms & \multicolumn{2}{c}{Mean: 0.011091} & \multicolumn{2}{c}{Median: 0.006086} \\
\quad 2000s large firms & \multicolumn{2}{c}{Mean: 0.001413} & \multicolumn{2}{c}{Median: 0.000923} \\
\quad 2010s small firms & \multicolumn{2}{c}{Mean: 0.010101} & \multicolumn{2}{c}{Median: 0.005458} \\
\quad 2010s large firms & \multicolumn{2}{c}{Mean: 0.000561} & \multicolumn{2}{c}{Median: 0.000438} \\
\bottomrule
\end{tabular}
\begin{figurenotes}
Panels A and B report parameters from the variance decomposition in equation \eqref{eq:var_decomp}, estimated separately for each decade $\times$ size group using market-adjusted returns with additive firm and year fixed effects. Large firms are those above the 95th percentile of market capitalization (16 issuers); small firms are those below (494 firms with at least one positive-milestone announcement; 510 issuers in the full sample, corresponding to 601 distinct firm names, since several subsidiaries share an issuer). Panel C reports the distribution of the estimated noise variance $\hat{\sigma}^2_{\varepsilon,it}$ from equation \eqref{eq:noise_var} across firm-year cells.
\end{figurenotes}
\end{center}
\end{table}

\subsection{Raw versus Expected CARs}
\label{subsec:raw_vs_expected}

The signal extraction disciplines the raw CARs in two ways. First, it shrinks each observation toward zero in proportion to the extent to which the total variance on announcement days is noise rather than signal. Second, it enforces the sign constraints implied by economics: expected CARs are non-negative for positive milestones and non-positive for discontinuations, regardless of the raw return on a given day.

\begin{table}[t!]
\centering
\caption{Distribution of Raw and Expected CARs (\%)}
\label{tab:sum_stat_car}
\begin{tabular}{llrrrrrrr}
\toprule
& & \multicolumn{5}{c}{Percentiles} \\
\cmidrule(lr){3-7}
Stage & Group & 10 & 25 & 50 & 75 & 90 & Mean & $N$ \\
\midrule
\multicolumn{9}{l}{\textit{Panel A: Raw CARs}} \\[3pt]
\multirow{2}{*}{Discovery}
 & Large & $-$2.809 & $-$1.225 & 0.052 & 1.309 & 3.121 & 0.088 & 1{,}378 \\
 & Small & $-$7.978 & $-$3.038 & $-$0.062 & 3.296 & 8.697 & 0.641 & 3{,}254 \\[4pt]
\multirow{2}{*}{Approval}
 & Large & $-$2.698 & $-$1.330 & 0.104 & 1.397 & 2.948 & 0.200 & 208 \\
 & Small & $-$6.570 & $-$1.359 & 0.807 & 3.736 & 11.521 & 3.887 & 349 \\[4pt]
\multirow{2}{*}{Discontinued}
 & Large & $-$3.130 & $-$1.629 & $-$0.091 & 0.775 & 4.363 & $-$0.118 & 39 \\
 & Small & $-$12.717 & $-$4.842 & $-$0.016 & 6.895 & 13.546 & $-$0.689 & 73 \\[6pt]
\multicolumn{9}{l}{\textit{Panel B: Expected CARs}} \\[3pt]
\multirow{2}{*}{Discovery}
 & Large & 0.012 & 0.014 & 0.023 & 0.955 & 1.492 & 0.523 & 1{,}378 \\
 & Small & 0.828 & 1.514 & 2.325 & 3.314 & 4.560 & 2.687 & 3{,}254 \\[4pt]
\multirow{2}{*}{Approval}
 & Large & 0.526 & 0.666 & 0.830 & 1.137 & 1.528 & 0.966 & 208 \\
 & Small & 0.902 & 1.389 & 2.210 & 4.132 & 7.884 & 4.581 & 349 \\[4pt]
\multirow{2}{*}{Discontinued}
 & Large & $-$1.192 & $-$0.829 & $-$0.003 & $-$0.002 & $-$0.001 & $-$0.361 & 39 \\
 & Small & $-$10.643 & $-$7.174 & $-$4.298 & $-$1.948 & $-$1.213 & $-$6.436 & 73 \\
\bottomrule
\end{tabular}
\begin{figurenotes}
Raw CARs (Panel A) and expected CARs (Panel B) for single-announcement firm-dates, computed using market-adjusted returns over a $[0,+2]$ trading-day window. Firm size is classified at the 95th percentile of median market capitalization. Expected CARs are non-negative by construction for positive announcements and non-positive for discontinuations. FDA application events are included in the estimation but omitted from this table for brevity; the full sample comprises $N = 5{,}301$ single-announcement firm-dates across 601 firms.
\end{figurenotes}
\end{table}

Table \ref{tab:sum_stat_car} documents the shrinkage. Large firms retain approximately 30.3\% of the raw CAR magnitude and small firms approximately 49.0\%, consistent with the higher signal-to-noise ratios for smaller firms in Panel B of Table \ref{tab:signal_extraction}. The asymmetry is most pronounced for terminations. Raw CARs for small firms span a wide range from $-12.7\%$ to $+13.5\%$ in the p10--p90 interval, reflecting the noisiness of any single day's return. In contrast, expected CARs are tightly concentrated and uniformly negative, as the model requires. For discovery events, where the median raw CAR is close to zero, the non-negativity prior exerts the most influence, and expected CARs can exceed the raw observation by a large margin.

\begin{table}[t!]
\centering
\caption{Raw CAR versus Expected CAR: Regression Comparison}
\label{tab:expected_vs_actual}
\begin{threeparttable}
\begin{tabular}{@{}lcc@{}}
\toprule
& Raw CAR & Expected CAR \\
\midrule
Discovery        & 0.0048**  & 0.0204*** \\
                 & (0.0019)  & (0.0027)  \\[4pt]
Approval         & 0.0251**  & 0.0323*** \\
                 & (0.0121)  & (0.0080)  \\[4pt]
Discontinuation  & $-$0.0049 & $-$0.0432*** \\
                 & (0.0105)  & (0.0131)  \\
\midrule
Observations     & 5{,}301   & 5{,}301   \\
Firms (clusters) & 601       & 601       \\
$R^2$            & 0.0041    & 0.1747    \\
\bottomrule
\end{tabular}
\end{threeparttable}
\begin{figurenotes}
Regression estimates with firm- and year-fixed effects. Bootstrapped standard errors in parentheses: clustered by firm for the Raw CAR column, and from a firm-date bootstrap with 100 resamples for the Expected CAR column. Both specifications use the same three announcement-type indicators on the $N = 5{,}301$ single-announcement firm-date sample (601 firms). The 42.6-fold increase in $R^2$ from Column (1) to Column (2) reflects the noise reduction achieved by signal extraction. * $p < 0.10$, ** $p < 0.05$, *** $p < 0.01$
\end{figurenotes}
\end{table}

The practical payoff is visible in the regression. Table \ref{tab:expected_vs_actual} compares OLS estimates using raw CAR (Column 1) with expected CAR as the dependent variable (Column 2). With raw CARs, the termination coefficient is insignificant, and all standard errors are wide. With expected CARs, all three coefficients are precisely estimated and significant at the 1\% level, and $R^2$ rises from 0.0041 to 0.1747, a 42.6-fold improvement. This indicates that the signal extraction is working as designed. The KPSS framework separates the announcement signal from day-to-day market noise, and the gain is largest precisely where noise is most problematic, i.e., for terminations, where a single bad day can swamp the true economic content of the event.

The dependent variable $\mathbb{E}[\text{CAR}_{it}]$ is 
a generated variable, and the standard errors here are obtained from 
a bootstrap. 
In particular, we bootstrap the entire pipeline at the firm-date level. Each of the 
$B = 100$ replicates resample the data, re-estimate the 
signal-extraction parameters, and recompute $\mathbb{E}[\text{CAR}_{it}]$ 
for every firm-date in that replicate, and re-runs the announcement 
regression.

\subsection{Multi-announcement-day\label{app:multiannouncement}}

Our baseline specification restricts attention to announcement days on which a firm makes a single pipeline announcement. CAR on a day with multiple announcements is a compound signal that cannot be cleanly mapped to a drug-stage-specific surprise without additional assumptions. Nonetheless, multi-announcement days constitute roughly 23\% of discovery events in the expanded sample ($1{,}370$ of $6{,}002$), so it is interesting to examine the sample selection.

To test whether it materially affects our estimates, we re-run the upstream signal-extraction step on the unrestricted sample. Multi-announcement days enter the design matrix through three separate columns---multi-positive, multi-negative, and multi-mixed---that absorb their variance without contaminating the single-event stage columns. The signal-to-noise ratios for the discovery, approval, and discovery-discontinuation stages, therefore, continue to be identified purely off single-announcement days.

Table \ref{tab:singlevsmulti_raw} provides a first check by comparing raw discovery-day CARs across the two subsamples. There is no systematic pattern suggesting that multi-announcement days carry fundamentally different information content than single-announcement days.
Table \ref{tab:singlevsmulti_deltas} reports the implied signal-to-noise ratios $\eta_{\disc}$ and $\eta_{\appr}$ for the two samples. The estimates suggest that the signal-to-noise ratios do not change with sample.

Table \ref{tab:singlevsmulti_means} reports the headline objects directly. Panel A shows the within-type means of the announcement-level expected CARs, and Panel B shows the cell-level dollar CARs by firm size that feed the valuation calculations in Section \ref{sec:valuation_results}. The within-type means in Panel A are similar across samples for Discovery and Approval; the discontinuation mean is somewhat smaller in absolute value under the unrestricted sample, reflecting the inclusion of compound-signal days in the multi-mixed category. The cell-level dollar CARs in Panel B differ by between 2\% and 13\% across samples, with the qualitative ranking and order of magnitude preserved.

Taken together, the three tables show that neither the signal-to-noise ratios nor the cell-level dollar CARs are sensitive to whether multi-announcement days are included or excluded. The single-announcement restriction is a conservative, analytically motivated choice, but it does not drive our results.

\begin{table}[t!!]
\centering
\caption{Raw discovery-day CARs: single-announcement vs.\ multi-announcement days}
\label{tab:singlevsmulti_raw}
\small
\begin{tabular}{llrrrrrr}
\toprule
 & & \multicolumn{3}{c}{Single-announcement} & \multicolumn{3}{c}{Multi-announcement} \\
\cmidrule(lr){3-5} \cmidrule(lr){6-8}
Decade & Size & $N$ & Mean & Median & $N$ & Mean & Median \\
\midrule
2000 & Small  & $1{,}198$ & $+0.01079$ & $-0.00082$ & $315$ & $+0.00473$ & $-0.00525$ \\
 & Large  & $626$   & $+0.00054$ & $-0.00034$ & $235$ & $+0.00569$ & $+0.00129$ \\
2010 & Small  & $2{,}056$ & $+0.00386$ & $-0.00058$ & $594$ & $+0.02080$ & $+0.00438$ \\
 & Large  & $752$   & $+0.00117$ & $+0.00074$ & $226$ & $-0.00453$ & $-0.00329$ \\
\midrule
\multicolumn{2}{l}{Total events} & \multicolumn{3}{r}{$4{,}632$} & \multicolumn{3}{r}{$1{,}370$} \\
\bottomrule
\end{tabular}
\begin{figurenotes}
Raw market-adjusted cumulative abnormal returns (MAR-CARs) on discovery announcement days, split by whether the firm made exactly one pipeline announcement on the event date (single-announcement) or more than one (multi-announcement). The sample covers the baseline mid specification over 2000--2019. Size classification follows the decade-specific percentile rank procedure described in Section \ref{sec:data}: large firms are at or above the 95th percentile of the within-decade market capitalization distribution. Means and medians are computed within each decade-size-subpopulation cell.
\end{figurenotes}
\end{table}

\begin{table}[t!!]
\centering
\caption{Signal-to-noise ratios under single-announcement vs.\ unrestricted sample}
\label{tab:singlevsmulti_deltas}
\small
\begin{tabular}{llcccc}
\toprule
 & & \multicolumn{2}{c}{$\eta_{\disc}$} & \multicolumn{2}{c}{$\eta_{\appr}$} \\
\cmidrule(lr){3-4} \cmidrule(lr){5-6}
Decade & Size & Single & All & Single & All \\
\midrule
2000 & Small  & $0.2070$ & $0.2075$ & $0.4599$ & $0.4601$ \\
 & Large  & $0.2268$ & $0.2280$ & $0.2319$ & $0.2330$ \\
2010 & Small  & $0.1888$ & $0.1892$ & $0.6051$ & $0.6053$ \\
& Large  & $0.0001$ & $0.0002$ & $0.2395$ & $0.2402$ \\
\bottomrule
\end{tabular}
\begin{figurenotes}
Implied signal-to-noise ratios $\eta = (e^{\gamma}-1)/e^{\gamma}$, computed from the MAR-based $\gamma$ estimates under each sample filter. The ratio $\eta$ maps raw CARs to expected CARs in the valuation formula: a value of $\eta$ close to one means most of the observed CAR is attributed to the announcement signal, while a value close to zero means most is attributed to noise. ``Single'' restricts estimation to days on which the firm makes exactly one pipeline announcement. ``All'' estimates $\gamma$ on the unrestricted sample, where multi-announcement days enter the design matrix through three separate columns (multi-positive, multi-negative, multi-mixed); the reported $\eta_{\disc}$ and $\eta_{\appr}$ are therefore still identified off single-announcement days only. Size classification and sample period are as in Table \ref{tab:singlevsmulti_raw}.
\end{figurenotes}
\end{table}

\begin{table}[t!!]
\centering
\caption{Stability of Expected CARs: Single vs.\ All Announcement Days}
\label{tab:singlevsmulti_means}
\begin{threeparttable}
\begin{tabular}{lcc}
\toprule
 & Single & All \\
\midrule
\multicolumn{3}{l}{\textit{Panel A: Mean Expected CAR by Announcement Type}} \\\addlinespace[2pt]
Discovery        & 0.0204 & 0.0184 \\
Approval         & 0.0323 & 0.0309 \\
Discontinuation  & -0.0432 & -0.0305 \\
\addlinespace
\multicolumn{3}{l}{\textit{Panel B: Average Change in Firm's value (millions of USD)}} \\\addlinespace[2pt]
$\overline{D}_{\appr}$ (Small) & 210.6 & 205.8 \\
$\overline{D}_{\appr}$ (Large) & 1,312.5 & 1,241.0 \\
$\overline{D}_{\disc}$ (Small) & 50.0 & 44.7 \\
$\overline{D}_{\disc}$ (Large) & 824.4 & 714.2 \\
\midrule
Observations
 & 5,301 & 6,799 \\
\bottomrule
\end{tabular}
\end{threeparttable}
\begin{figurenotes}
Panel A reports the within-type means of the announcement-level $\mathbb{E}(\text{CAR})$, under the 
single-announcement sample (firm-dates with exactly one pipeline 
announcement) and the unrestricted sample (all announcements, 
with multi-announcement days entering the upstream signal-extraction 
step through separate columns for multi-positive, multi-negative, 
and multi-mixed events). Panel B reports the average of the change in a firm's value  
$\overline{D} = \overline{\mathbb{E}(\text{CAR}) \cdot \MKTCAP}$ following an announcement, ordered 
by firm size, which are used in Tables \ref{tab:by_size}, 
\ref{tab:disease_heterogeneity}, and \ref{tab:bootstrap_decade}. 
Both panels show that the headline quantities are essentially 
identical across the two samples.
\end{figurenotes}
\end{table}

\subsection{Event-Study Plots around Announcements\label{app:eventstudy}}

\begin{figure}[ht!]
\caption{Cumulative Abnormal Returns around Announcements}
\centering
\includegraphics[width=\textwidth]{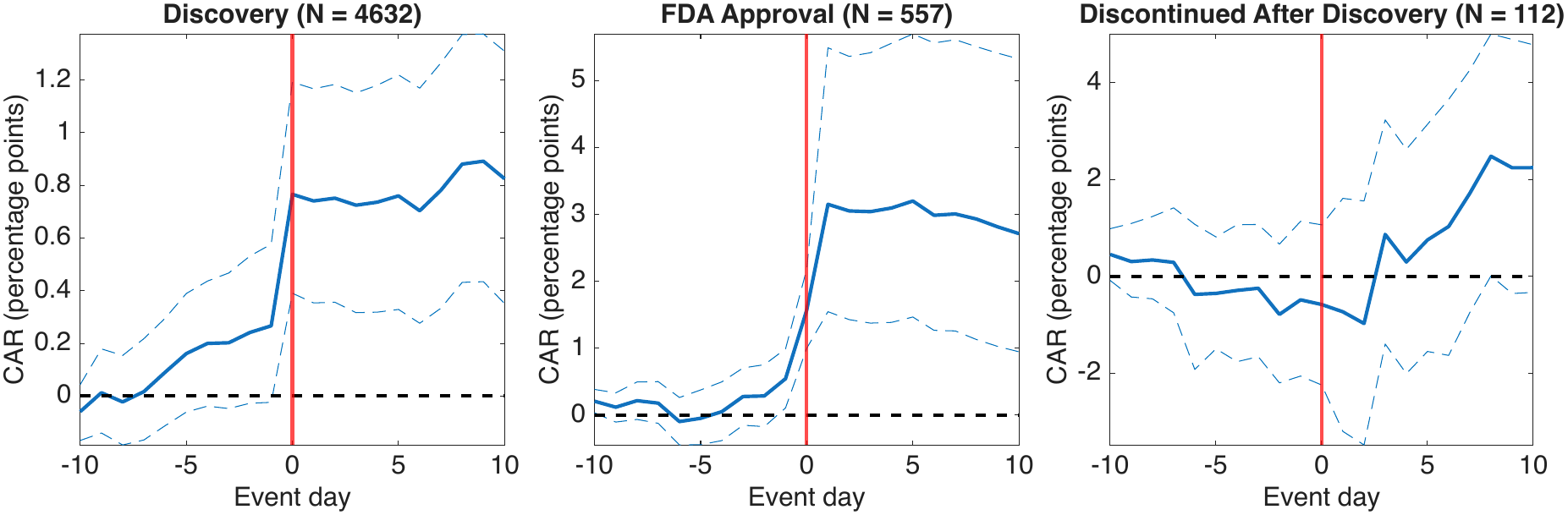}
\label{fig:eventstudy_main}
\begin{figurenotes}
Average raw cumulative abnormal returns (market-adjusted), in percentage points, from 10 trading days before to 10
trading days after the announcement (day 0, red vertical line), for the
three types of announcements. The sample is restricted to
firm-dates with a single announcement,  and dashed lines are pointwise 90\% bootstrap confidence intervals.
\end{figurenotes}
\end{figure}

Figure \ref{fig:eventstudy_main} plots average raw cumulative abnormal
returns from ten trading days before to ten trading days after each of the
three announcement types used in estimation, on the same single-announcement
sample as Section \ref{sec:car_delta}. The CARs shown here are the
market-adjusted returns of Step 1, \emph{before} the signal extraction of
Step 2.
 
For FDA approvals, the response is sharp and
essentially complete within the $[0,+2]$ estimation window: raw prices move
by 2.51 percentage points (90\% confidence interval [0.93, 4.74]),
consistent with the expected CAR of 3.23\%. Second, both types of positive
announcements show some ``trend" ahead of the announcement, of
0.27 percentage points for discovery (against an announcement response of
0.49, confidence interval [0.22, 0.79]) and 0.54 percentage points for
approvals. Our estimation uses data from only the $[0,+2]$ window, so, any
change in value before the event date is missed, which in turn implies smaller development costs.
 
Third, for discontinuations at the discovery stage, the raw response over
$[0,+2]$ is $-0.49$ percentage points, with a 90\% confidence interval of
$[-2.32, 1.13]$ that includes zero.
The pre-announcement path shows no statistically discernible trend. 
The modest response of raw CAR suggests that abandoning one early-stage compound among many, each
with a low probability of success, affects the firm's value only slightly. 
However, because this signal is small relative to the return volatility of
the small firms announcing discontinuation, the signal extraction of Step 2 is
necessary for our exercise. 
 
\section{Hazard Model Estimation \label{app:hazard_estimation}}

In this section, we provide the methodological foundation for the transition probabilities and expected discount factors used in the valuation exercise. Our goal is to estimate, for each stage of drug development, the distribution of the time it takes for a drug to advance to the next stage. We build from the data structure, introduce the competing-risks model, explain our estimation choices, and describe how the estimated hazards feed into the transition probabilities and discount factors reported in the main text.

\subsection{Data Structure and the Competing Risks}

For each drug-indication pair $i$ observed at development stage $j$, we observe a duration and an outcome. The duration is the time spent at stage $j$ before the drug exits. The outcome is one of three events: the drug advances to the next stage (success), is discontinued (failure), or disappears from the data without a recorded outcome, which we call no development reported (NDR). NDR is economically distinct from failure. It captures drugs that neither clearly advance nor clearly terminate, often because the developing firm stops reporting pipeline updates. Treating NDR as failure would overstate attrition, while treating it as a censored observation would ignore information about the drug's trajectory. Therefore, we treat it as a third competing risk in the main specification.

It is helpful to introduce some notation. The observed data for drug $i$ at stage $j$ take the form
\begin{equation}\label{eq:obs}
    T_{ij} = \min\bigl\{T^s_{ij},\; T^f_{ij},\; T^{ndr}_{ij}\bigr\},
    \qquad
    \kappa_{ij} = \arg\min\bigl\{T^s_{ij},\; T^f_{ij},\; T^{ndr}_{ij}\bigr\},
\end{equation}
where $T^s_{ij}$, $T^f_{ij}$, and $T^{ndr}_{ij}$ are the latent times to each event, $T_{ij}$ is the observed exit time, and $\kappa_{ij} \in \{s, f, ndr\}$ identifies which event occurred first. We right-censor observations that have not ended by the data download date (August 30, 2019). Our objective is to estimate the distributions of the three competing risks from $T_{ij}$ and $\kappa_{ij}$.

The identification of cause-specific hazards from data of the form \eqref{eq:obs} is well established in the competing risks literature \citep{FineGray1999, HonoreMuney2006, AbbringVandenberg2007}. Under the assumption that the latent exit times are conditionally independent given the observed covariates and unobserved heterogeneity, the observed time $T_{ij}$ and cause indicator $\kappa_{ij}$ together identify the cause-specific hazard for each event type. Intuitively, the cause indicator tells us which risk ``won'' the race, and the exit time tells us how long the race lasted. Together, they pin down the full hazard function for each cause. The main assumption is that conditional independence holds after controlling for covariates and firm-level frailty, which we model explicitly below.

\subsection{The Gompertz Competing Risks Model with Frailty}

For each cause $j \in \{s, f, ndr\}$ and development stage $k$, we model the hazard for drug $i$ with observed covariates $\mathbf{x}_i$ and unobserved firm-level frailty $\nu_i$ as a Gompertz proportional hazard
\begin{equation}\label{eq:gompertz_cond}
    h_j(t \mid \mathbf{x}_i, \nu_i)
    = \nu_i \cdot \exp\!\bigl(\gamma_j\, t + \beta_{0j}
      + \mathbf{x}_i' \boldsymbol{\beta}_j\bigr).
\end{equation}
The shape parameter $\gamma_j$ allows the baseline hazard to change with time spent in the stage. When $\gamma_j < 0$, a drug that has not yet exited becomes progressively less likely to do so. This captures the empirical reality that drugs lingering in a stage are increasingly unlikely to advance. When $\gamma_j > 0$, the hazard rises with duration, which is consistent with the interpretation that prolonged inactivity predicts eventual exit. The parameter $\beta_{0j}$ is the baseline log-hazard, $\boldsymbol{\beta}_j$ captures the effects of covariates, and $\nu_i$ is the firm-level frailty. Our covariates include the firm size percentile, nine therapeutic area indicators, and a decade indicator (2010s versus the 2000s reference period).

The frailty term $\nu_i$ captures unobserved firm-level heterogeneity that affects all risks simultaneously. A firm with high $\nu_i$ has uniformly elevated hazards across all causes, because it advances drugs faster but also fails and reports NDR at higher rates. We model $\nu_i$ as drawn from a gamma distribution with mean one and variance $\theta_j$, i.e., $\nu_i \sim \text{Gamma}(1,\, \theta_j)$. We estimate the variance parameter $\theta_j$ from the data and, for parsimony, include it only when it is statistically significant at the 5\% level.

The conditional cumulative hazard at the mean frailty $\nu_i = 1$ is
\begin{equation*}
    H_j^{\text{cond}}(t \mid \mathbf{x}_i)
    = \frac{\exp(\beta_{0j} + \mathbf{x}_i' \boldsymbol{\beta}_j)}{\gamma_j}
      \bigl[\exp(\gamma_j t) - 1\bigr].
\end{equation*}
To obtain population-averaged predictions, which are what we need for transition probabilities averaged over the distribution of firms, we integrate out the frailty. For gamma frailty with variance $\theta_j$, the marginal cumulative hazard is $H_j^{\text{marg}}(t) = \frac{1}{\theta_j} \ln\!\bigl(1 + \theta_j \times H_j^{\text{cond}}(t)\bigr)$, and the population-averaged overall survival probability is given by
\begin{equation}\label{eq:pop_surv}
    S^{\text{pop}}(t)
    = \exp\!\Bigl(-\textstyle\sum_j H_j^{\text{marg}}(t)\Bigr)
    = \prod_j \Bigl(1 + \theta_j \times H_j^{\text{cond}}(t)\Bigr)^{-1/\theta_j}.
\end{equation}
The marginal subdensity for cause $j$ is $f_j^{\text{marg}}(t) = h_j^{\text{marg}}(t) \cdot S^{\text{pop}}(t)$, and the cumulative incidence function is $\text{CIF}_j(T) = \int_0^T f_j^{\text{marg}}(t)\, dt$. The transition probability from stage $k$ to $k+1$ is
\begin{equation}\label{eq:trans_prob}
    p_{k+1|k}
    = \frac{\text{CIF}_s^k(T)}{\text{CIF}_s^k(T) + \text{CIF}_f^k(T)},
\end{equation}
which conditions on resolution by excluding the unresolved mass attributable to NDR and ongoing censoring.

We estimate the model independently at each development stage rather than jointly across the full pipeline, primarily because of data coverage. Many drugs enter our dataset at an intermediate stage; for instance, a drug may be observed in Phase II without a prior Phase I record, or in Phase III without an earlier history. Drugs are also licensed, acquired, or transferred between firms, and data collection through Cortellis can be incomplete for the earlier stages of older compounds. By pooling all drugs observed at a given stage regardless of prior history, we substantially increase the effective sample size at each transition. The estimation sample contains 13,766 drug-indication-stage observations across the five stages.\footnote{For sensitivity analysis, we jointly estimate all five stages
with a Bayesian multivariate frailty model that allows the cause-specific
hazards to be correlated across stages, imposing a single common
correlation parameter $\rho$ across stage pairs for tractability. We
cannot reject the null that $\rho = 0$. Estimation details are available
from the authors.}

Furthermore, the transition probabilities estimated here are intended to capture the market's belief about the biological and regulatory probability that a drug advances from one stage to the next. The key identifying assumption in our valuation exercise is that these probabilities are driven by the science of drug development, not by stock market reactions to announcements. Drug development timelines and success rates are more likely to be determined by biology, chemistry, and regulatory science, and the pharmaceutical industry has been accumulating data on these rates for decades.

Firm-level frailty plays an important role in this interpretation. By including a firm-level random effect in every hazard function, we allow for the possibility that some firms are systematically better or worse at navigating the development process. This may be due to unobserved factors such as scientific expertise, manufacturing capability, regulatory knowledge, or portfolio focus. Without frailty, these persistent firm-level differences would be absorbed into the error term, biasing the estimated covariate effects. In frailty models, the persistent component of firm performance is attributed to the latent quality term $\nu_i$, and the observed covariates capture the remaining variation.

\subsection{Estimation Results for Hazard Functions}

We flag duration outliers using stage- and outcome-specific thresholds. We identify right-tail outliers as observations exceeding the median $+ \ell \times$ MAD for $\ell \in \{3, 5\}$. We identify left-tail outliers using the 1st and 5th percentile cutoffs. We right-censor ongoing observations as of August 30, 2019. In our preferred baseline estimate, we exclude outliers at $\ell = 3$ MAD (right tail) and 1st percentile (left tail).

Table \ref{tab:frailty_params} reports the estimated frailty variances $\theta_j$ for each stage and outcome combination in the main specification. Frailty is statistically significant for the success hazard at Discovery ($\theta = 0.665$), Phase I ($\theta = 0.172$), and Phase II ($\theta = 0.214$). For the failure hazard, frailty is significant at Discovery ($\theta = 2.544$), Phase I ($\theta = 1.041$), Phase II ($\theta = 0.384$), and Phase III ($\theta = 0.535$). The especially large failure frailty at Discovery reflects the wide variation in the quality of preclinical evidence across drug candidates entering the pipeline. Some firms advance candidates with strong biological support, while others pursue more speculative targets. In contrast, frailty is indistinguishable from zero for the success hazard at Phase III and Application, and for the failure hazard at Application. This suggests that the observed covariates capture the remaining heterogeneity at late stages.

\begin{table}[t!!]
\centering
\caption{Estimated Frailty Variance ($\theta$) by Stage and Outcome}\label{tab:frailty_params}
\begin{tabular}{lccc}
\toprule
Stage & $\theta_{\text{Success}}$ & $\theta_{\text{Failure}}$ & $\theta_{\text{NDR}}$ \\
\midrule
Discovery       & 0.665  & 2.544  & 0.406 \\
Phase I         & 0.172  & 1.041  & 0.366 \\
Phase II        & 0.214  & 0.384  & 0.528 \\
Phase III       & 0.000  & 0.535  & 0.000 \\
Application & 0.000 & 0.000  & 0.000 \\
\bottomrule
\end{tabular}
\begin{figurenotes}
Frailty variance from the gamma shared frailty model, estimated separately for each stage $\times$ outcome. Values of 0.000 indicate that the likelihood ratio test of $\theta = 0$ was not rejected at the 5\% level (or the frailty model failed to converge), and we use a plain Gompertz instead. Main specification (NDR as competing risk, outliers $k=3$ + P1).
\end{figurenotes}
\end{table}

Table \ref{tab:gompertz_all} reports the full parameter estimates for the success, failure, and NDR hazards across all five stages. Several patterns are worth noting.

For the success hazard, the shape parameter $\gamma$ is negative at four of five stages, which confirms that drugs that linger in a stage are progressively less likely to advance. The exception is Phase III, where $\gamma \approx 0$, indicating that the success hazard is roughly constant with time in the stage. The firm size coefficient is positive and significant at Discovery ($0.642$, $p < 0.01$) and Phase III ($0.806$, $p < 0.05$), and marginally significant at Application ($0.698$, $p < 0.10$). To interpret the magnitude, a move from the bottom to the top of the firm size distribution roughly doubles the success hazard at Discovery ($\exp(0.642) \approx 1.90$). This is consistent with larger firms having greater regulatory expertise and commercial infrastructure at the stages where these resources are most consequential.

For the failure hazard, the shape parameter is positive at Discovery, Phase III, and Application, indicating that the risk of discontinuation rises with time in these stages. The firm-size coefficient on the failure hazard is generally small and insignificant, suggesting that larger firms do not systematically discontinue drugs at lower rates, conditional on observed characteristics. The 2010s decade indicator is negative and significant across all early stages, consistent with an industry-wide reduction in early-stage failure rates over time, potentially reflecting improved target selection and biomarker-driven development.

For the NDR hazard, the shape parameter $\gamma$ is positive and large at every stage. This is consistent with the interpretation that drugs lingering without progress are increasingly classified as NDR. The strong positive duration dependence in the NDR hazard is the reason for treating NDR as a competing risk rather than censoring. NDR is not a random exit but a systematic one, with a rising probability over time within a stage.

\subsection{Transition Probabilities and Expected Discount Factors}

\paragraph{Transition probabilities.} For each stage $k$, the conditional transition probability $p_{k+1|k}$ is the probability that a drug advances to the next stage, $k+1$, conditional on eventually resolving (either advancing or failing). We compute this from equation \eqref{eq:trans_prob} using the marginal cumulative incidence functions evaluated at a horizon of $T = 30$ years, beyond which remaining probability mass is negligible. We obtain the unconditional probability of reaching approval from discovery by chaining the five conditional probabilities, $p_{\appr|\disc} = \prod_{k=1}^{4} p_{k+1|k}$. More generally, the unconditional probability of reaching any downstream stage $m$ from any upstream stage $\ell < m$ is $\prod_{k=\ell}^{m-1} p_{k+1|k}$, which we use for valuation when computing the present value of cash flows at stage $m$ from the perspective of a drug at stage $\ell$.

\paragraph{Expected discount factor.} The expected discount factor for stage $k$, conditional on success, is $$   \mathbb{E}\bigl[\delta^{\tau_{k\mapsto k+1}} \bigr]
    = \frac{\int_0^T \delta^t \cdot f_s^{\text{marg,k}}(t)\, dt}
           {\text{CIF}_s^k(T)},$$
where $f_s^{\text{marg,k}}(t)$ is the marginal subdensity for success in stage $k$ from equation \eqref{eq:pop_surv}. We evaluate the integral numerically and verify convergence by comparing results at $T = 100$ years; the differences are negligible across all stages and discount factors.
The above formula gives the expected discounting over time spent at a single stage. To obtain the expected discount factor for the full path from stage $\ell$ to stage $m$, under the assumption that durations at successive stages are conditionally independent given covariates, we can chain the stage-level expected discount factors, $\mathbb{E}\bigl[\delta^{\tau_{\ell \to m}}\bigr] = \prod_{k=\ell}^{m-1} \mathbb{E}\bigl[\delta^{\tau_k} \bigr]$.\footnote{The chaining formula follows from the independence of stage durations. Write $\delta = e^{-r}$ so that $\delta^{\tau_k} = e^{-r\tau_k}$. The expected discount factor is the moment generating function (MGF) of $\tau_k$ evaluated at $-r$, i.e., $\mathbb{E}[\delta^{\tau_k}] = \mathbb{M}_{\tau_k}(-r)$. For two consecutive stages with durations $\tau_k$ and $\tau_{k+1}$ that are conditionally independent given covariates, the total duration is $\tau_k + \tau_{k+1}$. The MGF satisfies $\mathbb{E}\bigl[\delta^{\tau_k + \tau_{k+1}}\bigr] = \mathbb{M}_{\tau_k + \tau_{k+1}}(-r) = \mathbb{M}_{\tau_k}(-r) \times \mathbb{M}_{\tau_{k+1}}(-r) = \mathbb{E}\bigl[\delta^{\tau_k}\bigr] \times \mathbb{E}\bigl[\delta^{\tau_{k+1}}\bigr]$. Applying this argument inductively gives $\mathbb{E}[\delta^{\tau_{\ell \to m}}] = \prod_{k=\ell}^{m-1} \mathbb{E}[\delta^{\tau_k}]$.}

The baseline discount rate is composed of two parts. The first is the risk-free rate, proxied by the ten-year real interest rate from the FRED database (series REAINTRATREARAT10Y), which averages approximately 0.74\% per year over the 2010s. The second is a pharmaceutical risk premium of five percentage points, following \cite{KoijenPhilipsonUhlig2016}. Combining these two components yields $r_{\text{pharma}} \approx 5.74\%$ per year, or an annual discount factor of $\delta \approx 0.95$. We also compute discount factors using decade-specific risk-free rates to allow for the secular decline in real interest rates over the sample period.

The resulting expected discount factors, reported in Table \ref{tab:gompertz_inputs} of the main text, range from 0.868 at Phase III to 0.953 at Application. The similarity across firm sizes in Table \ref{tab:gompertz_inputs} indicates that the size advantage in unconditional approval rates operates through transition probabilities, not through faster development times.

\subsection{Robustness Specifications}

Our main specification makes two modeling choices: how to handle duration outliers and how to handle NDR observations. We assess sensitivity to both dimensions by estimating seven specifications that span the space of reasonable alternatives. The specifications differ from the main one in the following ways:

\begin{enumerate}
    \item \textit{Main specification.} We exclude duration outliers using MAD-based thresholds at $3$ on the right tail and the 1st percentile on the left tail. We treat NDR as a third competing risk and estimate three cause-specific hazards: success, failure, and NDR. This is the specification whose results are reported in the main text and in Tables \ref{tab:frailty_params}--\ref{tab:gompertz_all}.

    \item \textit{R1: Stricter left-tail outlier exclusion.} We move the left-tail threshold from the 1st to the 5th percentile, which removes a larger fraction of very short durations. The right-tail threshold and the NDR treatment are unchanged from the main specification. Very short durations may reflect data entry errors or instantaneous administrative transitions rather than genuine development activity.

    \item \textit{R2: Looser right-tail outlier exclusion.} We raise the right-tail MAD multiplier from $k = 3$ to $k = 5$, which retains more long-duration observations. The left-tail threshold remains at the 1st percentile. This specification checks whether the results are sensitive to excluding drugs with unusually long stage durations.

\end{enumerate}

\begin{enumerate}
  \setcounter{enumi}{3}
\item \textit{R3: Looser exclusion on both tails.} We combine the 5th-percentile left-tail threshold from R1 with the $k = 5$ right-tail threshold from R2. This is the most permissive outlier treatment and retains the broadest set of observations.

 \item \textit{R4: NDR treated as right-censored.} Instead of a competing risk, we treat NDR observations as right-censored at the last observed date. This reduces the model to two cause-specific hazards (success and failure), and is the most common approach in the applied duration literature. In contrast to the main specification, this approach treats information in NDR observations as uninformative about eventual outcomes.

 \item \textit{R5: NDR treated as failure.} We recode NDR observations as failure events. This is the most conservative treatment, equating prolonged inactivity with drug termination. The model again reduces to two cause-specific hazards. This specification provides an upper bound on the failure rate and a lower bound on the transition probabilities.

    \item \textit{R6: NDR excluded from the sample.} We drop observations that eventually exit via NDR entirely from the estimation sample. This is the most restrictive approach, and is appropriate if one believes that NDRs are systematically different from both success and failure in ways that contaminate the estimation of the other two hazards.

\end{enumerate}

Table \ref{tab:gompertz_robustness} reports the results. Panel A 
displays the firm-size coefficient on the success hazard at each 
development stage across all seven specifications. 
We note the following. 

First, the size effect is robust to outlier treatment. Across 
specifications R1--R3, which vary the outlier thresholds while holding 
the NDR treatment fixed, the coefficient at Discovery ranges from 
$0.592$ to $0.642$ and is significant at the 1\% level in every case. The coefficient at Phase III ranges from $0.806$ to $0.933$ and is 
significant at the 5\% level throughout. These are the two stages where 
the main text identifies the largest size advantage, and neither is 
sensitive to the degree of outlier trimming.

Second, the success hazard coefficients are identical 
across the three NDR treatments (Baseline, R4, R5) that use the same 
estimation sample. This follows from the cause-specific hazard 
structure. The success hazard is estimated from advancement events 
only, so reclassifying NDR exits as censored or as failures does not 
alter the success model. R6, which drops NDR observations entirely and 
reduces the sample from 13,766 to 9,689, is the only NDR variant that 
changes the success coefficients. Removing NDR observations sharpens 
the size effect at late stages: the Phase III coefficient rises from 
$0.806$ to $1.021$ and becomes significant at the 1\% level, and the 
Application coefficient rises from $0.698$ to $0.967$. This is 
consistent with NDR exits introducing noise that attenuates the 
estimated size advantage at the stages where regulatory expertise 
matters most.

\begin{landscape}
\thispagestyle{empty}
\begin{table}[t]
\caption{Gompertz Competing-Risks Hazard Estimates: All Outcomes\label{tab:gompertz_all}}
\scriptsize
\hspace{-0.5in}\setlength{\tabcolsep}{3pt}
\renewcommand{\arraystretch}{0.92}
\begin{tabular}{l *{15}{r}}
\toprule
& \multicolumn{3}{c}{\textit{Discovery}}
& \multicolumn{3}{c}{\textit{Phase I}}
& \multicolumn{3}{c}{\textit{Phase II}}
& \multicolumn{3}{c}{\textit{Phase III}}
& \multicolumn{3}{c}{\textit{Application}} \\
\cmidrule(lr){2-4}\cmidrule(lr){5-7}\cmidrule(lr){8-10}%
\cmidrule(lr){11-13}\cmidrule(lr){14-16}
& S & F & Ndr & S & F & Ndr & S & F & Ndr & S & F & Ndr & S & F & Ndr \\
\midrule
\addlinespace[3pt]
\multicolumn{16}{l}{\textit{Covariates}} \\[3pt]
Firm size       & 0.642** & -0.352 & -0.108
                & -0.225  &  0.425 &  1.019**
                &  0.040  &  0.074 &  0.341
                &  0.806* & -0.689 &  1.743*
                &  0.698$^{\dagger}$ & -1.270 & -2.041 \\
                & (0.221) & (0.365) & (0.152)
                & (0.242) & (0.450) & (0.343)
                & (0.313) & (0.282) & (0.338)
                & (0.393) & (0.510) & (0.701)
                & (0.362) & (0.819) & (1.666) \\[4pt]
Cardiovascular  & -0.158 & -0.021 &  0.037
                & -0.238 & -0.015 &  0.285
                & -0.400 &  0.235 &  0.003
                & -0.527$^{\dagger}$ & -0.448 &  0.262
                & -0.052 &  0.151 &  1.005 \\
                & (0.191) & (0.272) & (0.098)
                & (0.274) & (0.289) & (0.232)
                & (0.289) & (0.193) & (0.195)
                & (0.289) & (0.349) & (0.401)
                & (0.226) & (0.788) & (1.141) \\[4pt]
Gastrointestinal &  0.242* &  0.583** & -0.150$^{\dagger}$
                 &  0.336** & -0.068 &  0.188
                 &  0.100 &  0.374** &  0.163
                 &  0.207 &  0.299 &  0.159
                 &  0.003 &  0.487 &  0.218 \\
                 & (0.118) & (0.206) & (0.082)
                 & (0.126) & (0.198) & (0.159)
                 & (0.158) & (0.113) & (0.114)
                 & (0.185) & (0.214) & (0.345)
                 & (0.159) & (0.711) & (1.236) \\[4pt]
Immune disorders &  0.446*** &  0.260 & -0.167*
                 &  0.253* & -0.277 &  0.015
                 & -0.108 & -0.194 & -0.045
                 &  0.370$^{\dagger}$ & -0.258 &  0.293
                 & -0.047 & -0.024 &  0.066 \\
                 & (0.106) & (0.209) & (0.083)
                 & (0.120) & (0.194) & (0.157)
                 & (0.172) & (0.140) & (0.114)
                 & (0.190) & (0.251) & (0.378)
                 & (0.154) & (0.712) & (0.785) \\[4pt]
Infectious       &  0.214 &  0.024 & -0.049
                 &  0.208 &  0.151 &  0.164
                 &  0.086 & -0.403$^{\dagger}$ & -0.338*
                 &  0.342 & -0.748* &  0.322
                 &  0.536** & -0.182 &  0.506 \\
                 & (0.148) & (0.245) & (0.090)
                 & (0.172) & (0.264) & (0.197)
                 & (0.215) & (0.210) & (0.171)
                 & (0.216) & (0.354) & (0.385)
                 & (0.155) & (0.769) & (1.149) \\[4pt]
Inflammatory     & -0.031 & -0.145 & -0.016
                 &  0.141 & -0.117 & -0.179
                 & -0.285 & -0.038 &  0.002
                 & -0.278 &  0.506$^{\dagger}$ &  0.194
                 &  0.069 & -0.220 &  1.099 \\
                 & (0.139) & (0.218) & (0.079)
                 & (0.149) & (0.215) & (0.173)
                 & (0.183) & (0.156) & (0.138)
                 & (0.214) & (0.284) & (0.381)
                 & (0.158) & (0.652) & (0.968) \\[4pt]
Neoplasm         &  0.755*** &  0.015 & -0.432***
                 & -0.112 & -0.655** & -0.730***
                 & -1.134*** & -0.714*** & -0.942***
                 & -0.514** &  0.105 & -0.392
                 &  0.380* & -1.470 &  1.179 \\
                 & (0.117) & (0.207) & (0.086)
                 & (0.135) & (0.201) & (0.166)
                 & (0.170) & (0.132) & (0.121)
                 & (0.193) & (0.247) & (0.371)
                 & (0.156) & (0.932) & (1.047) \\[4pt]
Neurological     &  0.083 &  0.722** & -0.005
                 &  0.155 & -0.055 &  0.380*
                 & -0.476* &  0.210 & -0.235
                 & -0.015 &  0.222 &  0.369
                 & -0.103 & -0.668 &  1.820$^{\dagger}$ \\
                 & (0.145) & (0.211) & (0.088)
                 & (0.173) & (0.236) & (0.187)
                 & (0.223) & (0.155) & (0.151)
                 & (0.222) & (0.293) & (0.400)
                 & (0.189) & (0.862) & (1.024) \\[4pt]
Rare diseases    &  0.280** & -0.278 & -0.511***
                 &  0.393** & -0.111 & -0.092
                 &  0.105 & -0.211$^{\dagger}$ & -0.070
                 &  0.089 &  0.166 & -0.573
                 &  0.208 & -0.933 &  0.311 \\
                 & (0.102) & (0.219) & (0.097)
                 & (0.118) & (0.209) & (0.180)
                 & (0.157) & (0.125) & (0.112)
                 & (0.175) & (0.213) & (0.383)
                 & (0.134) & (1.038) & (0.892) \\[4pt]
Other            & -0.108 &  0.331 & -0.025
                 &  0.023 & -0.011 &  0.321$^{\dagger}$
                 & -0.311 &  0.009 & -0.338*
                 & -0.135 &  0.056 & -0.191
                 &  0.038 & -0.387 & -0.140 \\
                 & (0.152) & (0.225) & (0.088)
                 & (0.168) & (0.225) & (0.182)
                 & (0.208) & (0.160) & (0.152)
                 & (0.228) & (0.294) & (0.404)
                 & (0.171) & (0.737) & (1.089) \\[4pt]
2010s            & -0.328*** & -0.685*** & -0.341***
                 & -0.682*** & -0.408*** & -0.219*
                 & -0.487*** & -0.426*** & -0.961***
                 & -0.245* & -0.589*** & -1.128***
                 &  0.311** &  0.272 &  1.366$^{\dagger}$ \\
                 & (0.076) & (0.126) & (0.046)
                 & (0.085) & (0.116) & (0.096)
                 & (0.114) & (0.083) & (0.083)
                 & (0.117) & (0.154) & (0.216)
                 & (0.102) & (0.403) & (0.742) \\[4pt]
Constant         & -2.876*** & -3.377*** & -2.473***
                 & -1.246*** & -2.520*** & -3.550***
                 & -1.896*** & -2.011*** & -2.623***
                 & -2.490*** & -1.870*** & -5.376***
                 & -1.026** & -2.144* & -5.176** \\
                 & (0.185) & (0.295) & (0.127)
                 & (0.228) & (0.385) & (0.311)
                 & (0.302) & (0.250) & (0.273)
                 & (0.406) & (0.468) & (0.747)
                 & (0.360) & (0.920) & (1.742) \\[6pt]
\multicolumn{16}{l}{\textit{Parameters}} \\[3pt]
$\gamma$         & -0.3767*** &  0.0715** &  0.2791***
                 & -0.4371*** & -0.1295*** &  0.1481***
                 & -0.1967*** & -0.0641** &  0.3305***
                 & -0.0276 &  0.1175** &  0.4285***
                 & -0.2654*** &  0.2938*** &  0.6954*** \\
                 & (0.0264) & (0.0258) & (0.0085)
                 & (0.0316) & (0.0275) & (0.0159)
                 & (0.0357) & (0.0218) & (0.0142)
                 & (0.0344) & (0.0380) & (0.0374)
                 & (0.0461) & (0.0653) & (0.1035) \\[4pt]
$\theta$         &  0.6864 &  2.6018 &  0.4181
                 &  0.1779 &  1.0335 &  0.3891
                 &  0.2408 &  0.3842 &  0.6101
                 & --- &  0.5368 & ---
                 & --- & --- & --- \\
                 & (0.1233) & (0.4617) & (0.0732)
                 & (0.0817) & (0.3267) & (0.1415)
                 & (0.1394) & (0.1266) & (0.1911)
                 & & (0.2858) &
                 & & & \\[6pt]
\midrule
Obs.\ (per stage) & \multicolumn{3}{c}{6,438}
                 & \multicolumn{3}{c}{2,667}
                 & \multicolumn{3}{c}{3,101}
                 & \multicolumn{3}{c}{948}
                 & \multicolumn{3}{c}{612} \\
Events S / F / NDR
                 & 972 & 390 & 2,558
                 & 660 & 373 & 560
                 & 364 & 690 & 816
                 & 309 & 215 & 117
                 & 501 & 38  & 26  \\
\bottomrule
\end{tabular}
\begin{figurenotes}
Gompertz proportional-hazard estimates for three competing risks at each development stage. Column headers: S = success (advancement), F = failure (discontinuation), N = no development reported (NDR). We include gamma shared frailty at the firm level, where the likelihood-ratio test rejects $\theta = 0$ at the 5\% level; a dash indicates that we use the plain Gompertz instead. Firm size is the within-year percentile rank of market capitalization (0--1). Therapeutic area indicators are non-mutually exclusive; the omitted category is indications not classified into the nine listed areas. The 2010s indicator equals one for stage entries from 2010 onward. The Application F model (38 events, $p = 0.20$) and the Application N model (26 events, $p = 0.14$) have limited statistical power and should be interpreted with caution. Standard errors are clustered at the firm level and are presented in parentheses. $^{***}p<0.001$, $^{**}p<0.01$, $^{*}p<0.05$, $^{\dagger}p<0.1$. Main specification: NDR as competing risk, outliers excluded at $\ell = 3$ MAD (right tail) and 1st percentile (left tail). $N = 13,766$ drug-indication-stage spells.
\end{figurenotes}
\end{table}
\end{landscape}

\begin{table}[t!!]
\centering
\caption{Robustness of Gompertz Estimates Across Specifications}
\label{tab:gompertz_robustness}
\begin{threeparttable}
\setlength{\tabcolsep}{4pt}
\textit{Panel A: Firm Size Effect on Success Hazard ($\beta_{\text{size}}$)}\\
\medskip
\begin{tabular}{llr ccccc}
\toprule
& & & \multicolumn{5}{c}{Stage} \\
\cmidrule(lr){4-8}
Spec & Description & $N$ & Discovery & Phase I & Phase II & Phase III & Application \\
\midrule
Baseline & $k{=}3$, P1, NDR competing & 13,766
  & 0.642$^{**}$ & $-$0.225 & 0.040 & 0.806$^{*}$ & 0.698$^{\ddagger}$ \\
R1 & $k{=}3$, P5, NDR competing & 13,049
  & 0.600$^{**}$ & $-$0.243 & $-$0.042 & 0.851$^{*}$ & 0.507 \\
R2 & $k{=}5$, P1, NDR competing & 14,624
  & 0.623$^{**}$ & $-$0.174 & 0.200 & 0.885$^{*}$ & 0.694$^{\ddagger}$ \\
R3 & $k{=}5$, P5, NDR competing & 13,907
  & 0.592$^{**}$ & $-$0.197 & 0.127 & 0.933$^{*}$ & 0.504 \\
\addlinespace
R4 & $k{=}3$, P1, NDR censored & 13,766
  & 0.642$^{**}$ & $-$0.225 & 0.040 & 0.806$^{*}$ & 0.698$^{\ddagger}$ \\
R5 & $k{=}3$, P1, NDR as failure & 13,766
  & 0.642$^{**}$ & $-$0.225 & 0.040 & 0.806$^{*}$ & 0.698$^{\ddagger}$ \\
R6 & $k{=}3$, P1, NDR excluded & 9,689
  & 0.650$^{**}$ & 0.044 & 0.115 & 1.021$^{**}$ & 0.967$^{**}$ \\
\bottomrule
\end{tabular}
\bigskip
    \textit{Panel B: Estimates of Values and Costs, by NDR Specifications ($\delta = 0.95$, $p_{\disc} = 0$)}
\medskip
\begin{tabular}{ll cccc cccc}
\midrule
& & \multicolumn{4}{c}{Small Firms} & \multicolumn{4}{c}{Large Firms} \\
\cmidrule(lr){3-6} \cmidrule(lr){7-10}
Spec & NDR Treatment
  & $p_{\appr|\disc}$ & $\Pi_{\disc}$ & $V_{\disc}$ & $\mathbb{E}_\disc^{Opt}(C)$
  & $p_{\appr|\disc}$ & $\Pi_{\disc}$ & $V_{\disc}$ & $\mathbb{E}_\disc^{Opt}(C)$ \\
\midrule
Baseline & Competing
  & 0.066 &    88 &    50 &    38
  & 0.085 & 1,129 &   824 &   305 \\
R4 & Censored
  & 0.027 &    23 &    50 & $-$27
  & 0.036 &   316 &   824 & $-$508 \\
R5 & As failure
  & 0.003 &     2 &    50 & $-$48
  & 0.004 &    34 &   824 & $-$790 \\
R6 & Excluded
  & 0.034 &    41 &    50 &  $-$9
  & 0.052 &   831 &   824 &     7 \\
\bottomrule
\end{tabular}
 \end{threeparttable}
\begin{figurenotes}
Panel A reports the coefficient on firm size percentile (0--1 scale) in the Gompertz success hazard at each development stage, across all seven robustness specifications. $k$ denotes the MAD multiplier for right-tail outlier exclusion; P1 and P5 denote the 1st and 5th percentile left-tail cutoffs. R4--R6 vary the treatment of NDR exits; R1--R3 vary outlier thresholds with NDR as a competing risk throughout. The success hazard coefficients for R4 and R5 are identical to the Main specification because the success hazard is estimated from advancement events only and does not depend on how NDR exits are classified. R6 (NDR excluded) excludes 4,077 NDR observations, reducing the sample and sharpening the size effect at late stages. Panel B reports Gompertz-predicted unconditional approval probabilities, present values, and implied development costs under each NDR treatment (Main outlier specification throughout)—all dollar values in millions. $V_{\disc}$ is invariant to NDR treatment because it depends only on the discovery CAR. $^{**}p<0.01$, $^{*}p<0.05$, $^{\ddagger}p<0.1$.
\end{figurenotes}
\end{table}

Panel B translates the Gompertz estimates into implied valuations under 
each NDR treatment, holding outlier thresholds at the baseline 
specification throughout. The competing-risk treatment produces the 
highest transition probabilities and the only positive implied 
development cost for small firms. For small firms, the baseline 
unconditional approval probability is 6.6\%, yielding $\Pi_{\disc} = 
\$88$ million and an implied cost of \$38 million. For large firms, 
the corresponding figures are 8.5\%, \$1,129 million, and \$305 
million.
 
Censoring NDR observations (R4) lowers the unconditional approval 
probability to 2.7\% for small firms and 3.6\% for large firms, 
compressing $\Pi_{\disc}$ to \$23 million and \$316 million 
respectively, and pushing implied costs below zero in both cases. 
Treating NDR as failure (R5) is the most conservative treatment: 
approval probabilities fall to just 0.3\% and 0.4\% for small and 
large firms, with $\Pi_{\disc}$ collapsing to \$2 million and \$34 
million and implied costs again negative. Excluding NDR observations 
entirely (R6) falls between the competing-risk and censored 
treatments, yielding approval probabilities of 3.4\% and 5.2\%, 
$\Pi_{\disc}$ of \$41 million and \$831 million, and implied costs 
of $-$\$9 million for small firms and \$7 million for large firms.
 
The sensitivity is thus largely one-sided: the competing-risk 
specification produces the most optimistic transition probabilities 
and the only positive implied cost for small firms, while R4 and R5 
push estimates downward and produce negative implied costs in both 
size groups. This ordering is expected. The competing-risk treatment 
separates NDR from failure, allowing the failure hazard to be 
estimated from genuine discontinuation events alone. The alternatives 
either ignore the information in NDR exits (censoring), conflate them 
with failures, or discard them. Because NDR exits are empirically 
associated with prolonged inactivity rather than active termination, 
the competing-risk treatment is the most economically appropriate, 
and the positive implied cost under this specification should be 
regarded as the primary estimate.

\section{Sensitivity Analysis\label{app:additional}}

In this section, we examine the sensitivity of our baseline valuation
to three modeling choices: the treatment of NDR in the competing hazard
model, the discount rate, and the degree of market anticipation of
discovery announcements. The final
subsection reports fully disaggregated valuations by size, indication,
and decade for readers interested in the underlying cross-sectional
heterogeneity.

\subsection{Sensitivity to NDR Treatment}

The treatment of NDR is the most consequential modeling choice because
it affects the estimated transition probabilities and hence the values.
Our primary specification treats NDR as a third
competing risk, which is the most agnostic treatment. The three
alternatives (censoring NDR exits, coding them as failures, or
dropping them from the sample) all produce lower estimated transition
probabilities and therefore lower $\Pi_{\text{disc}}$.
Table \ref{tab:ndr_robustness} shows that treating NDR as failure
produces the most conservative values, with $\Pi_{\text{disc}}$ of
\$2.4 million for small firms and \$33.8 million for large firms,
implying negative $\mathbb{E}^{Opt}_{\disc}(C)$, which should be interpreted as zero.

\begin{table}[t!!]
\centering
\caption{Robustness of Valuations to NDR Treatment}
\label{tab:ndr_robustness}
\begin{threeparttable}
\begin{tabular}{llcccc}
\toprule
Size & NDR Treatment & $\Pi_{\appr}$ & $\Pi_{\disc}$ & $V_{\disc}$
     & $\mathbb{E}^{Opt}_{\disc}(C)$ \\
\midrule
Small & Competing (baseline) & 2,157 & 87.9 & 50.0 & 37.9 \\
Small & Censored             & 1,639 & 23.4 & 50.0 & $\dagger$ \\
Small & As failure           & 1,264 &  2.4 & 50.0 & $\dagger$ \\
Small & Excluded             & 2,164 & 41.1 & 50.0 & $\dagger$ \\
\addlinespace
Large & Competing (baseline) & 20,837 & 1,129.1 & 824.4 & 304.7 \\
Large & Censored             & 16,001 &   316.3 & 824.4 & $\dagger$ \\
Large & As failure           & 12,515 &    33.8 & 824.4 & $\dagger$ \\
Large & Excluded             & 26,745 &   831.0 & 824.4 & 6.6 \\
\bottomrule
\end{tabular}
\end{threeparttable}
\begin{figurenotes}
All values are in millions of US dollars, and we set $\delta = 0.95$
and $p_{\disc} = 0$. ``Competing'' treats NDR as a third
competing risk (baseline); ``Censored'' treats NDR exits
as right-censored; ``As failure'' recodes NDR exits as discontinuation
events; ``Excluded'' drops NDR observations from estimation.
$V_{\disc}$ is identical across treatments because it depends
only on the discovery announcement CAR, not on the hazard model.
$\mathbb{E}^{Opt}_{\disc}(C)$ is the development cost.
$^{\dagger}$ Negative point estimate; should be interpreted as zero.
\end{figurenotes}
\end{table}

Table \ref{tab:ndr_ind} extends the NDR sensitivity to individual
therapeutic indications. The baseline (competing-risk) specification
consistently produces higher cost estimates across indications, while
the as-failure treatment produces the lowest. The qualitative ranking
across indications is broadly stable across both firm sizes. For small
firms, neoplasms, infectious diseases, and rare diseases show positive
implied costs under most NDR treatments, whereas cardiovascular
indications are negative across all specifications. The same ranking
holds for large firms, where rare, neoplasm, and infectious diseases
again top the table under the baseline specification.

\begin{table}[ht!!]
\centering
\caption{NDR Sensitivity Analysis: Development Cost, by Indication\label{tab:ndr_ind}}
\begin{threeparttable}
\begin{tabular}{l *{2}{r} *{2}{r} *{2}{r} *{2}{r}}
\toprule
 & \multicolumn{2}{c}{Baseline}
 & \multicolumn{2}{c}{Censored}
 & \multicolumn{2}{c}{As Failure}
 & \multicolumn{2}{c}{Excluded} \\
\cmidrule(lr){2-3}\cmidrule(lr){4-5}
\cmidrule(lr){6-7}\cmidrule(lr){8-9}
Indication & Small & Large & Small & Large & Small & Large & Small & Large \\
\midrule
Rare             & 389 & 9,693 & 37 & 2,544 & $\dagger$ & $\dagger$ & 71 & 4,287 \\
Neoplasm         & 208 & 3,715 &  7 &   489 & $\dagger$ & $\dagger$ & 13 & 1,046 \\
Infectious       & 137 & 2,235 & 13 & $\dagger$ & $\dagger$ & $\dagger$ & 45 & 1,010 \\
Immune           &  81 &   789 & $\dagger$ & $\dagger$ & $\dagger$ & $\dagger$ & $\dagger$ &   66 \\
Inflammatory     &   4 &   308 & $\dagger$ & $\dagger$ & $\dagger$ & $\dagger$ & $\dagger$ &   86 \\
Gastrointestinal &   3 &    74 & $\dagger$ & $\dagger$ & $\dagger$ & $\dagger$ & $\dagger$ &  118 \\
Neurological     &  32 & $\dagger$ & $\dagger$ & $\dagger$ & $\dagger$ & $\dagger$ & $\dagger$ & $\dagger$ \\
Cardiovascular   & $\dagger$ & $\dagger$ & $\dagger$ & $\dagger$ & $\dagger$ & $\dagger$ & $\dagger$ & $\dagger$ \\
\bottomrule
\end{tabular}
\end{threeparttable}
\begin{figurenotes}
$\delta = 0.95$, $p_{\disc} = 0$. Small = small firms; Large = large
firms (top 5\% by market capitalization). Each column group reports
$\mathbb{E}^{Opt}_{\disc}(C) = \Pi_\disc - V_\disc$ (millions USD) under a different treatment of NDR: ``Baseline''
treats NDR as a third competing risk (primary specification);
``Censored'' treats NDR exits as right-censored; ``As Failure''
recodes NDR as discontinuation events; ``Excluded'' drops NDR
observations from estimation. $^{\dagger}$: negative point
estimate interpreted as zero.
\end{figurenotes}
\end{table}

Table \ref{tab:ndr_size_decade} shows that the NDR sensitivity by
size and decade is consistent with the aggregate pattern. For small
firms, the competing-risk specification produces positive implied costs
in both decades, while the as-failure treatment pushes all estimates
below zero. For large firms, only the 2010s+ estimate is positive
under the competing-risk specification; the 2000s estimate is below
zero across all NDR treatments.

\begin{table}[htbp]
\centering
\caption{NDR Sensitivity Analysis: Development Cost, by Firm Type and Decade}
\label{tab:ndr_size_decade}
\begin{threeparttable}
\begin{tabular}{llcccc}
\toprule
Size & Decade & Competing & Censored & As Failure & Excluded \\
\midrule
Small & 2000s  &    26 & $\dagger$ & $\dagger$ &        33 \\
Small & 2010s+ &    47 & $\dagger$ & $\dagger$ & $\dagger$ \\
\addlinespace
Large & 2000s  & $\dagger$ & $\dagger$ & $\dagger$ & $\dagger$ \\
Large & 2010s+ & 1,254 &       334 &         6 &       480 \\
\bottomrule
\end{tabular}
\end{threeparttable}
\begin{figurenotes}
$\delta = 0.95$, $p_{\disc} = 0$. Column definitions follow
Table \ref{tab:ndr_robustness}. All values in millions of US dollars.
$^{\dagger}$ Negative point estimate; should be interpreted as zero.
\end{figurenotes}
\end{table}

\subsection{Sensitivity to the Discount Factor and Discovery 
Anticipation\label{sec:sens_rho}}

Implied costs depend on two parameters that may plausibly differ across 
firm sizes: the discount factor $\delta$ and the probability that the 
market anticipated the discovery announcement, $p_{\disc}$. 
The case for varying $\delta$ rests on the notion of systematic risk. 
Large firms hold diversified pipelines whose cash flows co-vary with the 
aggregate market because pricing pressure, regulatory tightening, and 
reduced healthcare spending affect the entire portfolio simultaneously 
and tend to occur during downturns. This covariance is not diversifiable 
and commands a risk premium above the risk-free rate. Small firms, in 
contrast, are closer to a binary bet on a single program. So, their risks are   
more idiosyncratic and, in principle, diversifiable by investors, 
suggesting a discount rate nearer the risk-free rate. However, there may also be 
common systematic factors that affect firms of all sizes and affect the discount 
factor. While determining the size-dependence of $\delta$ is beyond the scope of 
this research, we vary $\delta$ to assess its effect on our estimates.

When it comes to the discovery probability, $p_{\disc}$, because small-firm 
discoveries typically emerge from preclinical programs with no prior 
public disclosure, $p_{\disc} = 0$ is plausible for this group. Large-firm pipelines, by contrast, are extensively 
covered by sell-side analysts, making discoveries at least partially 
predictable, so $p_{\disc} > 0$ is plausible for large firms. 
Table \ref{tab:sensitivity_scenarios} reports implied costs at parameter 
combinations motivated by these economic differences, varying $\delta$ 
alone for small firms and both $\delta$ and $p_{\disc}$ for large firms.

\begin{table}[t!]
\centering
\caption{Sensitivity Analysis: Development Costs}
\label{tab:sensitivity_scenarios}
\begin{threeparttable}
\begin{tabular}{lcc}
\toprule
Parameters & $\mathbb{E}_\disc^{Opt}(C)$ & {[95\% CI]} \\
\midrule
\multicolumn{3}{l}{\textit{Panel A: Small Firms} ($p_{\disc} \equiv 0$)} \\[6pt]
$\delta = 0.95$ (baseline) &  38 & [0,  91]\\
$\delta_{rf+5}$            &  32 & [0,  82]\\
$\delta = 0.99$            &  79 & [0, 155]\\[8pt]
\multicolumn{3}{l}{\textit{Panel B: Large Firms}} \\[6pt]
$\delta = 0.95$, $p_{\disc} = 0$ (baseline) & 305 & [0, 869]\\
$\delta = 0.95$, $p_{\disc} = 0.1$          & 213 & [0, 790]\\
$\delta = 0.95$, $p_{\disc} = 0.2$          &  99 & [0, 692]\\[4pt]
$\delta_{rf+5}$, $p_{\disc} = 0$            & 223 & [0, 754]\\
$\delta_{rf+5}$, $p_{\disc} = 0.1$          & 131 & [0, 676]\\
$\delta_{rf+5}$, $p_{\disc} = 0.2$          &  17 & [0, 580]\\
\bottomrule
\end{tabular}
\end{threeparttable}
\begin{figurenotes}
All values in millions of dollars. 
Implied cost is
$\mathbb{E}_\disc^{Opt}(C) = \Pi_{\disc}(\delta) - V_{\disc}(p_{\disc})$,
where $\Pi_{\disc}$ depends only on $\delta$ and $V_{\disc}$ depends
only on $p_{\disc}$. The two effects are therefore separable. Large-firm
discoveries may be partially anticipated ($p_{\disc} > 0$) given greater
pipeline visibility. Small-firm discoveries are treated as complete
surprises ($p_{\disc} = 0$). $\delta_{rf+5}$ uses the decade-specific
ten-year real interest rate plus a five percentage point pharmaceutical
risk premium. Confidence intervals are 95\% \cite{CoxShi2023} intervals,
$[0,\, \widehat{C} + 1.645\,\widehat{\sigma}_C]$, reflecting the
non-negativity restriction on development costs.
\end{figurenotes}
\end{table}

\paragraph{Discount factor.}
For small firms, implied costs range from \$32 million to \$79 million across the three discount rate scenarios, with the baseline estimate of \$38 million sitting near the lower end. The upper bound corresponds to $\delta = 0.99$, which treats small-firm drug cash flows as nearly risk-free. 

The discount factor $\delta$ affects $\Pi_\disc$ but not $V_\disc$,
which depends only on the discovery CAR. Table \ref{tab:sens_delta}
reports $V_\disc$, $\Pi_\disc$, and the implied cost
$\mathbb{E}^{Opt}_{\disc}(C)$ for three values of $\delta$, with
aggregate results by firm size and a further breakdown by decade. As
$\delta$ rises, future cash flows are discounted less, $\Pi_\disc$
increases monotonically, and implied costs rise with it. For both firm
sizes, implied costs are positive across all three discount factors in
the aggregate specifications, confirming that the main finding is not
driven by the choice of $\delta$. The one exception is large firms in
the 2000s, where the lower approval-stage CARs of that decade produce
a negative point estimate at every discount factor; these should be
read as zero.

\begin{table}[t!!]
\centering
\caption{Sensitivity to the Discount Factor}
\label{tab:sens_delta}
\begin{threeparttable}
\setlength{\tabcolsep}{5pt}
\begin{tabular}{l r *{2}{r} *{2}{r} *{2}{r}}
\toprule
 & & \multicolumn{2}{c}{$\delta = 0.95$}
   & \multicolumn{2}{c}{$\delta = 0.99$}
   & \multicolumn{2}{c}{$\delta_{rf+5}$} \\
\cmidrule(lr){3-4}\cmidrule(lr){5-6}\cmidrule(lr){7-8}
 & $V_{\disc}$
 & $\Pi_\disc$ & $\mathbb{E}^{Opt}_{\disc}(C)$
 & $\Pi_\disc$ & $\mathbb{E}^{Opt}_{\disc}(C)$
 & $\Pi_\disc$ & $\mathbb{E}^{Opt}_{\disc}(C)$ \\
\midrule
\textit{Small firms}
  &    50.0 &    87.9 &  37.9 &   128.6 &    78.7 &    82.3 &  32.3 \\
\quad 2000s
  &    39.7 &    66.1 &  26.3 &    92.7 &    52.9 &    58.5 &  18.8 \\
\quad 2010s+
  &    55.0 &   101.8 &  46.7 &   151.7 &    96.6 &    97.6 &  42.5 \\
\addlinespace
\textit{Large firms}
  &   824.4 & 1,129.1 & 304.7 & 1,609.8 &   785.5 & 1,047.3 & 222.9 \\
\quad 2000s
  & 1,577.7 & 1,054.3 & $\dagger$ & 1,456.7 & $\dagger$ &   938.4 & $\dagger$ \\
\quad 2010s+
  &    19.9 & 1,273.8 & 1,253.9 & 1,865.4 & 1,845.5 & 1,223.4 & 1,203.5 \\
\bottomrule
\end{tabular}
\end{threeparttable}
\begin{figurenotes}
$p_{\disc} = 0$. Primary NDR specification. All values in
millions of US dollars. $V_{\disc}$ does not vary with
$\delta$. $\Pi_\disc$ is the present value of gross profits
at discovery, computed by chaining transition probabilities and
expected discount factors across all five development stages.
$\mathbb{E}^{Opt}_{\disc}(C)$ is the development cost.
$\delta_{rf+5}$ is a decade-specific discount factor equal to
one over one plus the ten-year real interest rate plus a five
percentage point pharmaceutical risk premium
\citep{KoijenPhilipsonUhlig2016}. The aggregate rows pool across
decades; the indented rows disaggregate by decade of stage entry.
$^{\dagger}$ Negative point estimate; should be interpreted as zero.
\end{figurenotes}
\end{table}

\paragraph{Discovery anticipation.}
For large firms, implied costs are more sensitive to the assumed parameters, ranging from \$17 million to \$305 million (Table \ref{tab:sensitivity_scenarios}). The baseline of $\delta = 0.95$ and $p_{\disc} = 0$ yields \$305 million. Allowing for partial anticipation reduces this substantially: $p_{\disc} = 0.1$ brings the estimate to \$213 million, and $p_{\disc} = 0.2$ to \$99 million; combining the risk-adjusted rate $\delta_{rf+5}$ with $p_{\disc} = 0.1$ yields \$131 million.

The parameter $p_\disc \in [0, 1)$ governs the fraction of the drug's
value that the market has already incorporated before the discovery
announcement. A higher $p_\disc$ implies the announcement was more
anticipated, which increases $V_\disc$ and compresses the implied cost
toward zero. Table \ref{tab:sens_rho0_v1} reports $V_\disc$ across
the grid $p_\disc \in \{0, 0.1, 0.2, 0.3, 0.4, 0.5\}$ for both size groups, 
including small firms, for completeness. For small
firms, $V_{\text{disc}}$ rises from \$50.0 million at $p_\disc = 0$
to \$99.9 million at $p_{\text{disc}} = 0.5$. For large firms, the
corresponding values are \$824.4 million and \$1.649 billion. The
implied cost $\mathbb{E}^{Opt}_{\disc}(C) = \Pi_\disc - V_\disc$
remains positive for small firms throughout this range, and for large
firms up to $p_\disc \approx 0.2$, beyond which $V_\disc$ exceeds
$\Pi_\disc$ and the point estimate crosses zero. Given that discovery
announcements represent the first public signal identifying the specific candidate, values of $p_\disc$ above 0.2 are difficult to justify
economically. The positive cost result is therefore robust throughout
the plausible range.

\begin{table}[t!!]
\centering
\caption{Sensitivity of $V_\disc$ to Discovery Anticipation}
\label{tab:sens_rho0_v1}
\begin{threeparttable}
\begin{tabular}{lcccccc}
\toprule
Size & $p_{\disc} = 0$ & $p_{\disc} = 0.1$ & $p_{\disc} = 0.2$
     & $p_{\disc} = 0.3$ & $p_{\disc} = 0.4$ & $p_{\disc} = 0.5$ \\
\midrule
Small &  50.0 &  55.5 &   62.4 &    71.4 &    83.3 &    99.9 \\
Large & 824.4 & 916.0 & 1,030.5 & 1,177.7 & 1,374.0 & 1,648.8 \\
\bottomrule
\end{tabular}
\end{threeparttable}
\begin{figurenotes}
$\delta = 0.95$. Primary NDR specification. $V_\disc =
(\E[\text{CAR}_\disc] \times \MKTCAP_\disc) / (1 - p_{\disc})$,
where the numerator is the cell-mean dollar CAR at discovery, and the
denominator adjusts for the fraction of value already incorporated.
All values in millions of US dollars.
\end{figurenotes}
\end{table}

This sensitivity analysis also applies to the sales-based comparison of
Section~\ref{sec:valuation_results}. Table~\ref{tab:sales_comparison} capitalizes average annual sales as a perpetuity at
$\delta = 0.95$. Table~\ref{tab:sales_capitalization} reports two
alternatives: (i) a finite 12-year patent life at the same discount factor, which
lowers the capitalization factor from 20 to 9.2, and (ii) a perpetuity at
$\delta = 0.99$, which raises it to 100. The choice of capitalization
assumption scales all numbers proportionally and therefore leaves the
comparison between firm sizes unchanged: the ratio of large- to small-firm
sales-based $\Pi_{\appr}$ is 2.83 in all cases, since it depends only on the
underlying sales distributions. 
However, under alternative (i), sales-implied values fall by more than half, so our estimate of
$\Pi_{\appr}$ for large firms sits further above what sales alone would suggest, and under alternative (ii) it sits below. The baseline perpetuity at
$\delta = 0.95$ gives the intermediate value and is the one we report in the main
text.
 
\begin{table}[t!]
\centering
\caption{Sales-Based Profit under Alternative Capitalization Assumptions}
\label{tab:sales_capitalization}
\begin{threeparttable}
\small
\setlength{\tabcolsep}{5pt}
\begin{tabular}{l ccc ccc}
\toprule
& \multicolumn{3}{c}{Small firms} & \multicolumn{3}{c}{Large firms} \\
\cmidrule(lr){2-4}\cmidrule(lr){5-7}
Margin & 40\% & 45\% & 50\% & 40\% & 45\% & 50\% \\
\midrule
\multicolumn{7}{l}{\textit{Panel A: 12-year patent life (\(\delta = 0.95\), factor = 9.2)}} \\
\addlinespace
Mean drug &        1,707 &        1,921 &        2,134 &        4,706 &        5,294 &        5,882 \\
Median drug &          386 &          434 &          483 &        3,222 &        3,625 &        4,028 \\
p90 drug &        4,290 &        4,827 &        5,363 &        9,802 &       11,028 &       12,253 \\
\addlinespace
\multicolumn{7}{l}{\textit{Panel B: Perpetuity (\(\delta = 0.99\), factor = 100)}} \\
\addlinespace
Mean drug &       18,573 &       20,895 &       23,217 &       51,190 &       57,589 &       63,988 \\
Median drug &        4,201 &        4,726 &        5,252 &       35,054 &       39,436 &       43,818 \\
p90 drug &       46,670 &       52,504 &       58,338 &      106,632 &      119,961 &      133,290 \\
\midrule
\(N\) drugs & \multicolumn{3}{c}{184} & \multicolumn{3}{c}{122} \\
\bottomrule
\end{tabular}
\end{threeparttable}
\begin{figurenotes}
Sample and construction as in Table \ref{tab:sales_comparison}. Sales-based
\(\Pi_{\appr} = \text{margin} \times \text{avg annual sales} \times
\text{capitalization factor}\). The 12-year annuity factor is
\((1-\delta^{12})/(1-\delta)\); the perpetuity factor is \(1/(1-\delta)\).
Panel A replaces the perpetuity assumption of Table \ref{tab:sales_comparison}
with a finite 12-year patent life at the baseline discount factor; Panel B
retains the perpetuity assumption and raises the discount factor to
\(\delta = 0.99\). All values in millions of US dollars.
\end{figurenotes}
\end{table}

\subsection{Disaggregated Valuations}

Table \ref{tab:disagg_valuations} reports fully disaggregated
valuations by firm size, therapeutic indication, and decade. The
aggregate rows confirm the baseline results. For small firms, the
qualitative ranking across indications is consistent with the main
text: neoplasms and rare diseases carry the highest approval values
and implied costs, while cardiovascular indications show negative
implied costs in both decades, which should be read as zero,
consistent with their low approval probabilities and modest
approval-stage CARs. For large firms, the ranking across indications
is broadly similar, but approval values and implied costs are
substantially larger, driven by larger market capitalizations rather
than higher approval probabilities. The most striking entry is rare
diseases for large firms, where $\Pi_\appr$ exceeds \$110 billion in
both decades and the implied cost reaches \$9.5 billion in the 2010s+,
reflecting the combination of high approval CARs and large market
capitalizations in that indication-decade cell.

\begin{table}[t!!]
\centering
\caption{Disaggregated Drug Valuations by Indication, Firm Size, and Decade}
\label{tab:disagg_valuations}
\setlength{\tabcolsep}{4pt}
\renewcommand{\arraystretch}{0.95}
\begin{threeparttable}
\begin{tabular}{l *{2}{r} *{2}{r} *{2}{r} *{2}{r} *{2}{r}}
\toprule
 & \multicolumn{2}{c}{$p_{\appr|\disc}$}
 & \multicolumn{2}{c}{$\Pi_\appr$}
 & \multicolumn{2}{c}{$\Pi_\disc$}
 & \multicolumn{2}{c}{$V_\disc$}
 & \multicolumn{2}{c}{$\mathbb{E}^{Opt}_{\disc}(C)$} \\
\cmidrule(lr){2-3}\cmidrule(lr){4-5}\cmidrule(lr){6-7}
\cmidrule(lr){8-9}\cmidrule(lr){10-11}
 & Small & Large & Small & Large & Small & Large & Small & Large & Small & Large \\
\midrule
\textit{Aggregate} & & & & & & & & & & \\
\quad 2000s
  & 0.0680 & 0.0849 & 1,484 & 18,616 &  66 & 1,054 & 40 & 1,578 & 26 & $\dagger$ \\
\quad 2010s+
  & 0.0649 & 0.0843 & 2,590 & 24,397 & 102 & 1,274 & 55 &    20 & 47 & 1,254 \\
\addlinespace
\multicolumn{11}{l}{\textit{By Indication}} \\\addlinespace[2pt]
Cardiovascular & & & & & & & & & & \\
\quad 2000s
  & 0.0351 & 0.0492 &   727 & 12,562 & 17 & 410 & 46 & 1,768 & $\dagger$ & $\dagger$ \\
\quad 2010s+
  & 0.0328 & 0.0447 & 2,229 & 14,178 & 44 & 393 & 48 &    23 & $\dagger$ & 370 \\
\addlinespace
Gastrointestinal & & & & & & & & & & \\
\quad 2000s
  & 0.0673 & 0.0863 &   876 & 17,011 & 40 & 1,010 & 49 & 1,664 & $\dagger$ & $\dagger$ \\
\quad 2010s+
  & 0.0623 & 0.0842 & 1,778 & 14,565 & 69 &   786 & 62 &    20 & 7 & 766 \\
\addlinespace
Immune & & & & & & & & & & \\
\quad 2000s
  & 0.1098 & 0.1338 &   664 &  9,841 &  48 &   883 & 48 & 1,235 & 0 & $\dagger$ \\
\quad 2010s+
  & 0.1113 & 0.1331 & 2,833 & 19,790 & 192 & 1,642 & 72 &    20 & 120 & 1,622 \\
\addlinespace
Infectious & & & & & & & & & & \\
\quad 2000s
  & 0.1348 & 0.1727 & 1,882 & 32,072 & 170 & 3,828 & 38 & 2,082 & 132 & 1,746 \\
\quad 2010s+
  & 0.1217 & 0.1465 & 2,341 & 32,161 & 178 & 3,047 & 38 &    20 & 140 & 3,027 \\
\addlinespace
Inflammatory & & & & & & & & & & \\
\quad 2000s
  & 0.0757 & 0.0970 &   763 & 14,250 & 39 &   944 & 46 & 1,391 & $\dagger$ & $\dagger$ \\
\quad 2010s+
  & 0.0639 & 0.0957 & 1,729 & 19,441 & 69 & 1,195 & 64 &    21 & 5 & 1,174 \\
\addlinespace
Neoplasm & & & & & & & & & & \\
\quad 2000s
  & 0.0674 & 0.0884 & 5,743 &  58,933 & 244 & 3,337 & 36 & 1,322 & 208 & 2,015 \\
\quad 2010s+
  & 0.0644 & 0.0830 & 6,687 &  99,544 & 253 & 4,940 & 52 &    19 & 201 & 4,921 \\
\addlinespace
Neurological & & & & & & & & & & \\
\quad 2000s
  & 0.0463 & 0.0542 & 2,263 & 15,039 & 70 & 552 & 31 & 1,530 & 39 & $\dagger$ \\
\quad 2010s+
  & 0.0453 & 0.0513 & 3,090 & 19,272 & 87 & 626 & 55 &    20 & 32 & 605 \\
\addlinespace
Rare & & & & & & & & & & \\
\quad 2000s
  & 0.1120 & 0.1395 & 3,654 & 111,123 & 267 & 10,097 & 27 & 1,201 & 240 & 8,896 \\
\quad 2010s+
  & 0.1038 & 0.1352 & 8,479 & 114,780 & 534 &  9,478 & 79 &    21 & 456 & 9,457 \\
\addlinespace
Other & & & & & & & & & & \\
\quad 2000s
  & 0.0447 & 0.0634 &   915 & 15,299 & 27 & 642 & 44 & 1,559 & $\dagger$ & $\dagger$ \\
\quad 2010s+
  & 0.0408 & 0.0564 & 1,141 & 11,931 & 28 & 416 & 48 &    20 & $\dagger$ & 396 \\
\bottomrule
\end{tabular}
\end{threeparttable}
\begin{figurenotes}
$\delta = 0.95$, $p_{\disc} = 0$. Primary NDR specification.
All dollar values in millions of US dollars. $p_{\appr|\disc}$ is
the probability of reaching FDA approval from discovery, computed
by chaining stage-specific conditional probabilities from the
Gompertz models. $\Pi_{\appr}$ is the expected present value
of gross profits conditional on approval. $\Pi_{\disc}$ is the
risk-adjusted present value at discovery, equal to $p_{\appr|\disc} \times
\mathbb{E}[\delta^{\tau}] \times \Pi_{\appr}$. $V_{\disc}$ is
the market value of the drug program at discovery, estimated from
announcement CARs. $\mathbb{E}^{Opt}_{\disc}(C)$ is the implied
development cost. $^{\dagger}$ Negative point estimate; should
be interpreted as zero (see Section \ref{sec:valuation_results}).
The aggregate rows pool across all indications; the indented rows
disaggregate by decade of stage entry.
\end{figurenotes}
\end{table}

\section{Public vs. Private Firms\label{app:public_private}}

Our identification strategy relies on using the stock market events to estimate the change in the market value of a firm following an announcement, which restricts the estimation sample 
to drugs developed by publicly traded firms. In this appendix, we assess whether this restriction biases our estimated transition 
probabilities, and consequently our drug valuations, by comparing 
Gompertz competing-risk models estimated on the full Cortellis database 
with those estimated on the publicly traded subsample alone.

Table \ref{tab:selection_sample} summarizes the composition of both 
samples. After applying the same outlier exclusion criteria used in the 
main analysis, the full sample contains 56,025 drug-stage observations, 
of which 13,766 (24.6\%) correspond to publicly traded firms. Public 
firms advance drugs more quickly through the pipeline--median 
durations are shorter at every stage, most notably at discovery (2.49 
vs.\ 3.13 years)--and are disproportionately concentrated in 
oncology (39.7\% vs.\ 32.3\%) and rare diseases (12.5\% vs.\ 9.4\%), 
while private firms are more active in infectious diseases 
(12.9\% vs.\ 9.0\%) and neurological conditions (14.1\% vs.\ 11.5\%).

\begin{table}[ht!!]
\centering
\caption{Sample Composition: All Firms vs.\ Public and Private Firms}
\label{tab:selection_sample}
\begin{threeparttable}
\scalebox{0.9}{\begin{tabular}{lccc}
\toprule
 & Full Sample & Public & Private \\
\midrule
\textit{Observations by stage} \\
\quad Discovery        & 33,121 &  6,438 & 26,683 \\
\quad Phase I          &  8,444 &  2,667 &  5,777 \\
\quad Phase II         &  9,549 &  3,101 &  6,448 \\
\quad Phase III        &  3,215 &    948 &  2,267 \\
\quad Application &  1,696 &    612 &  1,084 \\
\addlinespace
\quad Total            & 56,025 & 13,766 & 42,259 \\
\addlinespace
\textit{Median duration (years)} \\
\quad Discovery        &   3.00 &   2.49 &   3.13 \\
\quad Phase I          &   2.29 &   1.89 &   2.49 \\
\quad Phase II         &   2.65 &   2.42 &   2.75 \\
\quad Phase III        &   2.22 &   2.14 &   2.25 \\
\quad Application &   0.83 &   0.78 &   0.87 \\
\addlinespace
\textit{Therapeutic area (\%)} \\
\quad Neoplasm         &  34.1  &  39.7  &  32.3  \\
\quad Neurological     &  13.5  &  11.5  &  14.1  \\
\quad Infectious       &  11.9  &   9.0  &  12.9  \\
\quad Immune           &  11.4  &  13.2  &  10.8  \\
\quad Gastrointestinal &  11.2  &  13.1  &  10.6  \\
\quad Rare             &  10.2  &  12.5  &   9.4  \\
\quad Inflammatory     &   9.9  &  10.5  &   9.7  \\
\quad Cardiovascular   &   6.5  &   5.4  &   6.8  \\
\bottomrule
\end{tabular}}
\end{threeparttable}
\begin{figurenotes}
Observations after applying the same duration 
outlier exclusion (MAD-based right tail, percentile-based left tail) 
used in the main analysis. We identify publicly traded firms by 
successful merger with the stock market data used in the CAR analysis. 
Therapeutic area percentages sum to more than 100\% because drugs may 
target multiple indications.
\end{figurenotes}
\end{table}

To isolate the effect of sample selection on transition dynamics, we 
estimate identical Gompertz competing-risk models on three samples: 
(i) the full Cortellis sample, (ii) the publicly traded subsample, and 
(iii) private firms only. All three models use the same specification 
--therapeutic area indicators and a decade dummy as covariates-- 
with NDR treated as a competing risk. We exclude firm size and shared 
frailty from these models because market capitalization is unavailable 
for private firms, ensuring that any difference in the estimated transition probabilities reflects sample composition rather than specification 
differences. For each sample, we compute the linear predictor at 
sample-mean covariates, numerically integrate the Gompertz 
cause-specific hazards over a 30-year horizon to obtain stage-specific 
transition probabilities $p_{k|k-1}$, and chain these across all five 
stages to obtain the cumulative probability of approval from discovery.

Table \ref{tab:selection_rho} presents the results. The most notable difference is at the discovery stage, where publicly traded firms have 
a substantially higher transition probability to Phase I (0.678 vs.\ 
0.571 in the full sample), likely reflecting both selection effects to the extent that  
firms with more promising pipelines are more likely to seek public 
equity financing, and resource advantages that enable faster 
advancement of early-stage candidates. Differences at later stages are 
smaller and tend to go in the opposite direction, with transition rates 
at Phase I through Phase III, modestly higher in the full sample. These 
stage-level differences largely offset when chained across all five 
stages: the cumulative probability of approval from discovery is 6.6\% 
in the full sample, 7.0\% in the publicly traded subsample, and 6.8\% 
among private firms. 
Selection into the publicly traded sample, therefore, does not materially 
affect the transition dynamics underlying our drug valuations.

\begin{table}[ht!!]
\centering
\caption{Transition Probabilities: All Firms vs.\ Public Firms}
\label{tab:selection_rho}
\begin{threeparttable}
\scalebox{0.9}{\begin{tabular}{lcccc}
\toprule
 & Full Sample & Public & Private & Difference (Full Sample $-$ Public) \\
\midrule
\textit{Stage-specific} \\
\quad Discovery $\to$ Phase I     & 0.571 & 0.678 & 0.542 & $-$0.107 \\
\quad Phase I $\to$ Phase II      & 0.609 & 0.586 & 0.623 & $+$0.023 \\
\quad Phase II $\to$ Phase III    & 0.363 & 0.328 & 0.384 & $+$0.035 \\
\quad Phase III $\to$ Application     & 0.576 & 0.564 & 0.579 & $+$0.012 \\
\quad Application $\to$ Approval      & 0.910 & 0.956 & 0.898 & $-$0.046 \\
\addlinespace
\quad $p_{\appr|\disc}$ & 0.066 & 0.070 & 
0.068 & $-$0.004 \\
\bottomrule
\end{tabular}}
\end{threeparttable}
\begin{figurenotes}
Transition probabilities from Gompertz 
competing-risk models (no frailty, no firm size) with therapeutic area 
dummies and a decade indicator, evaluated at sample-mean covariates. 
NDR is treated as a competing risk. Integration horizon $30$ years. 
The probability of success from discovery stage, $p_{\appr|\disc}$, is the product of all five stage-specific 
transition probabilities.
\end{figurenotes}
\end{table}
\end{appendix}

\end{document}